 \documentclass[final,1p,times]{elsarticle}

\usepackage{graphicx}
\usepackage[usenames,dvipsnames]{color}

\usepackage{dcolumn}
\usepackage{bm}
\usepackage{color}
\usepackage[normalem]{ulem}
\usepackage{empheq}

  \usepackage{amsmath} 
  \usepackage{amssymb} 
  \usepackage{amsthm}
  
\usepackage{mathtools}
\usepackage{color}

\newcommand{\be}{\begin{equation}}
\newcommand{\ee}{\end{equation}}
\newcommand{\bea}{\begin{eqnarray}}
\newcommand{\eea}{\end{eqnarray}}

\newcommand{\ihddt}{i\hbar\frac{\partial}{\partial t}}

\newcommand{\op}[1]{\hat{#1}}
\newcommand{\ave}[1]{\langle #1 \rangle}
\newcommand{\comm}[2]{\left[ #1,\,#2 \right]_{-}}
\newcommand{\trace}[1]{{\rm Tr}\left[ #1 \right]}

\newcommand{\absN}[1]{{| #1 \rangle}}

\newcommand{\sumA}{{\left(\alpha^\star|\hat{B} \right)}}
\newcommand{\sumAS}{{\left(\hat{B}^\dagger|\alpha \right)}}

\newcommand{\ex}[1]{{#1}_{\rm ex}}

\newcommand{\opNC}{{\op{N}_{\rm C}}}
\newcommand{\opNN}{{\op{N}_{\rm N}}}

\newcommand{\toX}{\xrightarrow{\rm ex}}
\newcommand{\NC}{N_{\rm C}}
\newcommand{\NN}{N_{\rm N}}
\newcommand{\nuN}{F_{\rm N}}
\newcommand{\sel}{{\rm Sel}}

\newcommand{\avex}[1]{{\langle #1 \rangle}_{\rm ex}}

\newcommand{\DI}[2]{\Delta{}I^{#1}_{#2}}
\newcommand{\Dstr}[2]{\Delta^{#1}_{#2}}
\newcommand{\seq}[1]{{\{#1\}}}
\newcommand{\srm}[2]{\{#1\, \big|\, #2\}}
\newcommand{\ssel}[2]{\{#1\}(#2)}

\newcommand{\ind}[1]{{[#1]}}
\newcommand{\Dcoll}[2]{{\cal D}^{#1}_{#2}}
\newcommand{\sS}[2]{S^{#1}_{#2}}
\newcommand{\sL}[2]{L^{#1}_{#2}}
\newcommand{\sQ}[2]{Q^{#1}_{#2}}
\newcommand{\sC}[2]{C^{#1}_{#2}}

\newcommand{\str}[1]{{\rm Str}\left[#1\right]}

\journal{Annals of Physics}

\begin{document}

\begin{frontmatter}



\title{Hyperbolic Bloch equations: \\atom-cluster kinetics of an interacting Bose gas}
\author{M.~Kira}

\address{Department of Physics, Philipps-University Marburg, Renthof 5, D-35032
Marburg, Germany}

\begin{abstract}
Experiments with ultracold Bose gases can already produce so strong atom--atom interactions that one can observe intriguing many-body dynamics between the Bose-Einstein condensate (BEC) and the normal component. The excitation picture is applied to uniquely express
the many-body state uniquely in terms of correlated atom clusters within the normal component alone. Implicit notation formalism is developed to {\it explicitly} derive the quantum kinetics of {\it all} atom clusters. The clusters are shown to build up sequentially, from smaller to larger ones, which is utilized to nonperturbatively describe the interacting BEC with as few clusters as possible. This yields the hyperbolic Bloch equations (HBEs) that not only generalize the Hartree-Fock Bogoliubov approach but also are  analogous to the semiconductor Bloch equations (SBEs). 
This connection is utilized to apply sophisticated many-body techniques of semiconductor quantum optics to BEC investigations. Here, the HBEs are implemented to determine how a strongly interacting Bose gas reacts to a fast switching from weak to strong interactions, often referred to as unitarity. The computations for $^{35}$Rb demonstrate that molecular states (dimers) depend on atom density, and that the many-body interactions create coherent 
transients on a 100\,$\mu$s time scale 
converting BEC into normal state via quantum depletion. 
\end{abstract}


\end{frontmatter}


\section{Introduction}
\label{sec:intro}

Ultrafast spectroscopy\cite{Zewial:1988,Steinmeyer:1999,Shah:1999,Jones:2000,Weiner:2000,Hentschel:2001,Huber:2001,Kira:2006b,Krausz:2009,Cundiff:2012} is originally a concept of optics that is based on exciting and probing matter with laser pulses. When such pulses switch on and off the light--matter interaction on a time scale much faster than the fundamental scattering processes, one can monitor\cite{Gobel:1990,Heberle:1994} 
and even control quantum processes\cite{Warren:1993,Heberle:1995,Hache:1997,Wehner:2001} on their natural time scales. In practice, one  typically uses picosecond to attosecond laser pulses depending on how fast the studied phenomena are. By monitoring the excitation dynamics that follows, one can directly investigate, e.g., the progress of chemical reactions\cite{Zewail:1994,Assion:1998}. 
Ideally, this method can reveal how one can control the coherent dynamics, governed by the Schr\"{o}dinger equation, in oder to influence the late-time phenomena despite the onset of inevitable 
relaxation and thermalization. This approach is particularly interesting in semiconductor optics\cite{Knorr:1992,Axt:1998,Khitrova:1999,Chemla:2001,Rossi:2002,Koch:2006,Kira:2006,Li:2006,Haug:2008,Smith:2010,Cundiff:2012,Malic:2013} and quantum optics\cite{Kira:2006b,Book:2011,Jahnke:2012} because semiconductor 
excitations can exhibit a great variety of quantum processes that proceed on different time scales. As hallmarks of genuine quantum kinetics, perturbative as well as thermodynamic approaches fail\cite{Cundiff:1994,Kira:1999,Chatterjee:2004,Khitrova:2006}, coherent control\cite{Warren:1993,Heberle:1995,Hache:1997,Wehner:2001} becomes possible, and quantum beats\cite{Gobel:1990,Heberle:1994} or even the quantum fluctuations of light\cite{Kira:2011,Mootz:2014,Hunter:2014} can influence generation of new states and configurations.

Ultrafast control of many-body systems offers the possibility to explore how elementary quantum events translate into macroscopic phenomena, which is probably the most challenging problem of contemporary physics. Therefore, it is most fruitful to engage an approach I call manybodypedia, a synergistic learning process where different aspects of many-body quantum physics are systematically analyzed from a point of view of seemingly different systems. 
We are very much experiencing this progress already: a lattice of ultracold atoms is applied to simulate \cite{Jaksch:1998,Greiner:2002,Bloch:2008,Bakr:2009} solid-state models, atomic Bose-Einstein condensation is being pursued\cite{Snoke:1990,Kasprzak:2006,Stolz:2012} with semiconductors, Efimov physics\cite{Efimov:1973} of nuclei are studied in interacting Fermi\cite{DeMarco:1999,Giorgini:2008,Castin:2010} as well as Bose\cite{Kraemer:2006,Stecher:2009,Williams:2009,Pollack:2009,Castin:2010,Wild:2012}
 gas of atoms, only a few examples to mention. 

In this context, atomic Bose-Einstein condensates (BECs)\cite{Anderson:1995,Bradley:1995,Davis:1995}  are perhaps the cleanest example manifesting macroscopic quantum phenomena. Since such a Bose gas contains interacting atoms, it clearly is interesting to study what kind of quantum kinetics results from the atom--atom interactions. 
In the spirit of manybodypedia, one can use the ideas of ultrafast spectroscopy because one can change the strength of atom--atom interactions faster than relaxation by sweeping the system through a Feshbach resonance\cite{Feshbach:1958,Stwalley:1976,Tiesinga:1993,Inouye:1998,Courteille:1998,Roberts:1998,Kohler:2006,Chin:2010} with an external magnetic field. 
In particular, recent experiments\cite{Donley:2001,Altin:2011,Makotyn:2014} in interacting Bose gas demonstrate that one can switch the interactions, on a time scale much faster than essential scattering rates. This means that one can start exploring analogous nonequilibrium many-body quantum kinetics as that performed in 
semiconductors using ultrafast spectroscopy.\cite{Book:2011} 
Obviously, the time scales of interacting Bose gas are very different from those encountered in ordinary ultrafast spectroscopy because fast Bose gas processes proceed on a submillisecond time scale due to extremely slow motion of ultracold atoms, in contrast to subpicosecond times for quantum processes in solids.

In this paper, I will develop a general framework to describe many-body quantum kinetics of interacting Bose gas following a fast switching of interactions. The introduced method is based on the cluster-expansion approach presented in Refs.~\cite{Wyld:1963,Fricke:1996,Kira:2006,Kira:2008,Book:2011} investigating ultrafast as well as quantum-optical semiconductor spectroscopy. 
Cluster-expansion based approaches are also used in quantum chemistry\cite{Cizek:1966,Bartlett:2007} in the form of the coupled-clusters approach that was originally developed to explore the many-body physics of nuclei\cite{Coester:1958,Coester:1960}. Therefore, it is fair to say that the cluster-expansion is a genuine manybodypedia approach. 
In general, it can be applied to systematically identify the correlated particle clusters within any interacting many-body system. Even more so, it can be systematically applied to include the dynamics among {\it all} particle clusters up to a user-defined particle number $N$. 
Physically, an $N$-particle clusters correspond to molecular states as well as correlated transition amplitudes. Therefore, the cluster expansion provides a natural way to identify stable cluster configurations in many-body systems.\cite{Book:2011}

Even though pairwise many-body interactions induce the unsolvable BBKGY hierarchy problem\cite{Book:2011}, $N$-particle clusters can build up only from lower level clusters sequentially, not directly as discussed in Refs.~\cite{Kira:2006b,Kira:2006,Book:2011,Mootz:2012} The cluster-expansion follows this formation dynamics systematically and it becomes a numerically feasible approach when the system can be described with few clusters. 
Since atoms inside a BEC are correlated to all orders\cite{Kira:Xpic}, it clearly is challenging to formulate an efficient cluster-expansion approach for an interacting Bose gas. I will show in Sec.~\ref{sec:Xpicture} that this can be realized by using the excitation picture, introduced already in Ref.~\cite{Kira:Xpic}, which focuses the analysis on the normal-component excitation around the BEC. 
This approach not only eliminates the need to follow the BEC-related clusters, but it reveals that the particle interactions create clusters to the normal-component in the same way as 
they would do in 
semiconductors. 

Following the terminology of Ref.~\cite{Book:2011}, quantum statistics refers to any representation that uniquely defines all quantum properties of the many-body system. For example, the density matrix and the Wigner function are equivalent representations of quantum statistics\cite{Kira:2008}. 
Formal aspects of dynamics related to quantum statistics are introduced in Sec.~\ref{sec:chi_dyn}. To derive the explicit quantum dynamics of {\it all} clusters with one derivation, I introduce the implicit notation formalism (INF) in Sec.~\ref{sec:INF}. The general cluster equations are presented in Sec.~\ref{sec:cluster_dyn} main consequences of cluster dynamics are presented in Sec.~\ref{sec:QK_doublets}.

In a Bose gas, interactions can convert BEC into normal-component atoms, and the process is often referred to as quantum depletion\cite{Xu:2006,Wuster:2007,Cui:2013} to contrast it with thermal depletion induced by thermal fluctuations. At the moment, it is not yet very well known how Bose BECs behave when interactions become strong enough. 
For example, when the BEC collapses in the so-called Bosenova experiments\cite{Donley:2001} within time $t_{\rm collapse}$, the traditional Hartree-Fock Bogoliubov approach (HFB)\cite{Baranger:1961,Goodman:1974,Mang:1975,Zaremba:1999,Bender:2003,Milstein:2003, Wuster:2005} predict up to 100\% too large $t_{\rm collapse}$. 
Detailed comparisons of HFB and the truncated Wigner approximation\cite{Sinatra:2002,Norrie:2006} (TWA) was performed in Ref.~\cite{Wuster:2007} with the conclusion that atom-atom correlations must be systematically included to the HFB to describe strongly interacting Bose gas. I will develop a general cluster-expansion based framework for quantum 
kinetics of clusters. These investigations yield the hyperbolic Bloch equations (HBE), introduced in Sec.~\ref{sec:QK_doublets} as the generalization to the HFB. 
They are analogous to the semiconductor Bloch equations\cite{Lindberg:1988,Haug:2008,Book:2011} which provide the backbone of understanding how electron--hole pair excitations of solids interact via the Coulomb interaction. I will show in Sec.~\ref{sec:numerics} that the HBEs are tailor made to analyze the nonequilibrium kinetics of quantum depletion in fast-switch experiments.

The strongest atom--atom interaction is observed at the so-called unitarity\cite{Cornish:2000,Braaten:2006,Giorgini:2008,Papp:2008,Navon:2011,Smith:2012,Wild:2012}. In Sec.~\ref{sec:numerics}, I will present numerical HBE solutions at the unitarity. I will show that Bose BEC can survive even at the unitarity and that one must include interaction effects beyond the Bogoliubov 
excitations\cite{StamperKurn:1999,Jin:1996,Mewes:1996,Utsunomiya:2008} to properly describe the quantum kinetics of the quantum depletion. 
Following the ideology of manybodypedia, I will also summarize learned aspect as well as many-body connections between semiconductors and interacting Bose gas. Throughout the derivations and analysis, I will use ``we'' instead of ``I'' to encourage reader's participation to the theory development.

\section{Interacting Bose gas in the excitation picture}
\label{sec:Xpicture}

We concentrate on describing the many-body aspects of the interacting Bose gas by introducing explicit, physically-motivated, simplifications. In typical experiments, atoms are confined inside a trap whose size is much larger than atom size. Therefore, the many-body interactions take place in a region where the atom gas remains essentially homogeneous. 
In addition, atoms may escape from the trap e.g., via the three-Body loss\cite{Burt:1997,Esry:1999,Kim:2004,Brodsky:2006,Daley:2009,Laurent:2013} that is physically analogous to the loss-generating Auger scattering\cite{Klimov:2000,Hader:2005,Robel:2009} in solids. However, the related scattering rates are typically in the range of milliseconds\cite{Makotyn:2014} even for strongly interacting atom gas, 
while the relevant atom--atom interactions\cite{Donley:2001,Altin:2011,Makotyn:2014}  proceed on a few microsecond timescale. Since we here want to fully concentrate on the many-body quantum kinetics, we analyze the quantum kinetics of atoms inside a homogeneous region, in the spirit of local-density approximation, and ignore the atom loss. 
The kinetics of inhomogeneous atom profile and atom loss has been succesfully included, e.g., with the HFB and TWA methods discussed above; this prior knowhow can be used as a solid basis to genralize the HBE approach further in the future.

\subsection{Explicit operator transformations}
\label{sec:Xtrafo}

The homogeneous Bose gas can most conveniently be described by introducing the boson operators $B_{\bf k}$ and $B^\dagger_{\bf k}$ for each atom having momentum $\hbar{\bf k}$ where ${\bf k}$ is the wave vector. These operators satisfy the usual boson commutation relations:
\begin{align}
  \comm{B_{\bf k}}{B^\dagger_{{\bf k}'}} = \delta_{{\bf k},{\bf k}'}\,,
  \qquad
  \comm{B_{\bf k}}{B_{{\bf k}'}} = 0 = \comm{B^\dagger_{\bf k}}{B^\dagger_{{\bf k}'}}
\label{eq:boson-comm}\,.
\end{align}
The BEC is then formed at the lowest momentum state $B_0$ while $B_{{\bf k} \neq 0}$ identify the normal-component atoms that have the kinetic energy 
\begin{align}
  E_{\bf k} = \frac{\hbar^2{\bf k}^2}{2m}
\label{eq:Kinetic}
\end{align}
of a free atom with mass $m$. In addition, we assume that the system is closed, both in energy and total atom number, such that we work within the microcanonical ensemble where the total atom number is fixed to ${\cal N}$.
We later on need a lowering-operator $\op{L}_0$ and rising-operator $\op{L}^\dagger_0$ for the BEC. They are defined by the following properties:
\begin{align}
  \op{L}^\dagger_0 \absN{n}_0 = \absN{n+1}_0\,,
  \qquad
  \op{L}_0 \absN{n+1}_0 = \absN{n}_0\,,
  \qquad
   \op{L}_0 \absN{0}_0 =0
\label{eq:L-operators}\,,
\end{align}
where $\absN{n}_0$ denotes a BEC Fock state that contains exactly $n$ atoms. We will also need the total number operator for the normal-component and BEC atoms
\begin{eqnarray}
  \opNN \equiv \sum_{{\bf k}\neq 0} B^\dagger_{\bf k} B_{\bf k}\,,
  \qquad
  \opNC \equiv {\cal N} - \opNN
\label{eq:opNN}\;,
\end{eqnarray}
respectively, where ${\cal N}$ is the total number of atoms, and we have removed the BEC-related ${\bf k} =0$ term from the sum.

As shown in Ref.~\cite{Kira:Xpic}, the BEC contains atom clusters to all orders, which makes the direct application of the cluster-expansion approach inefficient. One possibility is to neglect the BEC correlations, which is reasonable only for dilute BECs\cite{Kohler:2002,Kohler:2003} where quantum depletion is not appreciable. 
However, these correlations strongly modify the quantum depletion, and thus, must be fully included for the strongly interacting Bose gas, despite it seems a major complication for the cluster-expansion analysis. Solution to this problem is found in Ref.~\cite{Kira:Xpic} by converting the interacting Bose gas into the excitation picture that eliminates the ``trivial'' BEC clusters, providing a cluster-expansion friendly formulation. 
We will next outline the main consequences of this method in order to produce extreme convergence for the cluster-expansion approach.

The excitation picture follows from a non-unitary transformation 
\begin{align}
  \ex{\op{T}} \equiv \left( \op{L}_0 \right)^{\opNC}
\label{eq:T_ex}
\end{align}
which transforms all relevant operators 
\begin{align}
  \op{O}_{\rm ex} = \ex{\op{T}} \, \op{O} \, \ex{\op{T}}^\dagger
\label{eq:excitation-picture}
\end{align}
to the excitation picture. The explicit properties of $\ex{\op{T}}$ are worked out in Ref.~\cite{Kira:Xpic}; mathematically, $\ex{\op{T}}$ is not an unitary operator such that it produces a set of nontrivial transformations. However, the excitation picture rewards us with four very important properties which deliver us a cluster-expansion friendly formulation of the strongly interacting Bose gas. First, any expectation value,
\begin{eqnarray}
  \ave{\op{O}} \equiv \trace{\op{O}\,\op{\rho}} = \trace{\ex{O}\,\ex{\op{\rho}}} \equiv \ex{\ave{\ex{\op{O}}}}
\label{eq:ex-average}\;,
\end{eqnarray}
can be computed completely within the excitation picture where $\ex{\op{\rho}}$ ($\op{\rho}$) is the density matrix in the excitation (original) picture. Second, the excitation picture shifts $\op{\rho}$ to $\ex{\op{\rho}}$ that has vacuum state at the BEC while the normal component surrounding BEC remains excited. 
In other words, $\ex{\op{\rho}}$ only contains the normal-component atoms excited by the quantum depletion, which justifies the name -- the excitation picture. Third, since the BEC part of the $\ex{\op{\rho}}$ is a vacuum, all BEC cluster vanish in the excitation picture, which makes its cluster-expansion description trivial. Fourth,  also the quantum dynamics,
\begin{eqnarray}
  i \hbar \frac{\partial}{\partial t} \ex{\ave{\ex{\op{O}}}} 
  = \ex{\ave{\comm{\ex{\op{O}}}{\ex{\op{H}}}}}
\label{eq:HEM-ex}\;,
\end{eqnarray}
can be evolved within the excitation picture because $\ex{\op{H}}$ is the system Hamiltonian in the excitation picture. This last property allows us to systematically compute how clusters are sequentially built up to the normal component due to the quantum depletion.

\begin{figure}[t]
\includegraphics*[scale=0.5]{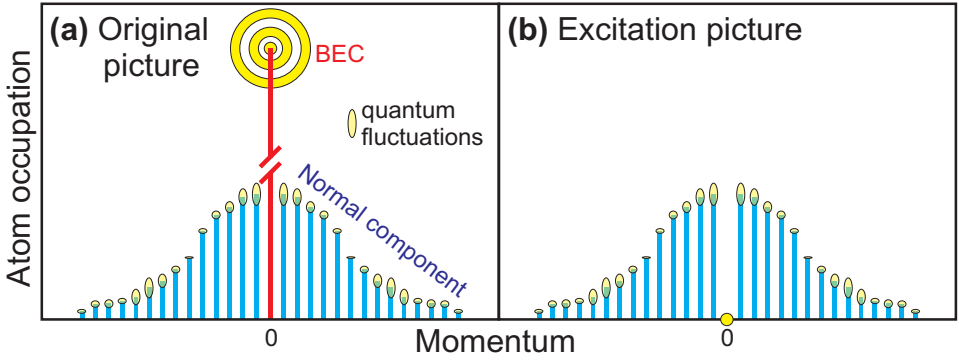}
\caption{State occupations of original vs.~excitation picture as a bar chart. The yellow areas symbolize the atom--atom correlations of each state. {\bf a} The original picture has a large, singular, BEC occupation (red bar) while the normal component (blue bars) is smoothly distributed. {\bf b} The excitation picture views only the excitations around the BEC that becomes a particle vacuum.
}\label{ExciPIC}
\end{figure}

Figure \ref{ExciPIC} illustrates the quantum-statistical implications of the original (left) and the excitation picture (right). Each of the bars denotes the occupation of atomic states as function of momentum. The ovals symbolize the fluctuations of the states. In the original picture, the BEC has a massive occupation and complicated fluctuations indicated by the set of concentral rings. 
The excitation picture views the many-body quantum statistics around the BEC such that the BEC appears as a vacuum with no clusters. At the same time, the normal-component occupation remains unchanged, which means that {\it all} nontrivial properties are encoded to the clusters of the normal component.

The BEC quantum statistics can be uniquely expressed in the expectation-value representation $\ave{[B^\dagger_0]^J B_0^K}$.\cite{Kira:2008} As explained in Ref.~\cite{Kira:Xpic}, it must be diagonal ($J=K$) and reduces into
\begin{eqnarray}
    \ave{[B^\dagger_0]^J B_0^K} \toX
    \delta_{J,K}\, \avex{\frac{\opNC!}{(\opNC -J+1)!}}
    =
    \delta_{J,K} \,\avex{\opNC!(\opNC -1)!\cdots (\opNC -J)!}
\label{eq:BEC-QS}\;,
\end{eqnarray}
in the excitation picture. Since $\opNC$ depends only on the normal component, the BEC quantum statistics is indeed defined by the excitation within the normal component. In other word, when the quantum kinetics of the normal component is known, we can directly identify quantum statistics of the BEC.

We also must consider a mix of BEC and normal-component operators
\begin{eqnarray}
  \ave{\op{O}^{J,J'}_{K,K'}} \equiv 
  \ave{[B^\dagger_0]^J B_0^K \, \op{O}^{J'}_{K'} } 
\label{eq:JJKK-average}\;,
\end{eqnarray}
where $\op{O}^{J'}_{K'}$ is a normal-component operator that contains $J'$ creation and $K'$ annihilation operators. For microcanonical systems, $\ave{\op{O}^{J,J'}_{K,K'}}$ exists only if the total number of creation and annihilation operators is equal, i.e.~we must have $J+J'=K+K'$ in the original picture\cite{Kira:Xpic}. 
For any other case, $\op{O}^{J,J'}_{K,K'}$ becomes a transition operator between the Hilbert spaces with a different number of atoms; the corresponding expectation values must vanish because such transitions are forbidden for microcanonical systems. In analogy to this, also an expectation value $\ave{[\op{L}^\dagger_0]^J \op{L}_0^K \, \op{O}^{J'}_{K'} }$ exists only if the condition $J+J'=K+K'$ is satisfied. Such expectation values have a simple and exact transformation\cite{Kira:Xpic}
\begin{eqnarray}
  \ave{[\op{L}^\dagger_0]^J \op{L}_0^K \, \op{O}^{J'}_{K'} } 
  \toX
  \left\{
    \begin{array}{cl}
     \avex{\op{O}^{J'}_{K'} }\,, & J+J' = K+K'
     \\
     0\,, & {\rm otherwise}
    \end{array}
  \right.  
\label{eq:JK-average}\;,
\end{eqnarray}
whenever $K$ is smaller than the atom number in the BEC. We observe that the normal-component operators $\op{O}^{J'}_{K'}$ remain unaltered and that the creation and annihilation operator numbers do not have to be equal. Generally, $\op{O}^{J'}_{K'}$ with $J'\neq K'$ define some coherent transition amplitude within the normal component. Equation \eqref{eq:JK-average} attaches a physical meaning to the corresponding $\avex{\op{O}^{J'}_{K'} }$ 
because it ties this coherence to a number change in the BEC such that the total number of atoms is not altered in the original picture. When $\op{L}_0$ is replaced by $B_0$ in Eq.~\eqref{eq:JK-average}, the transformation to the excitation picture becomes slightly more complicated, as shown in Ref.~\cite{Kira:Xpic}, but not needed here to complete the full cluster-expansion analysis.


\subsection{Clusters and homogeneity}
\label{sec:relevant clusters}

As shown in Ref.~\cite{Kira:2008}, an expectation with $J$ boson creation and $K$ boson annihilation operators is a $(J+K)$-particle operator. The expectation values with one boson operator then define {\it singlets} in the cluster-expansion sense. If a boson system is describable in terms of singlets, the system properties follow from a classical factorization where each $B_{\bf k}$ and $B^\dagger_{\bf k}$ is replaced by its expectation value  $\avex{B_{\bf k}}$ and $\avex{B^\dagger_{\bf k}}$, respectively, because then the boson operators behave like classical amplitudes. 
For example, the most classical light is described by a coherent amplitude, cf.~Ref.~\cite{Book:2011,Walls:2008} for further discussion. Contributions beyond the classical factorization define pure atom--atom correlations that can be identified recursively.

In this paper, we identify the correlations by using the excitation picture. As the first recursive step, one can start from two-atom expectation values:
\begin{eqnarray}
  \avex{B^\dagger_{\bf k} B_{{\bf k}'}} 
  &=&
  \avex{B^\dagger_{\bf k}}\avex{B_{{\bf k}'}}
  +
  \Delta\avex{B^\dagger_{\bf k} B_{{\bf k}'}} 
\nonumber\\
  \avex{B_{\bf k} B_{{\bf k}'}} 
  &=&
  \avex{B_{\bf k}}\avex{B_{{\bf k}'}}
  +
  \Delta\avex{B_{\bf k} B_{{\bf k}'}}
\label{eq:2-ple}\;,
\end{eqnarray}
where the first part is the classical factorization while the second part identifies the genuine two-atom correlation, i.e.~the {\it doublet}. The $N$-atom correlations can be defined following a straightforward recursive procedure\cite{Kira:2006,Book:2011}; once any given $N$-atom expectation value is expressed in terms of sum over all possible factorizations up to $(N-1)$-atom clusters, the remaining contribution defines the $N$-atom cluster. 
This approach is equivalent to applying the Wick's theorem\cite{Wick:1950}; a fully mathematical formulation of this intuitive cluster identification is given in Sec.~\ref{sec:systemH}. The three- and four-atom correlations are often referred to as {\it triplets} and {\it quadruplets}, respectively; the higher order clusters are simply called $N$-atom correlations.

For sufficiently large atom traps, the atom cloud extends over a much larger length scale than the atom size or the atom--atom interaction range. Therefore, the essential many-body effects of strongly interacting Bose gas can be understood by assuming homogeneous conditions implying that all expectation-value and correlation combinations of the field operators 
\begin{eqnarray}
  \op{\Psi}({\bf r}) = \frac{1}{{\cal L}^\frac{3}{2}}\sum_{\bf k} B_{\bf k}\, e^{i{\bf k}\cdot{\bf r}}
\label{eq:atom-field}\;,
\end{eqnarray}
and $\op{\Psi}^\dagger({\bf r})$ are translation invariant. This is satisfied only if the sum of creation-operator momenta is equal to the sum of annihilation-operator momenta within all expectation values and correlations. As discussed e.g.~in Ref.~\cite{Kira:2006,Book:2011}, this condition is equivalent to the random-phase approximation that is exact for homogeneous systems. Since the normal-component singlets $\avex{B_{\bf k}}$ have ${\bf k} \neq 0$, the corresponding singlets must vanish in homogeneous systems. 
Even though the BEC operators do not appear in the excitation picture, they follow a similar condition $\ave{B_0}=0$ in the original picture based on property \eqref{eq:BEC-QS}. In other words,  neither the BEC nor the normal-component atoms support the coherent amplitude, i.e.~$\ave{B_0} = \avex{B_{\bf k}}=0$ for the homogeneous microcanonical system studied here.

Transformation \eqref{eq:JK-average}, identification \eqref{eq:2-ple}, and homogeneity imply that the occupation of normal-component atoms,
\begin{eqnarray}
  \ave{B^\dagger_{\bf k} B_{\bf k}} \toX \avex{B^\dagger_{\bf k} B_{\bf k}}  = \Delta \avex{B^\dagger_{\bf k} B_{\bf k}} \equiv f_{\bf k}
\label{eq:occupation}\;,
\end{eqnarray}
is a pure doublet contribution. In the same way, also the transition amplitude
\begin{eqnarray}
 \ave{\op{L}^\dagger_0 \, \op{L}^\dagger_0 \, B_{\bf k} B_{-{\bf k}}} \toX \avex{B_{\bf k} B_{-{\bf k}}} 
 = \Delta \avex{B_{\bf k} B_{-{\bf k}}}
 \equiv s_{\bf k}   
\label{eq:SQZamp}\;,
\end{eqnarray}
is a doublet, but only in the excitation picture. The correlation pair, $(s_{\bf k},\,f_{\bf k})$, uniquely defines the pairwise atom correlations of the system in the excitation picture because the remaining 
$\Delta \avex{B^\dagger_{-{\bf k}} B^\dagger_{\bf k} } = s^\star_{\bf k}$ correlation is connected with $s_{\bf k}$ via complex conjugation. Later on, we also need a complete set of triplets
\begin{eqnarray}
  T^{{\bf k}',{\bf k}}_{1,2} \equiv \Delta \avex{B^\dagger_{{\bf k}+{\bf k}'} B_{{\bf k}'} B_{{\bf k}}}\,,
  \qquad
  T^{{\bf k}',{\bf k}}_{0,3} \equiv \Delta \avex{B_{-{\bf k}-{\bf k}'} B_{{\bf k}'} B_{{\bf k}}}
\label{eq:triplets}\;,
\end{eqnarray}
as well as quadruplets
\begin{eqnarray}
  Q^{{\bf q},{\bf k}',{\bf k}}_{2,2} 
  &\equiv& 
  \Delta \avex{B^\dagger_{{\bf k}} B^\dagger_{{\bf k}'} B_{{\bf k}'+{\bf q}} B_{{\bf k}-{\bf q}}}\,,
  \qquad
  Q^{{\bf q},{\bf k}',{\bf k}}_{1,3} 
  \equiv 
  \Delta \avex{B^\dagger_{{\bf k}+{\bf k}'+{\bf q}} B_{{\bf q}} B_{{\bf k}'} B_{{\bf k}}}\,,
\nonumber\\
  Q^{{\bf q},{\bf k}',{\bf k}}_{0,4} 
  &\equiv& 
  \Delta \avex{B_{-{\bf k}-{\bf k}'-{\bf q}} B_{{\bf q}} B_{{\bf k}'} B_{{\bf k}}}
\label{eq:quadruplets}\;,
\end{eqnarray}
where the subscripts label the number of creation (first index) and annihilation (second index) operators. It is straightforward to show that combinations $T^{{\bf k}',{\bf k}}_{2,1}$, $T^{{\bf k}',{\bf k}}_{3,0}$, $Q^{{\bf q},{\bf k}',{\bf k}}_{3,1}$ and $Q^{{\bf q},{\bf k}',{\bf k}}_{4,0}$ follow uniquely from the presented definitions when we apply a complex conjugation to the explicitly defined correlations. The quantum kinetics of {\it all} clusters is derived in Sec.~\ref{sec:QK_doublets}.

\begin{figure}[t]
\includegraphics*[scale=0.58]{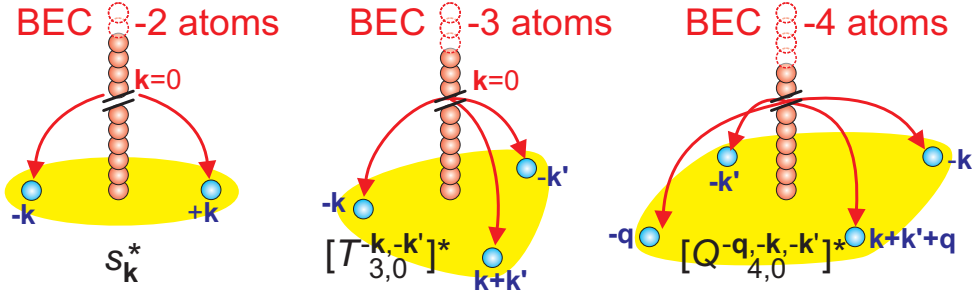}
\caption{Typical quantum-depletion processes in the original picture. The piled-up spheres symbolize the BEC and the blue spheres denote the normal-component atoms created by quantum depletion indicated by arrows; the dashed spheres denote atoms removed by quantum depletion. The left, middle, and right frames illustrate elementary processes creating atom doublets, triplets, and quadrauplets, respectively; the yellow area symbolizes the related correlations.}\label{Qdepletion}
\end{figure}

The physical meaning of correlations with an equal amount of boson creation and annihilation is rather simple to vision: clearly, $f_{\bf k}$ denotes the occupation of atoms at normal state ${\bf k}$ while $Q^{{\bf q},{\bf k}',{\bf k}}_{2,2}$ describes correlated scattering from states ${\bf k}-{\bf q}$ and ${\bf k}'+{\bf q}$ to ${\bf k}$ and ${\bf k}'$. In the special case of ${\bf q}=0$, $Q^{0,{\bf k}',{\bf k}}_{2,2}$ defines the correlated occupation for finding an atom at state ${\bf k}'$ while another atom is at ${\bf k}$. 
The other clusters with unequal number of boson operators have a slightly modified physical meaning, which we need to view in the original picture. Figure \ref{Qdepletion} schematically shows the correlations $s^\star_{\bf k}$ (up), $T^{-{\bf k}',-{\bf k}}_{3,0}$ (middle), and $Q^{-{\bf q},-{\bf k}',-{\bf k}}_{4,0}$, and the piled-up red spheres indicate the atoms of the BEC, arrows indicate the elementary quantum-depletion transitions that convert the BEC into normal-component atoms (blue spheres). The yellow areas symbolize the correlations that are induced within the normal component. The red dashed circles indicate the number of BEC atoms removed by quantum depletion.

Based on Eq.~\eqref{eq:SQZamp}, $s^\star_{\bf k}$ determines correlations of the transition $\op{L}_0 \, \op{L}_0 \, B^\dagger_{-{\bf k}} B^\dagger_{\bf k}$ which describes a process where the atom number of the BEC is lowered by two when two atoms are ejected to normal states ${\bf k}$ and $-{\bf k}$. This is of course an elementary process of quantum depletion that conserves the momentum. Note that quantum depletion cannot eject a single atom from a BEC to a normal component because that would violate the momentum conservation.  
Higher order quantum-depletion processes may eject three, four etc.~atoms from a BEC to the normal component. These processes are described by triplets and quadruplets presented in the middle and bottom frame of Fig.~\ref{Qdepletion}. More precisely, the triplets $T^{-{\bf k}',-{\bf k}}_{3,0}$ involve correlations related to the process $\op{L}_0 \,\op{L}_0 \, \op{L}_0 \, B^\dagger_{-{\bf k}} B^\dagger_{-{\bf k}'} \, B^\dagger_{{\bf k}+{\bf k}'}$ and the quadruplet contains the 
correlated part of the process  $\op{L}_0 \,\op{L}_0 \,\op{L}_0 \, \op{L}_0 \, B^\dagger_{-{\bf k}} \, B^\dagger_{-{\bf k}'}\, B^\dagger_{-{\bf q}} \, B^\dagger_{{\bf k}+{\bf k}'+{\bf q}}$, based on transformation \eqref{eq:JK-average} and homogeneity. 

Quantum depletion has also an inverse process where the normal-component atom scatter back into a BEC; the corresponding elementary processes are described by $s_{\bf k}$, $T_{0,3}$, $Q_{0,4}$ etc.~correlations. The quantum depletion and its inverse can also produce transitions within the normal state. 
For example $T^{{\bf k}',{\bf k}}_{1,2} \equiv \Delta \avex{B^\dagger_{\bf k} B^\dagger_{{\bf k}'} B_{{\bf k}+{\bf k}'}}$ decreases the BEC number by one while one normal-component atom scatters into a new state because it stems from the process $\op{L}_0 \, B^\dagger_{\bf k} B^\dagger_{{\bf k}'} B_{{\bf k}+{\bf k}'}$ in the original picture. This is not an elementary quantum-depletion process because it is assisted by a normal-component atom.

Once different clusters are generated to the normal component, they directly modify the BEC quantum statistics. Based on transformation \eqref{eq:BEC-QS}, we find the two lowest BEC expectation value to be
\begin{eqnarray}
  \ave{B^\dagger_0 B_0} 
  &\toX&
  \avex{\opNC} = {\cal N} - \sum_{{\bf k}\neq 0} f_{\bf k}  \equiv \NC\,,
\nonumber\\
  \ave{B^\dagger_0 B^\dagger_0 B_0B_0} 
  &\toX&
  \avex{\opNC(\opNC-1)}
  = 
  \NC^2+\NC
  +\sum_{{\bf k}\neq 0} \left(f_{\bf k}^2+|s_{\bf k}|^2 \right)
  + \sum_{{\bf k},{\bf k}'} Q^{0,{\bf k}',{\bf k}}_{2,2}
\label{eq:NC}\;,
\end{eqnarray}
that follows straightforwardly from the singlet-doublet-triplet-quadruplet factorization. For later use, we also identify the average BEC number fluctuations from
\begin{eqnarray}
  \Delta\NC^2 \equiv 
  \ave{\left(B^\dagger_0 B_0\right)^2}
  -
  \ave{B^\dagger_0 B_0}^2
  \equiv
  2 \NN\, c_{\rm shape} 
\label{eq:DNC}\;,
\end{eqnarray}
which follows from Eq.~\eqref{eq:NC} after identifying
\begin{eqnarray}
  c_{\rm shape} 
  \equiv 
  \frac{1}{2}
   + \frac{1}{2\NN}
    \sum_{{\bf k}\neq 0} 
    \left(
      f_{\bf k}^2+|s_{\bf k}|^2 
      + 
      \sum_{{\bf k}'} Q^{0,{\bf k}',{\bf k}}_{2,2}
    \right)
\label{eq:c_shape}
\end{eqnarray}
as the shape of the quantum depletion, which includes the quadruplet to the original definition in Ref.~\cite{Kira:Xpic}. Since $(f_{\bf k},s_{\bf k})$ excitation can have infinitely many different functional forms, the magnitude of $c_{\rm shape}$ quantifies how different excitation shapes influence the BEC fluctuations. The Bogoliubov excitations produce\cite{Kira:Xpic}  $c_{\rm shape}=\frac{3\pi}{8}\approx1.1781$ that is compared with a full singlet--doublet excitation in Sec.~\ref{sec:numerics}.

\subsection{Operator dynamics and quantum statistics}
\label{sec:systemH}

As the major consequences of the excitation picture, the BEC is reduced to a vacuum that has trivial quantum statistics and {\it all} transformed operators $\ex{\op{O}}$ are expressible in terms of the normal-component operators alone. Even though transformation \eqref{eq:excitation-picture} to the excitation picture has unusual nonunitary character, the expectation-value dynamics still follows from the usual Heisenberg equation of motion \eqref{eq:HEM-ex}. 
Due to these general aspects, $\ex{\op{O}}$ can be expressed it in terms of boson operators $B_{\bf k}$ and $B^\dagger_{\bf k}$ such that the quantum dynamics of $\avex{\ex{\op{O}}}$ can be determined straightforwardly when we know $\ihddt B_{\bf k} = \comm{B_{\bf k}}{\ex{\op{H}}}$ and $\ihddt B^\dagger_{\bf k} = \comm{B^\dagger_{\bf k}}{\ex{\op{H}}}$. As we identify the clusters within $\ex{\op{O}}$, based on the number of boson operators, we may directly determine the quantum kinetics of the clusters within the excitation picture. 
To perform this derivation efficiently, we start from the boson-operators dynamics resulting from $\ex{\op{H}}$. The exact $\ex{\op{H}}$ is derived in Ref.~\cite{Kira:Xpic}, and we next introduce a few modified quantities in order to summarize the boson-operator dynamics compactly. 

\begin{figure}[t]
\includegraphics*[scale=0.5]{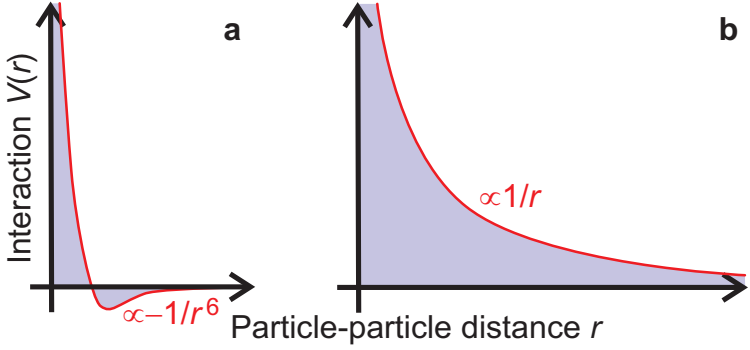}
\caption{Interactions in a Bose gas (left) vs.~a semiconductor (right) as function of interparticle distance.
 {\bf a} The atom--atom interactions have a repulsive core and a $-r^{-6}$ tail. {\bf b} Electrons and holes within a semiconductor experience a long-range Coulomb interaction.
}\label{Vpotentials}
\end{figure}

For ultracold atoms, the atom--atom interactions can be described with the so-called $s$-wave scattering\cite{Dalfovo:1999,Andersen:2004,Leggett:2001,Proukakis:2008, Bloch:2008,Blakie:2008,Dalibard:2011} where the pairwise interactions among atoms are described via a  potential $V({\bf r})=V(|{\bf r}|)$ that depends only on the inter-atom distance $|{\bf r}|$. A typical atom--atom interaction potential for $s$-wave scattering is shown in Fig.~\ref{Vpotentials}{\bf a}. 
For a small $|{\bf r}|$, the atomic $V({\bf r})$ is repulsive due to the electronic/atom core repulsion that dominates when the atoms are too close to each other. For a larger $|{\bf r}|$, atoms start to attract each other via the far-field dipole-dipole attraction. This yields a clear minimum at ${\bf r}_0$ that roughly determines the molecular bond length because $V({\bf r})$ eventually decays as ${\bf r}^{-6}$ due to the van der Waals force.
As a comparison, electronic excitations in solids experience only a monotonically decaying Coulomb interaction\cite{Book:2011} that decays as $|{\bf r}|^{-1}$. Therefore, the interacting particles of solid are coupled via a long-range interaction, unlike in the interacting Bose gas, as illustrated in Fig.~\ref{Vpotentials}{\bf b}.

In solids and atomic gases alike, interactions can bind particles to different particle clusters, despite the differences in the interaction range. In this context, bound atom pairs are referred to as dimers in analogy to excitons that are Coulomb bound electron--hole pairs\cite{Haug:2008,Book:2011}. Also higher order molecular states may exist as trimers (three atoms) and quadmers (four atoms). Clearly, quantum-depletion processes depicted in Fig.~\ref{Qdepletion} create dimers, trimers, quadmers, and so on, via the doublet ($s_{\bf k}$), 
triplet ($T_{0,3}$), and quadruplet ($Q_{0,4}$) correlations, respectively. The formation of triplets and quadruplets is intimately connected with the Efimov physics studied actively\cite{Kraemer:2006,Stecher:2009,Williams:2009,Pollack:2009,Castin:2010,Wild:2012} in atom gases, as discussed in Sec.~\ref{sec:Conclusions} and   \ref{app:triplets}. 
At the same time, such multimers have also connection to exotic quasiparticles of semiconductors, such as the dropletons\cite{Hunter:2014}.

In general, the atomic $V({\bf r})$ has a very short range due to the ${\bf r}^{-6}$  tail. In many relevant investigations, one may introduce a contact potential\cite{Dalfovo:1999,Giorgini:2008} where $V({\bf r})$ is replaced by a regularized $\delta$ function. In turns out, the cluster-expansion computations can be performed without this approximation; we use the full $V({\bf r})$ in its Fourier transformed form
\begin{eqnarray}
  V_{\bf q} \equiv \frac{1}{{\cal L}^3} \int d^3r\, V({\bf r}) e^{-i{\bf q} \cdot {\bf r}}
\label{eq:V-Fourier}\;.
\end{eqnarray}
where ${\cal L}^3$ is the quantization volume, and fully includes the short-range repulsion and the fast-decaying tail. Since the Coulombic interaction has a very long range in solids, the contact potential approximation does not describe, e.g., the semiconductor physics very well. Consequently, this quantitative difference between interacting atoms and semiconductors can clearly reveal different aspects of many-body physics when manybodypedia is pursued.

To classify the various contributions that appear in the operator dynamics, we introduce 
\begin{eqnarray}
  W^{[J]}_{\bf k} \equiv \NC^\frac{J}{2} \, V_{\bf k}\,,
  \qquad
  W^{[J]}_{{\bf k},\,{\bf k}'} \equiv \NC^\frac{J}{2} \,
  \left(
    V_{\bf k}+V_{{\bf k}'}
  \right)
\label{eq:W_def}
\end{eqnarray}
as an interaction matrix element that is weighted by different powers of the BEC occupation \eqref{eq:NC}. After $\ex{\op{H}}$ derived in Ref.~\cite{Kira:Xpic}, the central boson-operator dynamics is found to be
\begin{eqnarray}
  \ihddt B_{\bf k}
  &=&
 \tilde{E}_{\bf k}\, B_{\bf k} 
  + \sum_{{\bf q},{\bf k'}}\,W^{[0]}_{{\bf k}'-{\bf k}} B^\dagger_{{\bf q}-{\bf k}} B_{{\bf q}-{\bf k}'} B_{{\bf k}'}
\nonumber\\
  &&+W^{[2]}_{\bf k} B^\dagger_{-{\bf k}}
  +
  \sum_{{\bf k}'}
  \left[
    W^{[1]}_{{\bf k},{\bf k}'-{\bf k}}\, B^\dagger_{{\bf k}'-{\bf k}} B_{{\bf k}'}
    +
    W^{[1]}_{{\bf k}'}\, B_{{\bf k}-{\bf k}'} B_{{\bf k}'}
  \right]
\label{eq:BoccuD}\,,
\\
  \ihddt B^\dagger_{\bf k}
  &=&
  -\tilde{E}_{\bf k}\, B^\dagger_{\bf k} 
  - \sum_{{\bf q},{\bf k'}}\,W^{[0]}_{{\bf k}'-{\bf k}} 
   B^\dagger_{{\bf k}'} B^\dagger_{{\bf q}-{\bf k}'} B_{{\bf q}-{\bf k}}
\nonumber\\
  &&-W^{[2]}_{\bf k} B_{-{\bf k}}
  -
 \sum_{{\bf k}'}
  \left[
    W^{[1]}_{{\bf k},{\bf k}'-{\bf k}}\, B^\dagger_{{\bf k}'} B_{{\bf k}'-{\bf k}} 
    +
    W^{[1]}_{{\bf k}'}\, B^\dagger_{{\bf k}'} B^\dagger_{{\bf k}-{\bf k}'}
  \right]
\label{eq:B+occuD}\,.
\end{eqnarray}
Since the BEC effects are already included by the transformation \eqref{eq:excitation-picture}, the sums implicitly omit those wave vector combinations that produce a vanishing wave vector for the boson operators. 
We could exclude these terms from the sum explicitly, but the implicit exclusion does not produce unambiguity because neither $B_0$ nor $B^\dagger_0$ contribute to the excitation-picture expectation values. 
A similar omission of certain ${\bf k}$ combinations appears also in solids\cite{Book:2011}, but it has a different physical origin; due to the ionic background density, one must omit those interaction terms that yield a vanishing wave-vector argument for $V_{\bf k}$. This introduces yet another subtle difference between atomic and solid-state many-body systems, which interesting from the manybodypedia point of view.
Besides the explicit interaction terms, the operator dynamics contains also the renormalized kinetic energy
\begin{eqnarray}
  \tilde{E}_{\bf k} \equiv E_{\bf k} + N_C \, V_{\bf k}
  -
  \sum_{{\bf k}'\neq 0} V_{{\bf k}'}\, {\rm Re} \left[f_{{\bf k}'}+s_{{\bf k}'}\right] 
  -
  \frac{1}{\sqrt{\NC}} \sum_{{\bf k}',{\bf k}''} 
  V_{{\bf k}'} \,
  {\rm Re} 
  \left[
    T^{{\bf k}'',{\bf k}'}_{1,2}
  \right]
\label{eq:E-renormalized}\;,
\end{eqnarray}
where the $N_C \, V_{\bf k}$ contribution stems from the direct effect of the BEC while the remaining terms result from the BEC-number fluctuations. As discussed in   \ref{app:triplets}, the triplet $\frac{1}{\sqrt{\NC}}  T^{{\bf k}',{\bf k}}_{1,2}$ vanishes at limit $\NC \rightarrow 0$. Therefore, the effect of all renormalizations vanishes altogether for a vanishing $\NC$. 

Technically, Eqs.~\eqref{eq:BoccuD}--\eqref{eq:B+occuD} follow after we linearize the exact $\ex{\op{H}}$ with respect to the BEC-number fluctuations. As shown in Ref.~\cite{Kira:Xpic}, the excitation picture makes this approximation exact for large $\NC$ while it provides only a negligibly small overestimate of the interaction strength at low $\NC$. Since $W^{[1]}_{\bf k}$, $W^{[2]}_{\bf k}$, 
and energy renormalizations vanish at $\NC \rightarrow 0$, even a slight overestimate of this limit produces the correct many-body physics. Additionally, the cluster expansion includes the interactions to all orders \cite{Kira:2006,Book:2011}, which allows us to analyze strongly interacting Bose gas for a broad range of excitations, much beyond a perturbative treatment of interactions. 

Equations~\eqref{eq:BoccuD}--\eqref{eq:B+occuD} are now in a cluster-expansion friendly format because they form a closed set of normal-component operator dynamics. Within this structure, the $W^{[0]}$ contribution yields the standard hierarchy problem resulting from any pairwise interaction while the $W^{[1]}$ terms produces a new BEC-generated hierarchy. 
Both of them can be simultaneously dealt with the cluster-expansion, and we will show in Sec.~\ref{sec:sequential} that the quantum depletion generates the normal-component clusters sequentially. As a result, the excitation picture allows for a very efficient description of the interacting Bose gas in terms of very few atom clusters.

Additionally, we only need to consider the quantum statistics of the normal state because
it also defines the BEC properties via transformation \eqref{eq:BEC-QS}. Therefore, we introduce the normally-ordered characteristics function and the correlation-generating function\cite{Kira:2008},
\begin{eqnarray}
  \chi(\left\{ \alpha \right\}) 
  \equiv 
  \avex{
    e^{\sum_{\bf k} \alpha_{\bf k} \, B^\dagger_{\bf k}}
    \;\;
    e^{- \sum_{\bf k} \alpha^\star_{\bf k} \, B_{\bf k}}
   }  \equiv e^{\xi_N(\left\{ \alpha \right\})}\,,
\qquad
  \xi(\left\{ \alpha \right\})
  \equiv
  {\rm ln} 
    \left[ 
      \chi_N(\left\{ \alpha \right\})
    \right]
\label{eq:xi_N}\,,
\end{eqnarray}
respectively, both constituting a quantum statistics. We have used here a notation where the complex-valued variables $\alpha_{\bf k}$ ($\alpha^\star_{\bf k}$) are connected with normal-component operators $B^\dagger_{\bf k}$ ($B_{\bf k}$), while  $\left\{ \alpha \right\}$ refers to all of them. 
Since the BEC is formally absent in the excitation picture, we have not eliminated the BEC index from the sum over ${\bf k}$, but implicitly assume $\alpha_0=\alpha^\star_0=0$ to simplify the notation. 

To access the expectation values/correlations effortlessly, we introduce the differentials:
\begin{eqnarray}
  \partial_{\bf k} \equiv \frac{\partial}{\partial \alpha_{\bf k}}\,,\qquad
  \partial^\star_{\bf k} \equiv \frac{\partial}{\partial \alpha^\star_{\bf k}}
\label{eq:differentials}\;.
\end{eqnarray}
They generate any expectation value or atom-correlation via reduction formulas\cite{Kira:2008}:
\begin{eqnarray}
  \avex{
    B^\dagger_{{\bf k}_1} \cdots B^\dagger_{{\bf k}_J}
    B_{{\bf k}'_L} \cdots B_{{\bf k}'_1}
  }
  &=&
  (-1)^L
  \partial_{{\bf k}_1} \cdots \partial_{{\bf k}_J} \partial^\star_{{\bf k}'_L} \cdots \partial^\star_{{\bf k}'_1}  
  \left.
  \chi_N(\left\{ \alpha \right\}) \right|_{\{ \alpha =0 \}}
\label{eq:I-rep}\;,
\\
  \Delta\ave{
    B^\dagger_{{\bf k}_1} \cdots B^\dagger_{{\bf k}_J}
    B_{{\bf k}'_L} \cdots B_{{\bf k}'_1}
  }
  &=&
  (-1)^L
  \partial_{{\bf k}_1} \cdots \partial_{{\bf k}_J} \partial^\star_{{\bf k}'_L} \cdots \partial^\star_{{\bf k}'_1}  
  \left.
  \xi(\left\{ \alpha \right\}) \right|_{\{ \alpha =0 \}}
\label{eq:DI-rep}\;,
\end{eqnarray}
where $|_{\{ \alpha =0 \}}$ means that all $\alpha_{\bf k}$ indices are set to zero after the differentiation. In general, $\Delta \ave{\cdots}$ with $J$ creation and $K$ annihilation operators defines the amplitude of $(J+K)$-atom clusters that are truly correlated to each other within the many-body system. 
Since the excitation-picture $\chi_N$ and $\xi$ do not depend on $\alpha_0$, all nontrivial expectation values or correlations involving the BEC vanish in the excitation picture where the BEC becomes a vacuum state, see Fig.~\ref{ExciPIC}. Next, we start developing a general framework to solve the quantum dynamics of all atom clusters.

\section{Quantum kinetics}
\label{sec:chi_dyn}

Solving the quantum kinetics of the quantum statistics means that one simultaneously follows {\it all} quantum aspects of the interacting many-body systems. Since many-body problems are generally unsolvable, we first reformulate the problem into a form where the essential structure of the many-body problem becomes apparent. We analyze the quantum kinetics of $\chi$ to clarify the structure, and we start by streamlining the notation by introducing
\begin{eqnarray}
  \sumA \equiv \sum_{\bf k} \, \alpha^\star_{\bf k}\, B_{\bf k}\,,
  \qquad
  \sumAS \equiv \sum_{\bf k} \,  B^\dagger_{\bf k}\,\alpha_{\bf k}\,
\label{eq:sumA_AS}\;.
\end{eqnarray}
These sums do not contain $B_0$ that can formally be implemented by setting $\alpha_0$ to zero, as discussed above. We notice that these are the exponential arguments appearing in the characteristic function \eqref{eq:xi_N} that simplifies into
\begin{eqnarray}
  \chi_\alpha
  &=& \avex{e^\sumAS \, e^{-\sumA} }
  =  
  \sum_{J,K=0}^\infty C_{J,K}\,,
  \qquad
  C_{J,K} \equiv
  \frac{(-1)^K}{J! \,K!}
  \,
  \avex{\sumAS^J\,\sumA^K}  
\label{eq:chi_imp}\;,
\end{eqnarray}
after introducing an abbreviation $\chi_\alpha \equiv  \chi_N(\left\{ \alpha \right\})$, Taylor expanding the exponential functions, and identifying each individual term in the sum as $C_{J,K}$. 

The quantum kinetics of the $C_{J,K}$ contributions \eqref{eq:chi_imp} as well as the total $\chi_\alpha$ dynamics is derived in   \ref{app:chi_derivation} based on the operator dynamics \eqref{eq:BoccuD}--\eqref{eq:B+occuD}. The calculation produces an integro-differential equation
\begin{eqnarray}
  \ihddt \, \chi_\alpha
  \equiv
  \op{D}^{\rm tot}_\alpha \, \chi_\alpha
\label{eq:HEM-chiFIN1}\;,
\end{eqnarray}
where the differential operator can be organized in the orders of differentiation
\begin{align}
  \op{D}^{\rm tot}_\alpha
  &\equiv
  d^{(0)}_\alpha
  +
  \sum_{\bf k} 
  \left( 
    d^{(1,-)}_{{\bf k},\alpha} \, \partial^\star_{\bf k}
    -
    \left[d^{(1,+)}_{{\bf k},\alpha}\right]^\star \partial_{\bf k}
  \right)
   +
  \sum_{{\bf k},{\bf k}'}
  d^{(2)}_{{\bf k},{\bf k}',\alpha} \, \partial^\star_{\bf k}\, \partial_{{\bf k}'}
\nonumber\\
  &
  +
  \frac{1}{2} 
  \sum_{{\bf k},{\bf k}'}
    \left( 
    d^{(2,-)}_{{\bf k},{\bf k}',\alpha} \, \partial^\star_{\bf k}\, \partial^\star_{{\bf k}'}
    -
    \left[d^{(2,+)}_{{\bf k},{\bf k}',\alpha}\right]^\star \partial_{\bf k}\, \partial_{{\bf k}'}
  \right)
  -
    \sum_{{\bf k},{\bf k}',{\bf q}}
    \left( 
    d^{(3)}_{{\bf k},{\bf k}',\alpha} \, \partial_{{\bf k}+{\bf q}}\,  \partial^\star_{\bf k}\, \partial^\star_{{\bf k}'}
    -
    \left[d^{(3)}_{{\bf k},{\bf k}',\alpha}\right]^\star 
    \partial^\star_{{\bf k}+{\bf q}} \partial_{\bf k}\, \partial_{{\bf k}'}
  \right)
\label{eq:HEM-chiFIN2}
\end{align}
that contains the following constants:
\begin{align}
  &d^{(0)}_\alpha \equiv  \sum_{{\bf k}} W^{[2]}_{\bf k}\,
  \frac{\alpha^\star_{\bf k} \alpha^\star_{-{\bf k}}
    -
    \alpha_{\bf k} \alpha_{-{\bf k}}}{2}\,,
  &d^{(1,\pm)}_{{\bf k},\alpha}
  \equiv  
  \tilde{E}_{\bf k} \alpha^\star_{\bf k}
  +W^{[2]}_{\bf k}\,\alpha_{-{\bf k}}
  \pm 
  \frac{1}{2} \sum_{{\bf k}'} W^{[1]}_{{\bf k}-{\bf k}',{\bf k}'} \, \alpha^\star_{{\bf k}-{\bf k}'}\, \alpha^\star_{{\bf k}'}
\nonumber\\
  &d^{(2,\pm)}_{{\bf k},{\bf k}',\alpha}
  \equiv
  \sum_{\bf q} W^{[0]}_{\bf q} \alpha^\star_{{\bf k}-{\bf q}} \,  \alpha^\star_{{\bf k}'+{\bf q}} 
  \pm 
  W^{[1]}_{{\bf k},{\bf k}'} \, \alpha^\star_{{\bf k}+{\bf k}'}\,,
  &d^{(2)}_{{\bf k},{\bf k}',\alpha}
  \equiv
  W^{[1]}_{{\bf k}',{\bf k}-{\bf k}'} \alpha^\star_{{\bf k}-{\bf k}'} 
  +
  W^{[1]}_{{\bf k},{\bf k}'-{\bf k}} \alpha_{{\bf k}'-{\bf k}}\,,
\nonumber\\
  &d^{(3)}_{{\bf k},{\bf q},\alpha}
  \equiv
  W^{[0]}_{{\bf q}} \, \alpha_{{\bf k}-{\bf q}}
\label{eq:HEM-d_all}\;.
\end{align}
The $\chi$ dynamics \eqref{eq:HEM-chiFIN1}--\eqref{eq:HEM-d_all} is exact and defines a systematic starting point to solve the quantum kinetics of the interacting Bose gas. We observe that $\op{D}^{\rm tot}_\alpha$ contains a constant term together linear, quadratic, and cubic differentiations and a sum over the normal-component indices. 
It is clear that $\chi$ dynamics cannot be solved exactly due to the complexity of the integro-differential equation. Nonetheless, Eqs.~\eqref{eq:HEM-chiFIN1}--\eqref{eq:HEM-d_all} provide a convenient starting point to implement cluster-expansion-based approximations.

The dynamics of correlation generating function $\xi_\alpha \equiv \xi(\left\{ \alpha \right\})$ can be derived by inserting definition \eqref{eq:xi_N} into Eq.~\eqref{eq:HEM-chiFIN1}, producing
\begin{eqnarray}
  e^{\xi_\alpha} \ihddt \xi_\alpha
  =
  \op{D}^{\rm tot}_\alpha \,e^{\xi_\alpha}
  \quad \Leftrightarrow \quad
  \ihddt \xi_\alpha
  =
   e^{-\xi_\alpha}
  \op{D}^{\rm tot}_\alpha \, e^{\xi_\alpha}
\label{eq:HEM-xiSTART}\;.
\end{eqnarray}
The equivalence is valid because $e^{-\xi_\alpha}$ is always the proper inverse of $e^{\xi_\alpha}$. Since we know the explicit form of differentiation $\op{D}^{\rm tot}_\alpha$ via Eq.~\eqref{eq:HEM-chiFIN2}, we can now directly solve the quantum dynamics of the correlation-generating function. 
In general, $e^{-\xi_\alpha} \op{D}^{\rm tot}_\alpha \, e^{\xi_\alpha}$ can be viewed as a special transformation generated by 
$\xi_\alpha$, i.e., $e^{-\xi_\alpha}
  \op{D} \, e^{\xi_\alpha}$ defines the $\xi_\alpha$-transformation of an operator $\op{D}$.

In general,  $\op{D}^{\rm tot}_\alpha$ contains a constant term as well as the first-, second-, and third-order differentiation. Applying the $\xi_\alpha$ transformation \eqref{eq:HEM-xiSTART} to the constant part of $\op{D}^{\rm tot}_\alpha$ produces an unchanged constant contribution
\begin{eqnarray}
  \Xi^{(0)}_\alpha
  \equiv
  e^{-\xi_\alpha} d^{(0)}_\alpha \, e^{\xi_\alpha} = d^{(0)}_\alpha
  =
 \NC \sum_{{\bf k}} V_{\bf k}\,
  \frac{\alpha^\star_{\bf k} \alpha^\star_{-{\bf k}}
    -
    \alpha_{\bf k} \alpha_{-{\bf k}}}{2} 
\label{eq:HEM-xi-source}\;,
\end{eqnarray}
based on Eqs.~\eqref{eq:HEM-chiFIN2}--\eqref{eq:HEM-xiSTART} and \eqref{eq:W_def}. We see that $\Xi^{(0)}_\alpha$  does not depend on $\xi_\alpha$ such that it acts as a direct source that constantly generates $\xi_\alpha$. Since $\Xi^{(0)}_\alpha$ is linearly proportional to the product of BEC number $\NC$ and $V_{\bf k}$, we conclude that the {\it presence of BEC and interactions always excites correlations to the normal component}, i.e.~they initiate the quantum depletion.

To solve the full $\xi_\alpha$ dynamics that follows, we obviously need to $\xi_\alpha$ transform the remaining first-, second-, and third-order differentiations that appear in $\op{D}^{\rm tot}_\alpha$. 
In order to perform this efficiently, we transform generic  linear, quadratic, and cubic differentiations which we denote symbolically by $\partial_1$, $\partial_1 \partial_2$, and $\partial_1 \partial_2 \partial_3$, respectively. Their $\xi_\alpha$-transformations can be worked out straightforwardly, yielding
\begin{align}
   e^{-\xi_\alpha} \partial_1 \, e^{\xi_\alpha} &= \partial_1 \xi_\alpha\,, 
   \qquad\qquad\qquad
   e^{-\xi_\alpha}  \partial_1 \partial_2 \, e^{\xi_\alpha} 
   = 
   \partial_1 \partial_2 \,\xi_\alpha 
   + 
   \left[\partial_1 \, \xi_\alpha\right]\left[\partial_2 \,\xi_\alpha\right]
\nonumber\\
   e^{-\xi_\alpha}  \partial_1 \partial_2 \partial_3 \, e^{\xi_\alpha} 
   &= 
   \partial_1 \partial_2 \partial_3 \,\xi_\alpha 
   + 
   \left[\partial_1 \, \xi_\alpha\right]\left[\partial_2 \partial_3 \,\xi_\alpha\right]
   + 
   \left[\partial_2 \, \xi_\alpha\right]\left[\partial_1 \partial_2 \,\xi_\alpha\right]
    + 
   \left[\partial_3 \, \xi_\alpha\right]\left[\partial_1 \partial_2 \,\xi_\alpha\right]
\nonumber\\
   &+
    \left[\partial_1 \, \xi_\alpha\right]\left[\partial_2 \,\xi_\alpha\right]\left[\partial_3 \,\xi_\alpha\right]
\label{eq:HEM-xi-trafo}\;.
\end{align}
We observe that $\partial_1$ produces only contributions proportional to $\xi_\alpha$ while $ \partial_1 \partial_2$ yields  also terms with a quadratic $\xi_\alpha\, \xi_\alpha$ proportionality and $\partial_1 \partial_2 \partial_3$ results additionally in cubic terms, proportional to $\xi_\alpha\, \xi_\alpha\, \xi_\alpha$. 
We apply properties \eqref{eq:HEM-xi-trafo} when computing the $\xi_\alpha$ transform of the full $\op{D}^{\rm tot}_\alpha$ and sort out the constant, linear, quadratic, and cubic $\xi_\alpha$ contributions, yielding eventually 
\begin{eqnarray}
  \ihddt \xi_\alpha
  =
  \Xi^{(0)}_\alpha+\Xi^{(1)}_\alpha+\Xi^{(2)}_\alpha+\Xi^{(3)}_\alpha
\label{eq:HEM-xiClasses}\;,
\end{eqnarray}
where $\Xi^{(n)}_\alpha$ contributions are proportional to $(\xi_\alpha)^n$. The explicit constant, linear, quadratic, and cubic contributions are then
\begin{align}
  \Xi^{(0)}_\alpha
  &=
 \NC \sum_{{\bf k}} V_{\bf k}\,
  \frac{\alpha^\star_{\bf k} \alpha^\star_{-{\bf k}}
    -
    \alpha_{\bf k} \alpha_{-{\bf k}}}{2}\,,
\nonumber\\
  \Xi^{(1)}_\alpha
  &=
  \sum_{\bf k} 
  \left( 
    d^{(1,-)}_{{\bf k},\alpha} 
    \partial^\star_{\bf k}
    -
    \left(d^{(1,+)}_{{\bf k},\alpha}\right)^\star 
    \partial_{\bf k} 
  \right) \xi_\alpha
  +
   \sum_{{\bf k},{\bf k}'}
    d^{(2)}_{{\bf k},{\bf k}',\alpha} 
    \partial^\star_{\bf k}\, \partial_{{\bf k}'}\,\xi_\alpha
  +
    \sum_{{\bf k},{\bf k}'}
  \left(
    {\textstyle \frac{d^{(2,+)}_{{\bf k},{\bf k}',\alpha}}{2}}
    \partial^\star_{\bf k} \partial^\star_{{\bf k}'} 
    -
    {\textstyle
    \frac{\left(d^{(2,-)}_{{\bf k},{\bf k}',\alpha}\right)^\star}{2}
    }
    \partial_{\bf k} \partial_{{\bf k}'}
  \right)
  \xi_\alpha
\nonumber\\
 &+
      \sum_{{\bf k},{\bf k}',{\bf q}}
    \left(
    \left(d^{(3)}_{{\bf k},{\bf k}',\alpha}\right)^\star 
      \partial^\star_{{\bf k}+{\bf q}} \partial_{\bf k}\, \partial_{{\bf k}'} 
      -
      d^{(3)}_{{\bf k},{\bf k}',\alpha} \, 
      \partial_{{\bf k}+{\bf q}}\,  \partial^\star_{\bf k}\, \partial^\star_{{\bf k}'}\,\xi_\alpha 
  \right)
  \xi_\alpha 
\nonumber\\
  \Xi^{(2)}_\alpha
  &=
  \sum_{{\bf k},{\bf k}'}
  \left(
    d^{(2)}_{{\bf k},{\bf k}',\alpha} 
    \left[\partial^\star_{\bf k}\,\xi_\alpha\right] \left[\partial_{{\bf k}'}\,\xi_\alpha\right]
    +
    {\textstyle \frac{1}{2}}
    d^{(2,-)}_{{\bf k},{\bf k}',\alpha}
    \left[ \partial^\star_{\bf k}\, \xi_\alpha \right]
    \left[ \partial^\star_{{\bf k}'}\,\xi_\alpha \right]
    -
    {\textstyle \frac{1}{2}}
    \left(d^{(2,+)}_{{\bf k},{\bf k}',\alpha}\right)^\star
    \left[  \partial_{\bf k}\, \xi_\alpha \right]
    \left[ \partial_{{\bf k}'}\,\xi_\alpha \right]
  \right)
\nonumber\\
  &+
    \sum_{{\bf k},{\bf k}',{\bf q}}
    \left(d^{(3)}_{{\bf k},{\bf k}',\alpha}\right)^\star
    \left( 
      \left[\partial^\star_{{\bf k}+{\bf q}} \,\xi_\alpha \right]
      \left[\partial_{\bf k}\, \partial_{{\bf k}'} \,\xi_\alpha \right]
      +
      \left[\partial_{\bf k} \,\xi_\alpha \right]
      \left[\partial^\star_{{\bf k}+{\bf q}}\, \partial_{{\bf k}'} \,\xi_\alpha \right]
      +
      \left[ \partial_{{\bf k}'} \,\xi_\alpha \right]
      \left[\partial^\star_{{\bf k}+{\bf q}}\,\partial_{\bf k}\, \xi_\alpha \right]
    \right)
\nonumber\\
  &-
    \sum_{{\bf k},{\bf k}',{\bf q}}
    d^{(3)}_{{\bf k},{\bf k}',\alpha}
    \left( 
      \left[\partial_{{\bf k}+{\bf q}} \,\xi_\alpha \right]
      \left[\partial^\star_{\bf k}\, \partial^\star_{{\bf k}'} \,\xi_\alpha \right]
      +
      \left[\partial^\star_{\bf k} \,\xi_\alpha \right]
      \left[\partial_{{\bf k}+{\bf q}}\, \partial^\star_{{\bf k}'} \,\xi_\alpha \right]
      +
      \left[ \partial^\star_{{\bf k}'} \,\xi_\alpha \right]
      \left[\partial_{{\bf k}+{\bf q}}\,\partial^\star_{\bf k}\, \xi_\alpha \right]
    \right)  
\nonumber\\
  \Xi^{(3)}_\alpha
  &=
  \sum_{{\bf k},{\bf k}',{\bf q}}
  \left(
    \left(d^{(3)}_{{\bf k},{\bf k}',\alpha}\right)^\star 
    \left[
    \partial^\star_{{\bf k}+{\bf q}} \,\xi_\alpha \right]
    \left[\partial_{\bf k}\,\xi_\alpha  \right]
    \left[\partial_{{\bf k}'} \,\xi_\alpha \right]
    -
    d^{(3)}_{{\bf k},{\bf k}',\alpha} 
    \left[ \partial_{{\bf k}+{\bf q}}\,\xi_\alpha \right]
    \left[\partial^\star_{\bf k}\,\xi_\alpha \right]
    \left[\partial^\star_{{\bf k}'} \,\xi_\alpha \right]
  \right)
\label{eq:Xi-lin-quad-cub}\;.
\end{align}
where we have repeated result \eqref{eq:HEM-xi-source} for the sake of completeness; the $d$ coefficients are given by Eq.~\eqref{eq:HEM-d_all}. Equations \eqref{eq:HEM-xiClasses}--\eqref{eq:Xi-lin-quad-cub} define the exact quantum kinetics of the correlation-generating function, which we use as the general starting point to determine the correlation dynamics in the interacting Bose gas.

By comparing contributions \eqref{eq:Xi-lin-quad-cub} with Eq.~\eqref{eq:HEM-d_all}, we observe that $\Xi^{(0)}_\alpha+\Xi^{(1)}_\alpha$ is equal to the $\op{D}_{\rm tot} \, \xi_\alpha$ contribution such that it produces the same integro-differential structure as the $\chi_\alpha$ dynamics defined by Eqs.~\eqref{eq:HEM-chiFIN2}--\eqref{eq:HEM-d_all}. 
Consequently, the correlations and expectation values share an identical hierarchy problem which is, therefore, unavoidable also for the correlations. At the same time, the full $\xi_\alpha$ dynamics is structurally much more complicated than the $\chi_\alpha$ dynamics because Eq.~\eqref{eq:HEM-xiClasses} also contains new nonlinear contributions $\Xi^{(2)}_\alpha+\Xi^{(3)}_\alpha$. 
In Sec.~\ref{sec:INF}, we show that they do not produce a hierarchical coupling but only nonlinear scattering among already existing correlations. Since the hierarchy problem is the core hindrance of many-body solutions, these nonlinear contributions do not affect our overall 
capability to systematically solve the many-body quantum kinetics. However, the nonlinear contributions, make finding exact analytic solutions difficult, as is common for nonlinear equations. To develop viable numerical solution schemes, one obviously needs to truncate the hierarchy problem via a set of approximations that can be systematically improved. In Sec.~\ref{sec:cluster_structure}, we discuss how to implement a controlled truncation of the hierarchy problem by using the cluster-expansion approach.

\section{Implicit-notation formalism (INF)}
\label{sec:INF}

The quantum kinetics of a specific correlation can be directly projected from the $\xi_\alpha$ dynamics \eqref{eq:HEM-xiClasses}--\eqref{eq:Xi-lin-quad-cub} by using relation \eqref{eq:DI-rep}. A straightforward differentiation of the nonlinear $\Xi^{(n\ge2)}_\alpha$ contributions then produces a complicated correlation-dependent structure that yields a tedious bookkeeping problem as the cluster number increases. 
To project these contributions efficiently for each correlation, it is beneficial to develop a new approach --- the implicit-notation formalism (INF) --- that allows us to determine the quantum dynamics of {\it all} correlations with one derivation.

The INF identifies a specific correlation implicitly via
\begin{eqnarray}
  \DI{\seq{K}}{\seq{P}}
  \equiv
  \Delta\avex{
    B^\dagger_{{\bf k}_1} \cdots B^\dagger_{{\bf k}_K}
    B_{{\bf p}_P} \cdots B_{{\bf p}_1}
  }
\label{eq:imp-corr}\;,
\end{eqnarray}
that is uniquely defined by {\it state sequences}
\begin{eqnarray}
  \seq{K} \equiv {\bf k}_1,{\bf k}_2,\cdots,{\bf k}_K\,,
  \qquad
  \seq{P} \equiv {\bf p}_1,{\bf p}_2,\cdots,{\bf p}_P
\label{eq:imp-seq}\;,
\end{eqnarray}
denoting the $K$ and $P$ boson states involved. In general, both $K$ and $P$ have a dual meaning: they denote both the number of elements as bare variables $K$ or $P$ while they identify a state sequences $\seq{K}$ or $\seq{P}$ when they appear with braces $\seq{\cdots}$. Note that $K$ and $P$ do not have to be equal in the excitation picture, as discussed in Sec.~\ref{sec:relevant clusters}. 
The creation-operator sequences are positioned as superscripts whereas subscripts assign the annihilation operators. To increase the distinguishability of creation- and annihilation-operator sequences, we adopt a notation convention where $K$ ($P$) identifies a sequence of creation (annihilation) operators and the sequence elements are denoted by wave vectors ${\bf k}$ (symbols ${\bf p}$) defining the specific atomic states. 
Since we are performing the derivations in the excitation picture, none of these vectorial symbols vanishes because all of them refer to normal-component atoms with ${\bf k} \neq 0$ and ${\bf p} \neq 0$.

\subsection{Pointers in the INF}
\label{sec:INFpointers}

To make the INF particularly efficient, we also introduce {\it pointer sequences}
\begin{eqnarray}
  \ind{\lambda} \equiv j_1,j_2,\cdots,j_\lambda\,,
  \qquad
  \ind{\nu} \equiv l_1,l_2,\cdots,l_\nu
\label{eq:imp-ind}\;
\end{eqnarray}
whose elements $j_s$ and $l_s$ are integer-valued pointers. Later on, we will use $j_s$ and $l_s$ to identify which elements are selected from full state sequences $\seq{K}$ and $\seq{P}$, respectively. Analogous to definition \eqref{eq:imp-corr}, $\lambda$ and $\nu$ label both the pointer sequence as well as the number of elements. Since the pointer sequences are applied to select states from the state sequences, it is clear that the number of pointers should be lower than or equal to the number of elements in a particular state sequence, i.e.~$\lambda \le K$ and $\nu \le P$.

\begin{figure}[t]
\includegraphics*[scale=0.42]{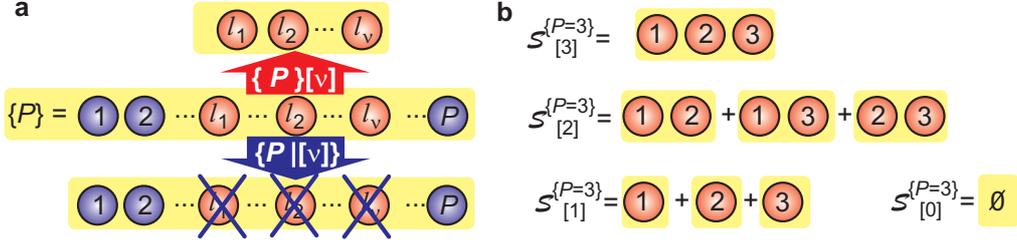}
\caption{Graphical representation of INF operations. Elements of state sequences are identified as spheres; the number denotes the specific pointer identifying the state. {\bf a} Pointer operations producing pointer selcetion ($\ssel{P}{\ind{\nu}}$, up) and ($\srm{P}{\ind{\nu}}$, down). {\bf b} Ordered pointer-selection sums for a sequence that contains three elements, according to Eq.~\eqref{eq:sum_def}.
}\label{fig:PointerOP}
\end{figure}

With the help of the pointers, we can introduce two classes of pointer operations: we can perform {\it element selection} via
\begin{eqnarray}
  \ssel{K}{\ind{\lambda}} \equiv {\bf k}_{j_1},{\bf k}_{j_2},\cdots,{\bf k}_{j_\lambda}\,,
\label{eq:selection}
\end{eqnarray}
where the outcome creates a new state sequence containing the elements corresponding to the pointers. The second pointer operator follows from the {\it element elimination}
\begin{eqnarray}
  \srm{K}{\ind{\lambda}} \equiv {\bf k}_1,\cdots,\xout{{\bf k}_{j_1}},\cdots,\xout{{\bf k}_{j_\lambda}},\cdots,{\bf k}_K\,,
\label{eq:elimination}\;,
\end{eqnarray}
where the outcome is a state sequence where elements ``selected'' by $\ind{\lambda}$ are removed (crossed out symbols) from the position identified by the pointers. Both the element selection and elimination are defined identically for $\seq{P}$ and illustrated schematically in Fig.~\ref{fig:PointerOP}{\bf a}; the spheres symbolize all states (middle sequence) within $\seq{P}$, 
the red spheres identify the states indicated by $\ind{\nu}$, the shaded yellow area denotes a state sequence, the result of the $\ssel{P}{\ind{\nu}}$ (arrow up) is illustrated by the upper sequence, and $\srm{P}{\ind{\nu}}$ outcome (arrow down) is depicted by the lower sequence. To streamline further derivations, we also introduce two more explicit forms,
\begin{eqnarray}
  \seq{K_j} \equiv \srm{K}{j}\,, \qquad   \seq{K_{j_1,j_2}} \equiv \srm{K}{j_1,j_2}
\label{eq:simple-elim}\;,
\end{eqnarray}
to denote the removal of a single state or two states, respectively. Also $\seq{P_l}$ and  $\seq{P_{l_1,l_2}}$ have analogous forms.

To make sure that the sequence operations \eqref{eq:selection}--\eqref{eq:elimination} are well-defined for all eventualities, we also need to determine how an empty set $\emptyset$ is interpreted in different instances. The empty set has zero elements, implying
\begin{eqnarray}
  \seq{0} \equiv \emptyset\,, \qquad \ind{0} \equiv \emptyset
\label{eq:imp-empty}\,,
\end{eqnarray}
for the state and pointer sequences, respectively. For state sequences, the $\emptyset$ argument must denote a correlation without the corresponding operators, i.e.
\begin{eqnarray}
  \DI{\emptyset}{\seq{P}}
  =
  \Delta\avex{
    B_{{\bf p}_P} \cdots B_{{\bf p}_1}
  }\,,
  \qquad
  \DI{\seq{K}}{\emptyset}
  =
  \Delta\avex{
    B^\dagger_{{\bf k}_1} \cdots B^\dagger_{{\bf k}_K}
  }\,,
  \qquad
  \DI{\emptyset}{\emptyset} \equiv 1
\label{eq:empty-corr}\;,
\end{eqnarray}
when definition \eqref{eq:imp-corr} is applied. The element selection \eqref{eq:selection} and elimination \eqref{eq:elimination} yield
\begin{eqnarray}
  \ssel{K}{\emptyset} = \emptyset\,,\qquad
  \srm{K}{\emptyset} = \seq{K}
\label{eq:empty-elim}\;,
\end{eqnarray}
respectively, when they have $\emptyset$ as an argument. These limiting cases are definied analogously for the annihilation-operator sequences; $K$ is simply replaced by $P$ in Eq.~\eqref{eq:empty-elim}. We also need to define what happens when either the element selection or the elimination exceeds the number of indices in the original sequence. We simply set the corresponding correlations to zero
\begin{eqnarray}
  \DI{\srm{K}{\ind{\lambda>K}}}{\seq{P}} =0\,, \quad \DI{\seq{K}}{\srm{P}{\ind{\nu>P}}} =0\,,
\quad
  \DI{\ssel{K}{\ind{\lambda>K}}}{\seq{P}} =0\,, \quad \DI{\seq{K}}{\ssel{P}{\ind{\nu>P}}} =0
\label{eq:too-many}\;
\end{eqnarray}
whenever either $\ind{\lambda}$ or $\ind{\nu}$ have too many indices compared to the number of elements within the sequences $\seq{K}$ or $\seq{P}$, respectively. 

\subsection{Differentiation in the INF}
\label{sec:INFdifferentiation}

The specific correlation dynamics can be obtained by applying the reduction formula \eqref{eq:DI-rep}. To express it with the INF, we introduce a {\it collective differentiation}
\begin{eqnarray}
  \Dcoll{\seq{K}}{\seq{P}}
  \equiv
  \partial_{{\bf k}_1} \cdots \partial_{{\bf k}_K} 
  \partial^\star_{{\bf p}_P} \cdots \partial^\star_{{\bf p}_1}  
\label{eq:D_collective}\;,
\end{eqnarray}
resembling definition \eqref{eq:imp-corr}. With the help of this and identification \eqref{eq:imp-corr}, the reduction formula \eqref{eq:DI-rep} casts into the form
\begin{eqnarray}
  \DI{\seq{K}}{\seq{P}}
  = \left. (-1)^P \Dcoll{\seq{K}}{\seq{P}}\, \xi_\alpha \right|_{\{\alpha=0\}} 
\label{eq:INF_reduction}\;.
\end{eqnarray}
In this context, $\{\alpha\}$ refers to the set of all $\alpha_{\bf k}$ and $\alpha^\star_{\bf p}$ indices connected with the normal-component states ${\bf k} \neq 0$ and ${\bf p} \neq 0$. 

Since $\xi_\alpha$ is an analytic function in the vicinity of $\{\alpha=0\}$, we may calculate the differentiation in any order, i.e.
\begin{eqnarray}
  \Dcoll{\seq{K}}{\seq{P}}
  =
  \Dcoll{\emptyset}{\seq{P}} \Dcoll{\seq{K}}{\emptyset}
  =
  \Dcoll{\seq{K}}{\emptyset}\Dcoll{\emptyset}{\seq{P}}
\label{eq:D_orders}\;,
\end{eqnarray}
where the $\emptyset$ argument means nonexistent contribution of the corresponding differentiation. More explicitly, we use
\begin{eqnarray}
  \Dcoll{\emptyset}{\seq{P}}
  = 
  \partial^\star_{{\bf p}_P} \cdots \partial^\star_{{\bf p}_1} 
\,,
  \qquad
  \Dcoll{\seq{K}}{\emptyset}
  =
  \partial_{{\bf k}_1} \cdots \partial_{{\bf k}_K}\,,
  \qquad
  \Dcoll{\emptyset}{\emptyset} \equiv 1
\label{eq:empty-diff}\;,
\end{eqnarray}
in analogy to Eq.~\eqref{eq:empty-corr}. We may further divide differentiations into two groups with selection and elimination operations \eqref{eq:selection}--\eqref{eq:elimination}, yielding
\begin{eqnarray}
  \Dcoll{\seq{K}}{\emptyset}
  =
  \Dcoll{\ssel{K}{\ind{\lambda}}}{\emptyset}\,\Dcoll{\srm{K}{\ind{\lambda}}}{\emptyset}\,,
\quad
  \Dcoll{\emptyset}{\seq{P}} 
  =
  \Dcoll{\emptyset}{\ssel{P}{\ind{\nu}}}\, \Dcoll{\emptyset}{\srm{P}{\ind{\nu}}}\,,
\quad
  \Dcoll{\seq{K}}{\seq{P}} 
  =
  \Dcoll{\ssel{K}{\ind{\lambda}}}{\ssel{P}{\ind{\nu}}}\,\Dcoll{\srm{K}{\ind{\lambda}}}{\srm{P}{\ind{\nu}}}
\label{eq:D_split}\;,
\end{eqnarray}
provided that the element numbers satisfy $\lambda \le K$ and $\nu \le P$.  These relations become indispensable when we evaluate the correlation dynamics with one derivation.

\subsection{Useful differentiation relations}
\label{sec:INFusefulD}

The explicit correlation dynamics can be identified by applying the $\Dcoll{\seq{K}}{\seq{P}}$ differentiation to 
the both sides of Eq.~\eqref{eq:HEM-xiClasses} and by setting $\{\alpha=0\}$ based on the reduction formula \eqref{eq:INF_reduction}. Equation \eqref{eq:HEM-d_all} implies that we must also differentiate the quadratic $\xi_\alpha^2$ and cubic $\xi^3_\alpha$ contributions which yield the most difficult part of the actual derivation. 
Additionally, Eqs.~\eqref{eq:HEM-d_all} and \eqref{eq:Xi-lin-quad-cub} suggest that we also need to deal with differentiation involving mixed products of $\alpha$-dependent polynomials and $\xi_\alpha$. Next, we work out how to evaluate both the nonlinear $\xi_\alpha$ as well as $\alpha$-polynomial contributions before we present the full dynamics in Sec.~\ref{sec:cluster_dyn}.

To solve the quadratic and cubic differentiation, we consider how generic quadratic $F_1 \, F_2$  and cubic $F_1\, F_2\,  F_3$ terms behave when differentiated; $F_1$, $F_2$, and $F_3$ are generic functions. We find in   \ref{app:relevant_D} that the differentiation can then be presented via
\begin{eqnarray}
  \Dcoll{\seq{K}}{\seq{P}}\, F_1\,F_2
  &=&
  \sum_\ind{\lambda}^K \sum_\ind{\nu}^P
  \left[
    \Dcoll{\ssel{K}{\ind{\lambda}}}{\ssel{P}{\ind{\nu}}}
    F_1
  \right]
  \left[
    \Dcoll{\srm{K}{\ind{\lambda}}}{\srm{P}{\ind{\nu}}}
    F_2
  \right]
\nonumber\\
  \Dcoll{\seq{K}}{\seq{P}}\, F_1\,F_2\,F_3
  &=&
  \sum_\ind{\lambda_1,\lambda_2}^K \sum_\ind{\nu_1,\nu_2}^P
  \left[
    \Dcoll{\ssel{K}{\ind{\lambda_1}}}{\ssel{P}{\ind{\nu_1}}}\,
    F_1
  \right]
  \left[
    \Dcoll{\ssel{K\big|\ind{\lambda_1}}{\ind{\lambda_2}}}{\ssel{P\big|\ind{\nu_1}}{\ind{\nu_2}}}\,
    F_2
  \right]
    \left[
    \Dcoll{\srm{\srm{K}{\ind{\lambda_1}}}{\ind{\lambda_2}}}{\srm{\srm{P}{\ind{\nu_1}}}{\ind{\nu_2}}}\,
    F_3
  \right]
\label{eq:INF-diff2and3}\;.
\end{eqnarray}
By using the pointer operations \eqref{eq:selection}--\eqref{eq:elimination}  as well as relation \eqref{eq:D_split}, the sums are defined to pick all {\it unique} combinations to select elements into separate differentiations from the  sequences. The {\it combinatorial sums} can be explicitly implemented by using the {\it ordered pointer-selection sum}:
\begin{eqnarray}
  \sel^K_{\ind{\lambda}}
  \equiv
  \sum_{j_1=1}^{K-\lambda+1} \sum_{j_2=1+j_1}^{K-\lambda+2} \cdots  \sum_{j_\lambda=1+j_{\lambda-1}}^{K}
  {\rm Sel}\left(\ind{\lambda} \rightarrow j_1,\cdots, j_\lambda \right)
\label{eq:sum_def}\;,
\end{eqnarray}
where ${\rm Sel}\left(\ind{\lambda} \rightarrow j_1,\cdots, j_\lambda \right)$ replaces pointer $\ind{\lambda}$ by a specific choice, i.e.~$j_1,\cdots, j_\lambda$, when it acts upon any $\ind{\lambda}$-dependent function. The $j_1,\cdots, j_\lambda$ sums generate all ordered pointer selections $j_1,\cdots, j_\lambda$ out of $K$ available indices only once. 
As a result, $\sel^K_{\ind{\lambda}}$ indeed generates a sum over all possible pointer selections when it acts upon $\ind{\lambda}$-dependent functions. Since $\lambda$ can refer to any number between $0$ and $K$, the general pointer selection becomes
\begin{eqnarray}
  &&
  \sum_{\ind{\lambda}}^K 
  \equiv 
  \sum_{\lambda=0}^K
  \sel^K_\ind{\lambda}\,,
  \qquad
  \sum_{\ind{\lambda_1,\lambda_2}}^K \equiv 
  \sum_{\lambda_1}^K \sum_{\lambda_2=0}^{K-\lambda_1} \sel^K_\ind{\lambda_1} \, \sel^{K-\lambda_1}_\ind{\lambda_2}
\label{eq:pointer_selK}\;,
\end{eqnarray}
for one and two selection groups, respectively. Analogously, $P$ selections follow directly from Eqs.~\eqref{eq:sum_def} via substitutions of symbols $K \rightarrow P$, $\lambda \rightarrow \nu$, and $j \rightarrow l$. Pointer selections $\sel^3_{\ind{3}}$, $\sel^3_{\ind{2}}$, $\sel^3_{\ind{1}}$, and $\sel^3_{\ind{0}}$ are schematically presented in Fig.~\ref{fig:PointerOP}{\bf b}. The selected index groups are collected within the yellow square while the index sums produce each of the selections only once. 

In general, each $\sel^K_{\ind{\lambda}}$ or $\sel^K_{\ind{\lambda_1}} \sel^{K-\lambda_1}_{\ind{\lambda_2}}$
produces
\begin{eqnarray}
  N^K_\lambda \equiv \frac{K!}{\lambda!\,(K-\nu)!}\,, \qquad
  N^K_{\lambda_1,\lambda_2} \equiv N^K_{\lambda_1} \, N^{K-\lambda_1}_{\lambda_1} 
  = \frac{K!}{\lambda_1!\,\lambda_2!\,(K-\lambda_1-\lambda_2)}
\label{eq:N_combinatorial}\;
\end{eqnarray}
terms, respectively, as defined by the binomial coefficient\cite{Bryant:1992}. With the help of these, one can also count the number of terms 
\begin{eqnarray}
  N\left( \sum_{\ind{\lambda}}^K  \right) = 2^K\,,
  \qquad
  N\left( \sum_{\ind{\lambda_1,\lambda_2}}^K \right) = 3^K
\label{eq:N_combinatorial-rest}\;,
\end{eqnarray}
within different combinatorial sums. Both of these follow directly from the binomial coefficient \eqref{eq:N_combinatorial} and the binomial theorem\cite{Arfken:1985}. Clearly, the number of selections ($2^3=8$) within Fig.~\ref{fig:PointerOP}{\bf b} satisfies Eq.~\eqref{eq:N_combinatorial-rest}.

The polynomial $\alpha$-contributions of Eqs.~\eqref{eq:HEM-xiClasses}--\eqref{eq:Xi-lin-quad-cub} have both linear $\alpha_{\bf k}$ and quadratic $\alpha^\star_{\bf k}$ contributions. When a linear contribution is differentiated together with $\xi_\alpha$, we obtain
\begin{eqnarray}
  \Dcoll{\seq{K}}{\seq{P}} \alpha_{\bf k} \, \xi_\alpha
  = 
  \sum_{j=1}^K \delta_{{\bf k},{\bf k}_j} \Dcoll{\seq{K_j}}{\seq{P}} \xi_\alpha
  +
  \alpha_{\bf k} \, \Dcoll{\seq{K}}{\seq{P}} \, \xi_\alpha
\label{eq:diff-alpha1}\;,
\end{eqnarray}
based on relation \eqref{eq:INF-diff2and3}. Since we always set all $\alpha_{\bf k}$ variables to zero after the differentiation, in order to identify correlations with relation \eqref{eq:INF_reduction}, the second term of Eq.~\eqref{eq:diff-alpha1} does not contribute. Therefore, we may use substitution 
\begin{eqnarray}
  \Dcoll{\seq{K}}{\seq{P}} \alpha_{\bf k} 
  \rightarrow
  \sum_{j=1}^K \delta_{{\bf k},{\bf k}_j} \Dcoll{\seq{K_j}}{\seq{P}}
\label{eq:diff-subst1}\;
\end{eqnarray}
when evaluating the explicit correlation dynamics from Eqs.~\eqref{eq:HEM-xiClasses}--\eqref{eq:HEM-xi-source} and \eqref{eq:Xi-lin-quad-cub}. The remaining linear and quadratic contributions follow from analogous substitutions:
\begin{align}
  &\Dcoll{\seq{K}}{\seq{P}} \alpha_{\bf k} \alpha_{{\bf k}'}
  \rightarrow
 \sum_{j_1=1}^K \sum_{j_2 \neq j_1}^K 
  \delta_{{\bf k},{\bf k}_{j_1}} \delta_{{\bf k}',{\bf k}_{j_2}} 
  \Dcoll{\seq{K_{j_1,j_2}}}{\seq{P}}
\nonumber\\
  &\Dcoll{\seq{K}}{\seq{P}} \alpha^\star_{\bf p} 
  \rightarrow
  \sum_{l=1}^P \delta_{{\bf p},{\bf p}_l} \Dcoll{\seq{K}}{\seq{P_l}} \,,
\qquad 
  \Dcoll{\seq{K}}{\seq{P}} \alpha^\star_{\bf p} \alpha^\star_{{\bf p}'}
  \rightarrow
 \sum_{l_1=1}^P \sum_{l_2 \neq l_1}^P 
  \delta_{{\bf p},{\bf p}_{l_1}} \delta_{{\bf p}',{\bf p}_{l_2}} 
  \Dcoll{\seq{K}}{\seq{P_{l_1,l_2}}}
\label{eq:diff-substREST}\;,
\end{align}
after applying relation \eqref{eq:simple-elim} and property \eqref{eq:INF-diff2and3} to have a shorter form for the state elimination.

\section{Quantum kinetics of atom clusters}
\label{sec:cluster_dyn}

To solve the quantum dynamics of correlations with one derivation, we follow next the straightforward strategy outlined at the beginning of Sec.~\ref{sec:INFusefulD}. More specifically, we apply the projection \eqref{eq:INF_reduction} to both sides of Eq.~\eqref{eq:HEM-xiClasses}, producing
\begin{eqnarray}
  \ihddt \DI{\seq{K}}{\seq{P}}
  &=&
  (-1)^P
  \left.
  \left(
    \Dcoll{\seq{K}}{\seq{P}}\, \Xi^{(0)}_\alpha 
    +
    \Dcoll{\seq{K}}{\seq{P}}\, \Xi^{(1)}_\alpha 
    +
    \Dcoll{\seq{K}}{\seq{P}}\, \Xi^{(2)}_\alpha 
    +
    \Dcoll{\seq{K}}{\seq{P}}\, \Xi^{(3)}_\alpha 
 \right)\right|_{\{\alpha=0\}}
 \nonumber\\
  &\equiv&
  \sS{\seq{K}}{\seq{P}}+\sL{\seq{K}}{\seq{P}}+\sQ{\seq{K}}{\seq{P}}+\sC{\seq{K}}{\seq{P}}
\label{eq:DI-dyn-start}\;,
\end{eqnarray}
where we have identified the source, linear, quadratic, and cubic contribution to be
\begin{eqnarray}
  &&\sS{\seq{K}}{\seq{P}}
  \equiv
  (-1)^P\left. \Dcoll{\seq{K}}{\seq{P}}\, \Xi^{(0)}_\alpha \right|_{\{\alpha=0\}}\,, 
  \quad 
  \sL{\seq{K}}{\seq{P}}
  \equiv
  (-1)^P\left. \Dcoll{\seq{K}}{\seq{P}}\, \Xi^{(1)}_\alpha \right|_{\{\alpha=0\}}\,,
\nonumber\\
  &&\sQ{\seq{K}}{\seq{P}}
  \equiv
  (-1)^P\left. \Dcoll{\seq{K}}{\seq{P}}\, \Xi^{(2)}_\alpha \right|_{\{\alpha=0\}}\,, 
  \quad 
  \sC{\seq{K}}{\seq{P}}
  \equiv
  (-1)^P\left. \Dcoll{\seq{K}}{\seq{P}}\, \Xi^{(3)}_\alpha \right|_{\{\alpha=0\}}\,,
\label{eq:DI-source}
\end{eqnarray}
respectively. Each of the terms can be evaluated straightforwardly by using the INF operations and differentiations identified in Sec.~\ref{sec:chi_dyn}.

\subsection{Source of atom-cluster generation}
\label{sec:cluster_source}

Obviously, the simplest contribution follows from the source $\sS{\seq{K}}{\seq{P}}$. With the help definitions \eqref{eq:HEM-xi-source} and \eqref{eq:DI-source}, it becomes
\begin{eqnarray} 
  \sS{\seq{K}}{\seq{P}}
  =
  (-1)^P \sum_{{\bf k}} {\textstyle \frac{\NC\,V_{\bf k}}{2}}\, 
  \left(
    \Dcoll{\seq{K}}{\seq{P}}\, \alpha^\star_{\bf k} \alpha^\star_{-{\bf k}} 
    -
    \Dcoll{\seq{K}}{\seq{P}}\, \alpha_{\bf k} \alpha_{-{\bf k}} 
  \right)_{\{\alpha=0\}}
\label{eq:xi-source0-dyn1}\;.
\end{eqnarray}
By implementing differentiation-relation \eqref{eq:diff-substREST}, we obtain
\begin{eqnarray}
  \sS{\seq{K}}{\seq{P}}
  =
  (-1)^P
  \sum_{{\bf k}} {\textstyle \frac{\NC\,V_{\bf k}}{2}}\,
  \left(
    \sum_{l_1=1}^P \sum_{l_2 \neq l_1}^P 
    \delta_{{\bf k},{\bf p}_{l_1}} \delta_{-{\bf k},{\bf p}_{l_2}} 
    \Dcoll{\seq{K}}{\seq{P_{l_1,l_2}}} 
    -
    \sum_{j_1=1}^K \sum_{j_2 \neq j_1}^K 
    \delta_{{\bf k},{\bf k}_{j_1}} \delta_{-{\bf k},{\bf k}_{j_2}} 
    \Dcoll{\seq{K_{j_1,j_2}}}{\seq{P}} 
  \right)_{\{\alpha=0\}}
\label{eq:xi-source0-dyn2}\;.
\end{eqnarray}
Since the differentiation acts on a constant, i.e.~1, it produces a vanishing result in all cases except when it has empty sets as arguments, according to relation \eqref{eq:empty-corr}. Therefore, the first term exists only for $K=0$ and $P=2$ while the second term must have $K=2$ and $P=0$. By applying these conditions to Eq.~\eqref{eq:xi-source0-dyn2}, we eventually find
\begin{eqnarray}
  \sS{\seq{K}}{\seq{P}}
  =
  \delta_{K,0}\,\delta_{P,2} \,\delta_{{\bf p}_1+{\bf p}_2,0}\,
  \NC\,V_{{\bf p}_1} 
  -
  \delta_{K,2}\,\delta_{P,0}\,\delta_{{\bf k}_1+{\bf k}_2,0} \,
  \NC\,V_{{\bf k}_1}
\label{eq:xi-source0-FIN}\;,
\end{eqnarray}
after implementing the Kronecker delta functions under the sums.

We see that contribution \eqref{eq:xi-source0-FIN} exists only for correlations having either two annihilation ($K=0$ and $P=2$) or creation operators ($K=2$ and $P=0$) among atoms whose total momentum adds up to zero. Therefore, we only need to consider how correlation
\begin{eqnarray}
  s_{\bf k} 
  \equiv \DI{\emptyset}{{\bf k},-{\bf k}}
  = \Delta\avex{B_{\bf k} B_{-{\bf k}}}
\label{eq:squeezing_s}
\end{eqnarray}
and its complex conjugate $s^\star_{\bf k} = \Delta\avex{B^\dagger_{-{\bf k}} B^\dagger_{\bf k} }$ are affected by the source term; when expressing $s_{\bf k}$, we have chosen $\seq{P}={\bf k},-{\bf k}$ with momenta that produce a nonvanishing source when implemented to Eq.~\eqref{eq:xi-source0-FIN}. Physically, $s_{\bf k}$ describes the amplitude of correlations among two normal-component atoms and it is directly related to the two-mode squeezing of boson fields\cite{Walls:2008} analyzed in Sec.~\ref{sec:bidirectional}; Its connection with the anomalous density\cite{Zaremba:1999,Milstein:2003,Wuster:2005} and the Bogoliubov excitations\cite{StamperKurn:1999,Jin:1996,Mewes:1996,Utsunomiya:2008} is discussed in Secs.~\ref{sec:HBE} and \ref{sec:Bogoliubov}, respectively.

By combining Eqs.~\eqref{eq:DI-dyn-start} and \eqref{eq:xi-source0-FIN}--\eqref{eq:squeezing_s}, we find that
\begin{eqnarray}
  \left.\ihddt s_{\bf k}\right|_{\rm S} \equiv \sS{\emptyset}{{\bf k},-{\bf k}} = \NC\,V_{{\bf k}} 
\label{eq:s_source}
\end{eqnarray}
is driven directly by $\NC\,V_{{\bf k}}$. In other words, whenever the system contains a BEC occupation and a nonvanishing $V_{\bf k}$, interactions directly generate the two-atom correlation amplitude. Therefore, $\sS{\seq{K}}{\seq{P}}$ indeed acts as a source that excites atoms from the BEC to the normal component of the system, giving rise to the quantum depletion. Since such a source exists only for the $s_{\bf k}$ correlation, the BEC always emits two correlated atoms to states with opposite momenta. 
This elementary process corresponds to the left diagram of Fig.~\ref{Qdepletion} in the original picture where the peak signifies the BEC and the quantum depletion (arrows) converts two BEC atoms into two normal-component atoms having wave vectors $+{\bf k}$ and $-{\bf k}$. In the excitation picture, only the creation of two normal-component atoms appears because this picture focuses the investigation around the BEC that formally remains a particle vacuum. 

\subsection{Full correlation dynamics}
\label{sec:cluster_kinetics}

Also the linear, quadratic, and cubic contributions follow after a derivation analogous to that performed above for the source. To illustrate the new calculation-technical aspects, we perform here explicitly only the derivation for the simplest quantum dynamics, involving the renormalized kinetic energy. By using definitions \eqref{eq:HEM-d_all}, \eqref{eq:Xi-lin-quad-cub}, and \eqref{eq:DI-source}, we find that the kinetic energy contributes only to
\begin{eqnarray}
  \left.\sL{\seq{K}}{\seq{P}}\right|_{\rm kin}
  &\equiv&
  (-1)^P
  \sum_{\bf k} 
  \tilde{E}_{\bf k}
  \left( 
     \Dcoll{\seq{K}}{\seq{P}}\,
     \alpha^\star_{\bf k}
    \left[\partial^\star_{\bf k} \xi_\alpha\right]
    -
    \Dcoll{\seq{K}}{\seq{P}}\,
     \alpha_{\bf k} 
    \left[\partial_{\bf k} \, \xi_\alpha \right]
  \right)_{\{\alpha=0\}}
\label{eq:HEM-Lterm1}\;,
\end{eqnarray}
after some reorganization of terms. Transformation \eqref{eq:diff-subst1} converts this expression into
\begin{eqnarray}
  \left.\sL{\seq{K}}{\seq{P}}\right|_{\rm kin}
  &\equiv&
  (-1)^P
  \sum_{\bf k} 
  \tilde{E}_{\bf k}
  \left(
    \sum_{l=1}^P \delta_{{\bf k},{\bf p}_l} 
    \left[
      \Dcoll{\seq{K}}{\seq{P_l}}\, \partial^\star_{\bf k} \xi_\alpha
    \right]
    -
    \sum_{j=1}^K \delta_{{\bf k},{\bf k}_j} 
    \left[
      \Dcoll{\seq{K_j}}{\seq{P}}\, \partial_{\bf k} \, \xi_\alpha 
    \right]
  \right)_{\{\alpha=0\}}
\nonumber\\
&=&
  (-1)^P
  \sum_{\bf k} 
  \tilde{E}_{\bf k}
  \left(
    \sum_{l=1}^P \delta_{{\bf k},{\bf p}_l} 
    \left[
      \Dcoll{\seq{K}}{\seq{P_l},{\bf k}}\,\xi_\alpha
    \right]
    -
    \sum_{j=1}^K \delta_{{\bf k},{\bf k}_j}
    \left[
      \Dcoll{\seq{K_j},{\bf k}}{\seq{P}} \, \xi_\alpha 
    \right]
  \right)_{\{\alpha=0\}}
\label{eq:HEM-Lterm2}\;,
\end{eqnarray}
where the second step follows after we have combined the differentials. By enforcing the Kronecker delta functions we find
\begin{eqnarray}
  \left.\sL{\seq{K}}{\seq{P}}\right|_{\rm kin}
&=&
    \sum_{l=1}^K 
    \tilde{E}_{{\bf p}_l}\,
    (-1)^P
    \left[
      \Dcoll{\seq{K}}{\seq{P_l},{\bf p}_l}\,\xi_\alpha
    \right]_{\{\alpha=0\}}
    -
    \sum_{j=1}^K \tilde{E}_{{\bf k}_j}\,
    (-1)^P
    \left[
      \Dcoll{\seq{K_j},{\bf k}_j}{\seq{P}} \, \xi_\alpha 
    \right]_{\{\alpha=0\}}
\label{eq:HEM-Lterm3}\;.
\end{eqnarray}
Since $\seq{K_j},{\bf k}_j$ and ${\seq{P_l},{\bf p}_l}$ define the same sequence as $\seq{K}$ and $\seq{P}$, respectively, reduction formula  \eqref{eq:INF_reduction} identifies the same correlation $\DI{\seq{K}}{\seq{P}}$ in all terms, leading to the expected kinetic-energy expression
\begin{eqnarray}
  \left.\sL{\seq{K}}{\seq{P}}\right|_{\rm kin}
&=&
  \left(
    \sum_{l=1}^K 
    \tilde{E}_{{\bf p}_l}\,
    -
    \sum_{j=1}^K \tilde{E}_{{\bf k}_j}\
  \right)
  \DI{\seq{K}}{\seq{P}}
\label{eq:HEM-LtermFIN}
\end{eqnarray}
that is proportional to the difference of kinetic energies of creation and annihilation operators that appear in the correlation $\DI{\seq{K}}{\seq{P}}$.

The remaining linear, quadratic, and cubic contributions follow with an analogous steps. The generic quantum-kinetics of clusters then becomes
\begin{eqnarray}
  \ihddt \DI{\seq{K}}{\seq{P}}
  &=&
  \sS{\seq{K}}{\seq{P}}+\sL{\seq{K}}{\seq{P}}+\sQ{\seq{K}}{\seq{P}}+\sC{\seq{K}}{\seq{P}}
\label{eq:clusterDYN}\;.
\end{eqnarray}
The explicit definition of the $\DI{\seq{K}}{\seq{P}}$ correlation is given by Eq.~\eqref{eq:imp-corr}. For the sake of completeness, we repeat here the source term \eqref{eq:xi-source0-FIN}
\begin{eqnarray}
  \sS{\seq{K}}{\seq{P}}
  =
  \delta_{K,0}\,\delta_{P,2} \,\delta_{{\bf p}_1+{\bf p}_2,0}\,
  \NC\,V_{{\bf p}_1} 
  -
  \delta_{K,2}\,\delta_{P,0}\,\delta_{{\bf k}_1+{\bf k}_2,0} \,
  \NC\,V_{{\bf k}_1}
\label{eq:source-FIN}\;.
\end{eqnarray}
Besides the kinetic energy \eqref{eq:HEM-LtermFIN}, the linear contributions consist of
\begin{align}
  \sL{\seq{K}}{\seq{P}}
&=
  \left(
    \sum_{l=1}^P 
    \tilde{E}_{{\bf p}_l}\,
    -
    \sum_{j=1}^K \tilde{E}_{{\bf k}_j}\
  \right)
  \DI{\seq{K}}{\seq{P}}
  +
  \sum_{l=1}^P W^{[2]}_{{\bf p}_l} \DI{\seq{K},-{\bf p}_l}{\seq{P_l}}
  -
  \sum_{j=1}^K W^{[2]}_{{\bf k}_j} \DI{\seq{K_j}}{\seq{P},-{\bf k}_j}
\nonumber\\
&+
  \sum_{\bf q} \sum_{l=1}^P
  \left(
    {\textstyle \frac{W^{[1]}_{{\bf q},{\bf p}_l-{\bf q}}}{2}} \,  \DI{\seq{K}}{\seq{P_l},{\bf q},{\bf p}_l-{\bf q}}
    +
    W^{[1]}_{{\bf q},{\bf p}_l} \,  \DI{\seq{K},{\bf q}}{\seq{P_l},{\bf p}_l+{\bf q}}
    +
    W^{[0]}_{{\bf q}} 
    \sum_{{\bf q}'}
    \DI{\seq{K},{\bf q}+{\bf q}'}{\seq{P_l},{\bf p}_l+{\bf q},{\bf q}'}
  \right)
\nonumber\\
&-
  \sum_{\bf q} \sum_{j=1}^K
  \left(
    {\textstyle \frac{W^{[1]}_{{\bf q},{\bf k}_j-{\bf q}}}{2}} \,  \DI{\seq{K_j},{\bf q},{\bf k}_j-{\bf q}}{\seq{P}}
    +
    W^{[1]}_{{\bf q},{\bf k}_j} \,  \DI{\seq{K_j},{\bf k}_j+{\bf q}}{\seq{P},{\bf q}}
    +
    \sum_{{\bf q}'}
    W^{[0]}_{{\bf q}'} \,  \DI{\seq{K_j},{\bf k}_j+{\bf q},{\bf q}'}{\seq{P},{\bf q}+{\bf q}'}
  \right)
\nonumber\\
&+
  \sum_{l_1=1}^{P-1} \sum_{l_2=l_1+1}^P
  \left(
    W^{[1]}_{{\bf p}_{l_1},{\bf p}_{l_2}} \,  \DI{\seq{K}}{\seq{P_{l_1,l_2}},{\bf p}_{l_1}+{\bf p}_{l_2}}
    +
    \sum_{\bf q}
     W^{[0]}_{{\bf q}} \,  \DI{\seq{K}}{\seq{P_{l_1,l_2}},{\bf p}_{l_1}+{\bf q},{\bf p}_{l_2}-{\bf q}}
  \right)
\nonumber\\
&-
  \sum_{j_1=1}^{K-1} \sum_{j_2=j_1+1}^K
  \left(
    W^{[1]}_{{\bf p}_{j_1},{\bf p}_{j_2}} \,  \DI{\seq{K_{j_1,j_2}},{\bf k}_{j_1}+{\bf k}_{j_2}}{\seq{P}}
    +
    \sum_{\bf q}
    W^{[0]}_{{\bf q}} \,  \DI{\seq{K_{j_1,j_2}},{\bf k}_{j_1}+{\bf q},{\bf k}_{j_2}-{\bf q}}{\seq{P}}
  \right)
\label{eq:linear-FIN}\;.
\end{align}
The sum terms that appear vanish whenever the upper limit is smaller than one. The quadratic contributions become
\begin{align}
  \sQ{\seq{K}}{\seq{P}}
&=
  \sum_{l=1}^P \sum_\ind{\lambda}^K\sum_\ind{\nu}^{P-1} \sum_{\bf q} 
  \left[
    {\frac{W^{[1]}_{{\bf q},{\bf p}_l-{\bf q}}}{2}} \,  
    \DI{\ssel{K}{\lambda}}{\ssel{P_l}{\nu},{\bf q}}\,  
    \DI{\srm{K}{\lambda}}{\srm{P_l}{\nu},{\bf p}_l-{\bf q}}
    +
    W^{[1]}_{{\bf q},{\bf p}_l} 
    \,\DI{\ssel{K}{\lambda},{\bf q}}{\ssel{P_l}{\nu}}
    \,\DI{\srm{K}{\lambda}}{\srm{P_l}{\nu},{\bf p}_l+{\bf q}}
    \right. 
\nonumber\\
&+
  \left.
  W^{[0]}_{{\bf q}}
  \sum_{{\bf q}'} 
  \left(
    \DI{\ssel{K}{\lambda},{\bf q}+{\bf q}'}{\ssel{P_l}{\nu}}
    \DI{\srm{K}{\lambda}}{\srm{P_l}{\nu},{\bf p}_l+{\bf q},{\bf q}'}
    +
    \DI{\ssel{K}{\lambda}}{\ssel{P_l}{\nu},{\bf q}'}
    \DI{\srm{K}{\lambda},{\bf q}+{\bf q}'}{\srm{P_l}{\nu},{\bf p}_l+{\bf q}}
    +
    \DI{\ssel{K}{\lambda}}{\ssel{P_l}{\nu},{\bf p}_l+{\bf q}}
    \DI{\srm{K}{\lambda},{\bf q}+{\bf q}'}{\srm{P_l}{\nu},{\bf q}'}
  \right)
  \right]
\nonumber\\
&-
  \sum_{j=1}^K \sum_\ind{\lambda}^{K-1} \sum_\ind{\nu}^P \sum_{\bf q} 
  \left[
    \frac{W^{[1]}_{{\bf q},{\bf k}_j-{\bf q}}}{2} 
    \,  \DI{\ssel{K_j}{\lambda},{\bf q}}{\ssel{P}{\nu}}
     \,  \DI{\srm{K_j}{\lambda},{\bf k}_j-{\bf q}}{\srm{P}{\nu}}
    +
    W^{[1]}_{{\bf q},{\bf k}_j} 
    \,  \DI{\ssel{K_j}{\lambda}}{\ssel{P}{\nu},{\bf q}}
    \,  \DI{\srm{K_j}{\lambda},{\bf k}_j+{\bf q}}{\srm{P}{\nu}}
    \right.
\nonumber\\
&+
  \left.
    W^{[0]}_{{\bf q}}
    \sum_{{\bf q}'}
    \left(
      \DI{\ssel{K_j}{\lambda}}{\ssel{P}{\nu},{\bf q}+{\bf q}'}
      \DI{\srm{K_j}{\lambda},{\bf k}_j+{\bf q},{\bf q}'}{\srm{P}{\nu}}
      +
      \DI{\ssel{K_j}{\lambda},{\bf q}'}{\ssel{P}{\nu}}
      \DI{\srm{K_j}{\lambda},{\bf k}_j+{\bf q}}{\srm{P}{\nu},{\bf q}+{\bf q}'}
      +
      \DI{\ssel{K_j}{\lambda},{\bf k}_j+{\bf q}}{\ssel{P}{\nu}}
      \DI{\srm{K_j}{\lambda},{\bf q}'}{\srm{P}{\nu},{\bf q}+{\bf q}'}
    \right)
  \right]
\nonumber\\
&+
  \sum_{l_1=1}^{P-1} \sum_{l_2=l_1+1}^P
  \sum_\ind{\lambda}^{K} \sum_\ind{\nu}^{P-2}  
  \sum_{\bf q}
     W^{[0]}_{{\bf q}} \,  
     \DI{\ssel{K}{\lambda}}{\ssel{P_{l_1,l_2}}{\nu},{\bf p}_{l_1}+{\bf q}}
     \DI{\srm{K}{\lambda}}{\srm{P_{l_1,l_2}}{\nu},{\bf p}_{l_2}-{\bf q}}
\nonumber\\
&-
  \sum_{j_1=1}^{K-1} \sum_{j_2=j_1+1}^K
  \sum_\ind{\lambda}^{K-2} \sum_\ind{\nu}^P 
  \sum_{\bf q} 
   W^{[0]}_{{\bf q}} \,  
   \DI{\ssel{K_{j_1,j_2}}{\lambda},{\bf k}_{j_1}+{\bf q}}{\ssel{P}{\nu}}
   \DI{\srm{K_{j_1,j_2}}{\lambda},{\bf k}_{j_2}-{\bf q}}{\srm{P}{\nu}}
\label{eq:quadratic-FIN}\;.
\end{align}
Like for the linear contributions, the $j$, $j_1$, $j_2$, $l$, $l_1$, and $l_2$ sums contribute only if their upper limit is greater than or equal to one. The permutation sums contribute only if the upper limit is greater than or equal to zero. The cubic contribution to the correlation dynamics follows from
\begin{align}
  \sC{\seq{K}}{\seq{P}}
&=
    \sum_{l=1}^P
    \sum_\ind{\lambda_1,\lambda_2}^K
    \sum_\ind{\nu_1,\nu_2}^{P-1}
    \sum_{{\bf q},{\bf q}'}
    W^{[0]}_{{\bf q}} \,
    \DI{\ssel{K}{\lambda_1},{\bf q}+{\bf q}'}{\ssel{P_l}{\nu_1}}\,
    \DI{\srm{K}{\lambda_1}(\lambda_2)}{\srm{P_l}{\nu_1}(\nu_2),{\bf p}_l+{\bf q}}\,
    \DI{\srm{K\big|{\lambda_1}}{\lambda_2}}{\srm{P_l\big|{\nu_1}}{\nu_2},{\bf q}'}
\nonumber\\
&-
  \sum_{j=1}^K
  \sum_\ind{\lambda_1,\lambda_2}^{K-1}
  \sum_\ind{\nu_1,\nu_2}^{P}
  \sum_{{\bf q},{\bf q}'}
    W^{[0]}_{{\bf q}}  \,  
    \DI{\ssel{K_j}{\lambda_1}}{\ssel{P}{\nu_1},{\bf q}+{\bf q}'}
    \DI{\ssel{ K_j \big| \lambda_1 }{\lambda_2},{\bf q}'}{\ssel{ P \big| \nu_1}{\nu_2}}
    \DI{\srm{K_j\big|{\lambda_1}}{\lambda_2},{\bf k}_j+{\bf q}}{\srm{P\big|{\nu_1}}{\nu_2}}
\label{eq:cubic-FIN}\;.
\end{align}
The first term contributes only for $P \ge 1$ whereas the last term contributes only for $K \ge 1$. Equations \eqref{eq:clusterDYN}--\eqref{eq:cubic-FIN} can be elaborated straightforwardly for any correlation with the ordered pointer selection sums \eqref{eq:sum_def}. 
The renormalized kinetic energy can be read out from identification \eqref{eq:E-renormalized} while the explicit BEC atom-number dependence follows from the matrix elements \eqref{eq:W_def}. 

The cluster-dynamics \eqref{eq:clusterDYN}--\eqref{eq:cubic-FIN} can, in principle, be derived using the explicit cluster factorization, discussed in Sec.~\ref{sec:relevant clusters}. Then, an $N$-atom cluster is technically defined by the difference of $N$-particle expectation value and all of its factorizations into clusters that have a lower rank than $N$. Only the doublet dynamics can be derived rather effortlessly by using this brute force identification when solved directly with the Heisenberg equation of motion \eqref{eq:HEM-ex}; 
the emerging three- and four-atom expectation values are naturally factorized exactly into singlets, doublets, triplets, and quadruplets. It is straightforward to show that INF Eqs.~\eqref{eq:clusterDYN}--\eqref{eq:cubic-FIN} indeed reproduce this doublet dynamics exactly. However, already evaluating the triplet dynamics with an explicit scheme becomes significantly more difficult. In general, the technical difficulty of the explicit derivation grows exponentially with the cluster number. 
As the major advantage, the INF delivers the exact quantum kinetics of {\it all} clusters with one derivation. Especially, {\it Eqs.~\eqref{eq:clusterDYN}--\eqref{eq:cubic-FIN} set an exact and pragmatic foundation to investigate true many-body properties of the interacting Bose gas with atom-cluster correlations}.

\subsection{Structure of atom-cluster dynamics}
\label{sec:cluster_structure}

As shown in Sec.~\ref{sec:chi_dyn}, the expectation-value and correlation dynamics share an identical hierarchy problem while the correlations are also modified by the nonlinear contributions \eqref{eq:quadratic-FIN}--\eqref{eq:cubic-FIN}. Of course, this additional complication would be a major drawback in any cluster-expansion-based calculation effort, unless there was a fundamental difference between the very nature of expectation values and correlations: 
$C$-particle correlations exist only if $C$ atoms are clustered together, whereas a system with $\NN$ normal-component atoms produces nonvanishing expectation values $\ave{N}$ containing up to $\NN$ creation and/or annihilation operators. Since the hierarchy problem nontrivially couples the existing $\ave{N}$, the expectation-value representation provides  neither clear nor controllable approximations. 
One also cannot solve all of the relevant $\ave{N}$ because the sheer number of elements needed grows exponentially with $N \le \NN$, while $\NN$ is typically very large. At the same time, atom clustering is weak in the normal component, as long as it does not produce macroscopic occupations as a BEC would. More physical justifications follow at the end of this section. If this claim is true, only a tiny subset of $(C\ll \NN)$-atom correlation equations describe the actual quantum kinetics, 
which yields a natural and systematic truncation of the hierarchy problem. This truncation has already been demonstrated to deliver an accurate and efficient description of, e.g., many-body as well as quantum-optical effects in various solid-state systems.\cite{Kira:1999b,Ell:1999,Lee:1999,Kira:2006,Kira:2008,Book:2011,Mootz:2012} Next, we will investigate how the cluster-expansion truncation can be implemented to systematically describe the quantum kinetics of the strongly interacting Bose gas.

We start by performing a purely structural analysis of the full cluster dynamics \eqref{eq:clusterDYN}--\eqref{eq:cubic-FIN}; for this purpose, we introduce a structure operation
\begin{eqnarray}
  \str{\DI{\seq{K}}{\seq{P}}} \equiv \Dstr{K}{P}
\label{eq:structure}\;,
\end{eqnarray}
where $\Dstr{K}{P}$ denotes a correlation with $K$ creation and $P$ annihilation operators; in total, $\Dstr{K}{P}$ is a $(K+P)$-atom correlation. We also assume that $\str{\Xi^{(n)}_\alpha}$ counts how many times a given $\Dstr{K}{P}$ appears within $\Xi^{(n)}_\alpha$ writing the count number in front of each $\Dstr{K}{P}$. 
The actual counts can be constructed by converting each combinatorial sum into the number of combinations by using Eq.~\eqref{eq:N_combinatorial}. For the sake of clarity, we also suppress expressing the interaction-matrix elements in $\str{\Xi^{(n)}_\alpha}$ and add all terms as positive integers, just to identify the structure. 

With this simple set of rules, we find that the source-term \eqref{eq:source-FIN} has a structure
\begin{eqnarray}
  \str{\sS{\seq{K}}{\seq{P}}}
  =
  \delta_{K,0}\,\delta_{P,2} 
  +
  \delta_{K,2}\,\delta_{P,0} 
\label{eq:source-STR}\;.
\end{eqnarray}
This shows that only $\Dstr{2}{0}$ and $\Dstr{0}{2}$ are directly generated from the BEC. The explicit correlation and driving term is identified by Eq.~\eqref{eq:s_source}, which shows that there is indeed only one term driving the two-atom correlations. By applying the same  projection to the linear contribution \eqref{eq:linear-STR}, we eventually find the linear structure
\begin{align}
  \str{\sL{\seq{K}}{\seq{P}}}
&=
  P(P-1)\,\Dstr{K}{P-1}
  +
  K(K-1)\,\Dstr{K-1}{P}
  +
  \left(
    P^2+K^2
  \right)\Dstr{K}{P}
  +
  P\,\Dstr{K+1}{P-1}
\nonumber\\
  &+
  K\,\Dstr{K-1}{P+1}
  +
  (P+K)\,\Dstr{K}{P+1}
  +
  (P+K)\,\Dstr{K+1}{P}
  +
  (P+K)\,\Dstr{K+1}{P+1}
\label{eq:linear-STR}\;,
\end{align}
after having rearranged the terms in the order of increasing cluster number. The last three contributions correspond to the hierarchy problem because they couple a $C=K+P$ atom correlation to $(C+1)-$ and $(C+2)$-atom correlations.

The quadratic contributions \eqref{eq:quadratic-FIN} produce the following structure
\begin{align}
  \str{\sQ{\seq{K}}{\seq{P}}}
&=
  \sum_{\lambda=0}^K \sum_{\nu=0}^{P-1} 
  N^K_\lambda \,P\,  N^{P-1}_\nu
  \left[
    \frac{1}{2} \,\Dstr{\lambda}{\nu+1}\,\Dstr{K-\lambda}{P-\nu} 
    +
    \Dstr{\lambda+1}{\nu}\,\Dstr{K-\lambda}{P-\nu} 
    +
    \Dstr{\lambda+1}{\nu}\,\Dstr{K-\lambda}{P-\nu+1} 
    +
    2 \,\Dstr{\lambda}{\nu+1}\,\Dstr{K-\lambda+1}{P-\nu} 
  \right] 
\nonumber\\
&+
  \sum_{\lambda=0}^{K-1} \sum_{\nu=0}^{P} 
  K\, N^{K-1}_\lambda  N^{P}_\nu
  \left[
    \frac{1}{2} \,\Dstr{\lambda+1}{\nu}\,\Dstr{K-\lambda}{P-\nu} 
    +
    \Dstr{\lambda}{\nu+1}\,\Dstr{K-\lambda}{P-\nu} 
    +
    \Dstr{\lambda}{\nu+1}\,\Dstr{K-\lambda+1}{P-\nu} 
    +
    2 \,\Dstr{\lambda+1}{\nu}\,\Dstr{K-\lambda}{P-\nu+1} 
  \right]   
\nonumber\\
&+
    \sum_{\lambda=0}^K \sum_{\nu=0}^{P-2} 
  N^K_\lambda \,P(P-1)\,  N^{P-2}_\nu
  \Dstr{\lambda}{\nu+1}\,\Dstr{K-\lambda}{P-\nu-1} 
   +
    \sum_{\lambda=0}^{K-2} \sum_{\nu=0}^{P} 
  K(K-1)\,N^{K-2}_\lambda \,N^{P}_\nu
  \Dstr{\lambda+1}{\nu}\,\Dstr{K-\lambda-1}{P-\nu}    
\label{eq:quadratic-STR}\;.
\end{align}
As discussed in Sec.~\ref{sec:relevant clusters}, singlets $\Dstr{1}{0}$ and $\Dstr{0}{1}$ cannot exist in the system, which eliminates the effect of $(\lambda,\nu)$ terms that appear at the borders of the summations. With this information, we conclude that the quadratic contribution contains at most a $(K+P)$-atom correlation, 
such that it does not induce an additional hierarchical coupling. In the same way, the structure of cubic contributions \eqref{eq:cubic-STR} becomes
\begin{align}
  \str{\sC{\seq{K}}{\seq{P}}}
&=
    \sum_{\lambda_1=0}^K \sum_{\lambda_2=0}^{K-\lambda_1}
    \sum_{\nu_1=0}^{P-1} \sum_{\nu_2=0}^{P-\nu_1-1}
    N^K_{\lambda_1,\lambda_2}\,
    P\,
    N^{P-1}_{\nu_1,\nu_2} \,
    \Dstr{\lambda_1+1}{\nu_1}
    \Dstr{\lambda_2}{\nu_2+1}
    \Dstr{K-\lambda_1-\lambda_2}{P-\nu_1-\nu_2}
\nonumber\\
&+
    \sum_{\lambda_1=0}^{K-1} \sum_{\lambda_2=0}^{K-\lambda_1-1}
    \sum_{\nu_1=0}^{P} \sum_{\nu_2=0}^{P-\nu_1}
    K\,N^{K-1}_{\lambda_1,\lambda_2}\,
    N^{P}_{\nu_1,\nu_2} \,
    \Dstr{\lambda_1}{\nu_1+1}
    \Dstr{\lambda_2+1}{\nu_2}
    \Dstr{K-\lambda_1-\lambda_2}{P-\nu_1-\nu_2}
\label{eq:cubic-STR}\;.
\end{align}
Since singlets do not exist, quadratic contributions contain only contributions up to $(K+P-2)$-atom clusters. Therefore, neither quadratic not cubic terms yield an additional hierarchy problem for the correlations. In other words, only the linear part \eqref{eq:linear-STR} yields a hierarchy problem.

The BEC atoms are, by definition, stationary with respect to each other such that they remain clustered together forever due to the lack of relative movement. Consequently, it is only natural that BEC exhibits correlated atom clusters to all orders even when interactions do not provide additional binding, as shown in Ref.~\cite{Kira:Xpic}. In contrast to the BEC, the normal-component atoms move randomly with respect to each other in thermodynamical equilibrium, which tends to break up any cluster group that is not bound by atom--atom interactions. 
Therefore, a weakly interacting Bose gas cannot sustain atom clusters in its normal component, which at its best becomes plasma where the atoms move as mean-field particles. Since the excitation picture allows us to describe the entire system in terms of normal-component correlations, the quantum dynamics indeed involves only low-$C$ atom clusters. Once interactions become strong enough, atom clusters can be sustained also in the normal component. Next, we 
will study the general cluster-formation laws, based on the structure of cluster dynamics derived above. 

\subsection{Sequential build up of clusters}
\label{sec:sequential}

Recent experiments\cite{Donley:2001,Altin:2011,Makotyn:2014} have demonstrated that atom--atom interactions can be abruptly switched on from weakly to strongly interacting Bose gas. 
These experiments are prepared such that essentially all of the atoms initially occupy the BEC when the Bose gas is still weakly interacting. Therefore, the $\Delta^J_K$ correlations vanish before the switch-on, based on the discussion at the end of Sec.~\ref{sec:cluster_structure}. In this situation, the jump to the strong interactions abruptly initiates the quantum depletion that starts to excite the $\Dstr{K}{P}$ clusters. 
More specifically, fast switching to the strong interactions directly excite atom doublets to the normal component through the source contribution \eqref{eq:source-FIN} while higher-order $\Dstr{K}{P}$ clusters are generated by further interactions.
Conceptually, such atomic experiments resemble ultrafast excitations in semiconductors\cite{Khitrova:1999,Chemla:2001,Rossi:2002,Koch:2006,Kira:2006,Li:2006,Haug:2008,Smith:2010,Cundiff:2012} where a femtosecond laser pulse abruptly generates electron--hole pairs that subsequently\cite{Kira:2006} and sequentially\cite{Kira:2006b,Mootz:2012,Book:2011} cluster into new quasiparticles\cite{Mootz:2014,Hunter:2014} due to the Coulomb interaction. 

To determine the dynamical structure of cluster generation in the switch-on experiments, we analyze next the quantum kinetics $C\equiv K+P$ atom clusters $\Dstr{K}{P}$ experience by identifying the cluster number $C_{\rm seq}\equiv \tilde{K}+\tilde{P}$ for each correlation $\Dstr{\tilde{K}}{\tilde{P}}$ 
that appears on the right hand side of Eqs.~\eqref{eq:source-STR}--\eqref{eq:cubic-STR}. This allows us to pinpoint precisely how the clusters within linear, quadratic, and cubic contributions sequentially generate the $C$-atom cluster dynamics. 

The source contribution \eqref{eq:source-STR}  creates {\it directly} only $(C=2)$-atom correlations $s_{\bf k}$ as discussed in connection with Eq.~\eqref{eq:s_source}. This indicates that the interactions {\it directly} drive only the doublet clusters to the system while the higher-order clusters must build up later in time via the linear, quadratic, and/or cubic contributions. 
By counting $C_{\rm seq}$ numbers that appear in Eq.~\eqref{eq:linear-STR}, we conclude that the linear contributions couple the $C$-atom correlations to clusters
\begin{eqnarray}
  C_{\rm seq}^{\rm lin} = C-1,\, C,\, C+1,\,C+2
\label{eq:Gen_lim_lin}\;.
\end{eqnarray}
This confirms again that the linear contributions yield a hierarchical coupling to clusters with one and two more clusters than the original $C$-atom correlation has. However, the hierarchical coupling does not {\it create} clusters but mediates the influence of higher-order clusters --- if they exist  --- to the clusters that contain either one or two atoms less. The correlation {\it generation} beyond doublets follows from the coupling to 
the $C-1$ clusters.
In other words, if $C_{\rm seq}^{\rm lin}$ cluster already exists, the linear contribution generates a $(C_{\rm seq}^{\rm lin} +1)$-atom correlation, given time to do that. Consequently, the source-generated doublets proliferate as $C \rightarrow C+1$ due to the linear terms, which is observed as a sequential build up of clusters, in full analogy to the observations in the ultrafast spectroscopy of semiconductors.\cite{Kira:2006b,Mootz:2012,Mootz:2014,Hunter:2014}

The sequential build up is further modified by the quadratic and cubic contributions. To determine what is the minimum level of atom clusters needed to generate $C$-particle clusters, we determine the largest cluster number $C_{\rm seq}^\lambda$ that appears in the quadratic ($\lambda={\rm quad}$) or cubic ($\lambda={\rm cub}$)  cluster products of Eqs.~\eqref{eq:quadratic-STR}--\eqref{eq:cubic-STR}.
By taking into account that singlets do not exist in the system, as shown in Sec.~\ref{sec:HBE}, we find that a $C$-atom correlation is generated by quadratic contributions only if one of the
\begin{eqnarray}
  C_{\rm seq}^{\rm quad} = {\textstyle \frac{C}{2},\,\frac{C}{2}+1,\cdots, C}
\label{eq:Gen_lim_Q}
\end{eqnarray}
atom correlations already exists in the system. If $\frac{C}{2}$ is fractional, it should be rounded up to the closest integer above it. In the same way, we find from Eq.~\eqref{eq:cubic-STR} that the cubic contributions generate a $C$-atom correlation only if
\begin{eqnarray}
  C_{\rm seq}^{\rm cub} = {\textstyle \frac{C+2}{3},\,\frac{C+2}{3}+1,\cdots,C }
\label{eq:Gen_lim_C}\;
\end{eqnarray}
already exist in the system. As above, $\frac{C+2}{3}$ is to be rounded to the nearest integer.
We observe again that neither the quadratic nor the cubic terms produce a hierarchy problem such that they only induce further generation of clusters. At the same time, the {\it cluster generation is always is sequential} because the generation process involves only a finite range of clusters \eqref{eq:Gen_lim_lin}--\eqref{eq:Gen_lim_C}, which makes instantaneous generation of all clusters impossible. For example, 1000-atom clusters are generated only if $\frac{1000+2}{3}=334$-atom clusters already exists in the system.
This allows us to introduce a very natural, and systematic, truncation to the hierarchy problem; one needs to evaluate Eqs.~\eqref{eq:source-FIN}--\eqref{eq:cubic-FIN} only up to the the $C$-particle cluster as long as the sequential evolution does not overpass it. Based the cluster-expansion experiences in semiconductor optics\cite{Kira:1999,Aoki:1999,Khitrova:1999,Kira:2001,Hoyer:2003,Chatterjee:2004,Koch:2006,Kira:2006,Kira:2006b,Smith:2010,Book:2011}, a very low number of clusters can explain the quantum kinetics over an extended period of time\cite{Mootz:2012,Kira:2006,Book:2011,Hunter:2014}, as explained below. 

To assess which correlations can be sequentially generated from the existing $C_{\rm seq}$-atom correlations, we determine the maximum $C$ allowed by Eqs.~\eqref{eq:Gen_lim_lin}--\eqref{eq:Gen_lim_C}. 
We simply solve the largest $C$ generated as function of $C_{\rm seq}^\lambda$ from the linear to cubic relations \eqref{eq:source-FIN}--\eqref{eq:cubic-FIN} and denote $C_{\rm seq}^\lambda$  by $C_{\rm seq}$ and the generated $C$-atom cluster by $C_{\rm gen}$. We then find that an existing $C_{\rm seq}$-particle correlation may generate correlations up to $C_{\rm seq}+1$ for the linear, $2C_{\rm seq}$ for the quadratic, and $3C_{\rm seq}+2$ for the cubic contributions. The largest of these implies the generation of
\begin{eqnarray}
  C_{\rm gen} = 3C_{\rm seq}-2
\label{eq:C_max}
\end{eqnarray}
clusters, for $C_{\rm seq} \ge 2$ existing clusters. The doublets are generated directly from the BEC and once they exist, they may generate atom clusters up to $C_{\rm gen}=4$. Once quadruplets emerge, they may generate atom clusters up to $C_{\rm gen}=10$, which in turn may induce correlations up to $C_{\rm gen}=28$, and so on, based on rule \eqref{eq:C_max}.

In general, Eq.~\eqref{eq:C_max} implies that the number of relevant clusters grows exponentially during the sequential build up. However, the proliferation of correlations has additional aspects that make the typical build up slow. The exponential growth in cluster number follows exclusively from the nonlinear contributions. 
As shown in Ref.~\cite{Kira:2006}, they contain, e.g., a Boltzmann-scattering like structure where in and out scattering of correlations compete, which makes the overall correlation formation slower than the individual scattering events. Due to many possible routes to generate and annihilate correlations, the correlations build up slowly via diffusive multi-step processes. Additionally, the quadratic and cubic terms induce screening of interactions as shown in Ref.~\cite{Kira:2006} because many-body clusters tend to find arrangements where a bound atom cluster becomes less reactive than its constituent atoms.
As the simplest example, atom--atom interactions are usually much stronger than molecular dimer-dimer interactions. Therefore, the formation of clusters yields a further slowing down of the formation of even higher-order cluster. Consequently, the diffusive scattering, screening, and clustering aspects can intuitively explain why quantum kinetics of many-body systems can be efficiently described by few clusters over an extended period of time. 
These aspects are identified in   \ref{app:triplets} where the triplet dynamics is explicitly studied.

Alternatively, one can use the interaction strenght --- here $V_{\bf k}$  --- as a perturbative parameter and show that clusters systematically include contributions much beyond the perturbation expansion. For example, the screening effects produce an infinite series in terms of $V_{\bf k}$ powers, whereas it is fully described by only one class of cluster dynamics\cite{Kira:2006}. 
Therefore, a cluster-based analysis is a systematic, nonperturbative, approach that can be used in regimes where perturbative approaches fail. More specifically, cluster dynamics up to $C$ clusters accurately describes the quantum kinetics until the time when substantial amount of correlations are formed beyond the $C$-particle cluster.
Even though the exact time ranges cannot be {\it a priori} defined, one can determine the time ranges of validity in pragmatic computations by adding one more cluster to the computation and then determine at which time the next correlation level becomes excited, cf.~Ref.~\cite{Mootz:2012} for simple examples. Of course, such an analysis may become extremely strenuous on computational resources, but they are directly implementable via the general cluster equations \eqref{eq:clusterDYN}--\eqref{eq:cubic-FIN} if need be.

\begin{figure}[t]
\includegraphics*[scale=0.65]{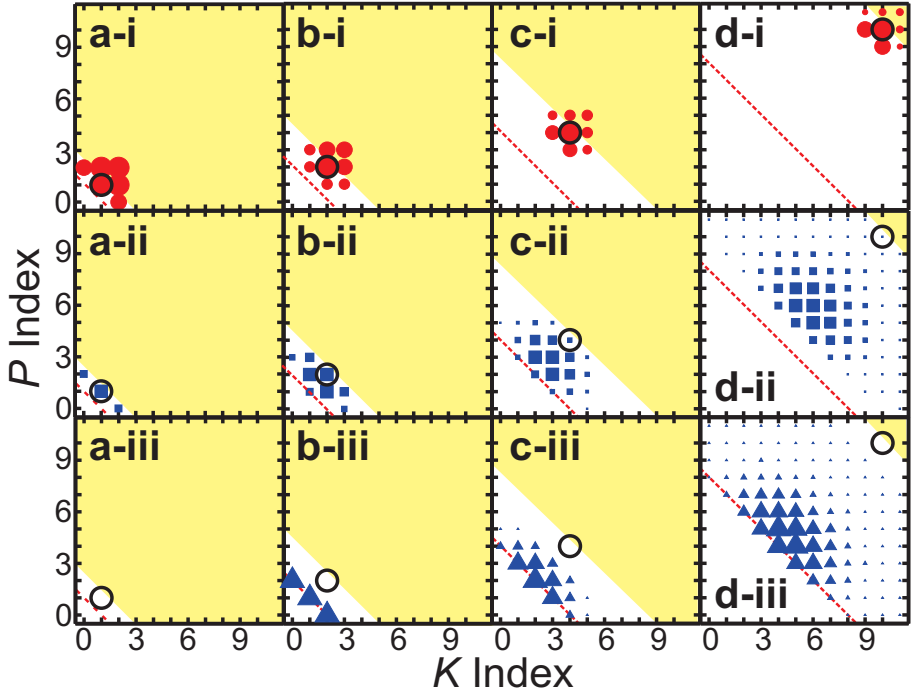}
\caption{Sequential build up of clusters based on the cluster-coupling structure \eqref{eq:linear-STR}--\eqref{eq:cubic-STR}. The filled circles, squares, and triangles denote the highest-rank cluster appearing in linear (i), quadratic (ii), and cubic (iii) contributions, respectively. The area of the symbol is proportional to the relative number of clusters. The open circle indicates the original correlation whose dynamics is followed. The shaded area indicates the limit of hierarchical coupling and the red-dashed line indicates the lowest clusters needed to initiate scattering. The structure of {\bf a} $\Delta^1_1$, {\bf b} $\Delta^2_2$, {\bf c} $\Delta^4_4$, {\bf d} $\Delta^{10}_{10}$ correlation dynamics is shown.
}\label{Cnumbers}
\end{figure}

To illustrate the sequential as well as diffusive aspects of the cluster dynamics, Fig.~\ref{Cnumbers} presents the structure of correlations dynamics $\Dstr{J}{J}$ for {\bf a} $J=1$, ${\bf b}$ $J=2$, ${\bf c}$ $J=4$, and ${\bf d}$ $J=10$. More specifically, the contributing $\Dstr{\tilde{J}}{\tilde{K}}$ correlations are shown for linear \eqref{eq:linear-STR}, quadratic \eqref{eq:quadratic-STR}, and cubic \eqref{eq:cubic-STR} contributions as circles, squares, and triangles, respectively. The area of each symbol is proportional to the relative number of terms within each contribution class, based on prefactors in Eqs.~\eqref{eq:Gen_lim_lin}--\eqref{eq:Gen_lim_C}. 
For the quadratic and cubic terms, the symbol is assigned to that $\Dstr{\tilde{J}}{\tilde{K}}$ which has the largest cluster number $C_{\rm seq}=\tilde{J}+\tilde{K}$ within the product of clusters. As discussed above, the symbol then defines which correlation must exist in the system to generate the analyzed $C$-atom cluster.
The open circle indicates the $\Dstr{J}{J}$ correlation whose dynamics is presented. The shaded area indicates the regime of the hierarchical coupling while the dashed line shows the $C_{\rm gen}$-criterion \eqref{eq:C_max}.

Also Fig.~\ref{Cnumbers} confirms that only the linear contributions (frames i) produce a hierarchical coupling because only they contain clusters within the shaded area, confirming results \eqref{eq:Gen_lim_lin}--\eqref{eq:Gen_lim_C}. For $\Dstr{1}{1}$, many correlations participate in the hierarchical coupling based on the large areas of the circles. At the same time, only three classes of correlations exist in quadratic contributions whereas the cubic terms vanish for $\Dstr{1}{1}$ (frames a). Once the $\Dstr{1}{1}$ doublets are generated, they may generate quadruplets $\Dstr{2}{2}$ based on the $C_{\rm gen}$-criterion \eqref{eq:C_max}. 
Figure \ref{Cnumbers}{\bf b} shows a similar hierarchical coupling as for the doublet, while the number of quadratic (frames ii) and cubic terms (frames iii) increases. This increasing trend continues as we follow the correlations $\Dstr{4}{4}$ and $\Dstr{10}{10}$ structure in Fig.~\ref{Cnumbers}{\bf c} and {\bf d}, respectively; these terms emerge next in the sequential build up based on the $C_
{\rm gen}$-criterion \eqref{eq:C_max}. The number of nonlinear correlation-product classes is indeed growing, which is consistent with the scattering and diffusion arguments given above to substantiate the extended time ranges before higher-order correlations are built up. For example, the number of cubic contributions follows from Eq.~\eqref{eq:cubic-STR}; $\Dstr{1}{1}$ has zero, $\Dstr{2}{2}$ eight, $\Dstr{4}{4}$ 3648, and $\Dstr{10}{10}$  $2.38\times 10^9$ cubic terms at the largest symbol of frames iii. Consequently, the cluster dynamics becomes indeed largely diffused among multiple cluster combinations as the cluster number becomes larger.

\begin{figure}[t]
\includegraphics*[scale=0.5]{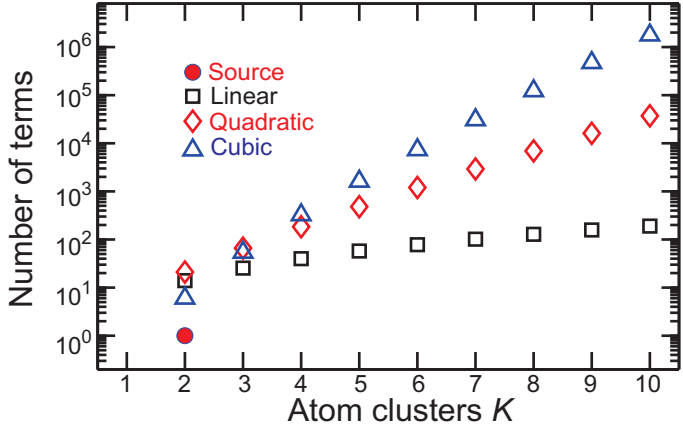}
\caption{Total number of cluster terms as function of atom-cluster number. More specifically, Eq.~\eqref{eq:S_number} is evaluated for $\Delta^K_0$ as function of $K$. Linear contributions denoted by square, quadratic by diamond, and cubic by triangle; The filled sphere indicates the one and only source term.}\label{atomCnumbers}
\end{figure}

To quantify the cluster-number diffusion, we can simply compute the total number of clusters that appear in Eqs.~ \eqref{eq:source-STR}--\eqref{eq:cubic-STR}. By applying relations \eqref{eq:N_combinatorial}--\eqref{eq:N_combinatorial-rest}, we find that the $\Dstr{K}{0}$ correlation dynamics contains
\begin{eqnarray}
  \left\{
  \begin{array}{l}
    \#S^{K}_{0} = \delta_{K,2}
    \\
    \#L^{K}_{0} = 2K^2+3K
  \end{array}
  \right.\,,
  \qquad
  \left\{
  \begin{array}{l}
    \#Q^{K}_{0} = \left(K^2+8K\right)2^{K-2}
    \\
    \#C^{K}_{P} = K \, 3^{K-1}
  \end{array}
  \right.\,,
  \qquad K\ge 2
\label{eq:S_number}\;,
\end{eqnarray}
terms in total due to the source ($\#S$), linear ($\#L$), quadratic ($\#Q$), and cubic ($\#C$) contributions, respectively. Figure \ref{atomCnumbers} shows how the total cluster number behaves for a $K$-particle cluster  $\Dstr{K}{0}$ as function of $K$. As discussed above, only the $K=2$ doublet is driven by the source (circle). We also see that the linear (squares), quadratic (diamonds), and cubic (triangles) terms produce a monotonic increase as function of $K$. 
As a general trend, both the quadratic and cubic terms are spread among very many configurations at high $K$ while the cubic contributions eventually become most abundant. For example, 10-atom correlations are influenced by more than $10^6$ different cubic correlation combinations. This is again an indication that the correlation dynamics involves many competing processes and correlation diffusion. 

\section{Quantum kinetics of two-atom correlations}
\label{sec:QK_doublets}

As shown in Sec.~\ref{sec:relevant clusters}, a homogeneous system has only two kinds of doublets: the occupation 
\begin{eqnarray}
  f_{\bf k}
  \equiv
  \DI{\bf k}{\bf k}
  = \Delta \avex{B^\dagger_{\bf k} \, B^\dagger_{\bf k}}
\label{eq:atom-occupation}\;,
\end{eqnarray}
as well as the transition amplitude $s_{\bf k}$ and its complex conjugate $s^\star_{\bf k}$ that follow from definition \eqref{eq:squeezing_s}. A sign inversion in Eq.~\eqref{eq:squeezing_s}, ${\bf k} \rightarrow -{\bf k}$, produces an inversion symmetry 
\begin{eqnarray}
  s_{-{\bf k}}
  = 
  \Delta \avex{B_{-{\bf k}} \, B_{{\bf k}}}
  =
  \Delta \avex{B_{\bf k} \, B_{-{\bf k}}}
  =
  s_{+{\bf k}}
\label{eq:atom-sqz}\;,
\end{eqnarray}
with respect to the momentum because boson annihilation operators commutate. Since $f_{\bf k}$ contains creation and annihilation operators, it is not automatically inversion symmetric. However, if $f_{-{\bf k}}$ and $f_{+{\bf k}}$ were not equal, a different number of atoms would propagate into the positive than to the negative direction, implying a net atom current within the trap. We analyze next cases where the macroscopic current vanishes such that also $f_{\bf k}$ is inversion symmetric. 
Therefore, the doublets $f_{\bf k}$, $s_{\bf k}$ and $s^\star_{\bf k}$ are uniquely defined by ${\bf k}$ values given in the the half space of all possible ${\bf k}$ values. We assume next that ${\bf k} > 0$ formally denotes such a half space where a set of all $B_{+{\bf k}}$ and $B_{-{\bf k}}$ assign independent boson operators that uniquely span the complete set of states.

\subsection{Quadrature fluctuations of bidirectional atoms}
\label{sec:bidirectional}

To give a physical interpretation  for an $(f_{\bf k},\,s_{\bf k})$ pair, we introduce the {\it bidirectional basis} via a unitary transformation
\begin{eqnarray}
  D_{{\bf k},\pm} 
  \equiv 
  \frac{e^{-i\theta_{{\bf k}}}
  }{\sqrt{2}}
  \left( 
    B_{+{\bf k}} 
    \pm 
    B_{-{\bf k}}
  \right)
  \qquad
  \Leftrightarrow
  \qquad
  B_{\pm{\bf k}} 
  =
  \frac{
  e^{+i\theta_{{\bf k}}}
  }{\sqrt{2}} 
  \left(D_{{\bf k},+} 
    \pm 
    D_{{\bf k},-} 
  \right)
\label{eq:D_unitary}\;,
\end{eqnarray}
where $\theta_{\bf k}$ is an adjustable phase and ${\bf k} >0$. Since $D_{{\bf k},\pm}$ contains an equal mix of forward ($+{\bf k}$) and backward ($-{\bf k}$) propagating atoms, it establishes a basis of bidirectional atoms. It is straightforward to show that bidirectional $D_{{\bf k},+}$ and $D_{{\bf k},-}$ form a complete set of independent bosonic operators that produce a diagonalization
\begin{eqnarray}
  \Delta \avex{D^\dagger_{{\bf k},\sigma}\, D_{{\bf k}',\sigma'}}
  =
  \delta_{{\bf k}',{\bf k}}\,\delta_{\sigma,\sigma'}\,
  f_{\bf k}\,,
\qquad
  \Delta \avex{D_{{\bf k},\sigma}\, D_{{\bf k}',\sigma'}}
  =
  \sigma\,\delta_{{\bf k}',{\bf k}}\,\delta_{\sigma,\sigma'}\,
  |s_{\bf k}|
\label{eq:D_expect}\;,
\end{eqnarray}
of doublets with respect to both the momentum and the sign index $\sigma=\pm$. This means that the doublets are also statistically independent when presented in the bidiagonal basis. To simplify the notation in forthcoming analysis, we have fixed $\theta_{\bf k}$ so that $s_{\bf k}\,e^{-2i\theta_{\bf k}} = |s_{\bf k}|$ yields a positive and real-valued 
squeezing amplitude. By substituting transformation \eqref{eq:D_unitary} into Eq.~\eqref{eq:atom-field}, the field operator becomes
\begin{eqnarray}
  \op{\Psi}({\bf r}) = \sqrt{\frac{2}{{\cal L}^3}} \sum_{{\bf k}>0} e^{i\theta_{\bf k}}
  \left( 
    {\rm cos} \, {\bf k} \cdot {\bf r} \; D_{{\bf k},+}
    +
    i\,
    {\rm sin} \,{\bf k} \cdot {\bf r} \; D_{{\bf k},-}
  \right)
\label{eq:atom-fieldCS}\;.
\end{eqnarray}
We see that $D_{{\bf k},+}$ corresponds to the cosine while  $D_{{\bf k},-}$ is connected with the sine function. This is consistent with $D_{{\bf k},\pm}$ defining the bidirectional basis because cosine and sine are the bidirectional components of the plane waves.

\begin{figure}[t]
\includegraphics*[scale=0.55]{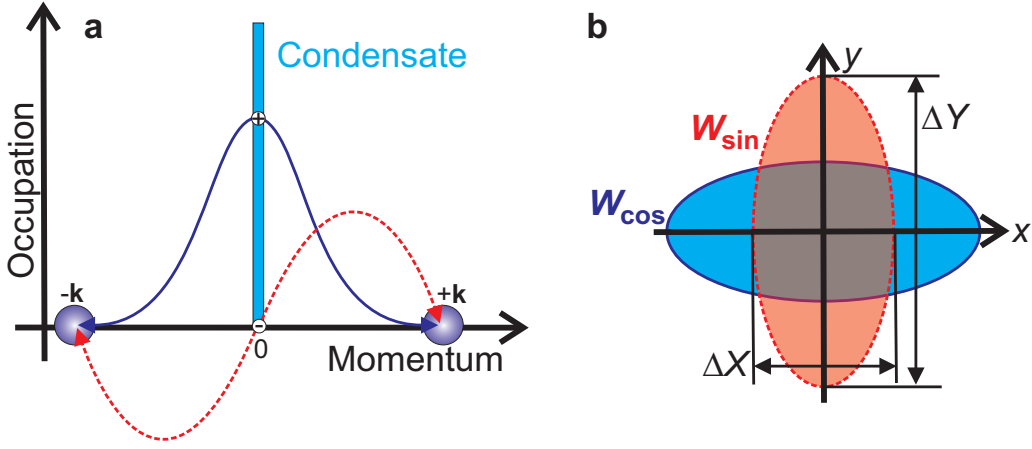}
\caption{Quantum depletion and squeezing amplitude. {\bf a} Elementary process of quantum depletion moves two atoms from a BEC (bar) to two normal-component atoms (spheres at $\pm{\bf k}$). The symmetric and antisymmetric processes are identified by cosine- (blue) and sine-shaped (red) arrows. {\bf b} The quadrature fluctuations in the phase space are shown for the corresponding cosine- ($\sigma=+$, blue ellipse) and sine-mode ($\sigma=-$, red ellipse), based on Eq.~\eqref{eq:XY_quadrature}.
}\label{HBEsqueezing}
\end{figure}

Based on the diagonality \eqref{eq:D_expect}, atoms at any given bidirectional state are not statistically mixed with any other bidirectional state. At the same time, the quantum depletion first creates doublets, dictated by the sequential build up identified in Sec.~\ref{sec:sequential}; the explicit source term follows from Eqs.~\eqref{eq:xi-source0-FIN}--\eqref{eq:s_source}. 
Due to the diagonality \eqref{eq:D_expect} and the explicit sine--cosine basis \eqref{eq:atom-fieldCS}, we conclude that the quantum depletion must eject atoms from the BEC to the normal component as bidirectional atoms defined by either cosine or sine mode.  Figure \ref{HBEsqueezing}{\bf a} illustrates schematically how the quantum depletion converts a BEC atom to a cosine- (solid arrows) and sine-atom mode (dashed arrows) from the BEC (bar).

In general, quantum properties related to any bosonic operator $B$ can always be represented via a phase-space distribution\cite{Walls:2008,Book:2011}, such as the Wigner function $W(\alpha)$ where $\alpha = x + i y$ is a canonical phase-space coordinate associated with $B \leftrightarrow \alpha$. The $\alpha$ dependence of $W(\alpha)$ determines roughly the distribution of bosonic amplitudes of the state. Due to the Heisenberg uncertainty principle, $W(\alpha)$ has always some spread in the phase space constituting the quantum fluctuations.
For a many-body boson system, $W$ becomes a high-dimensional object because each state introduces a new $\alpha$ coordinate to $W$. Next, we analyze which general constraints apply to the normal component based on the doublet properties. Since doublets of bidirectional modes are statistically independent, a product of single-particle $W(\alpha)$ of cosine and sine modes determines the doublet-related fluctuations of the entire many-body system. 
In other words, we may investigate individual doublet fluctuations in $W(\alpha)$ in oder to understand the quantum fluctuations of the many-body system.

To analyze the bidirectional mode's quantum fluctuations in terms of $\alpha=x+iy$, we introduce the quadrature operators,
\begin{eqnarray}
  \left\{
    \begin{array}{c}
      \op{X}_{{\bf k},\sigma} \equiv \frac{1}{2}\left(D_{{\bf k},\sigma} +  D^\dagger_{{\bf k},\sigma} \right)
      \\
      \op{Y}_{{\bf k},\sigma} \equiv \frac{1}{2i}\left(D_{{\bf k},\sigma} -  D^\dagger_{{\bf k},\sigma} \right)
    \end{array}
  \right.  
\label{eq:XY_quadrature}\;,
\end{eqnarray}
for each bidirectional mode. Physically, $\op{X}_{{\bf k},\sigma}$ and $\op{Y}_{{\bf k},\sigma}$ determine observables related to the real and imaginary part of $D_{{\bf k}, \sigma} = \op{X}_{{\bf k},\sigma} + i \op{Y}_{{\bf k},\sigma}$, respectively. 
In other words, their quantum properties are assigned by the real and imaginary parts of the phase-space coordinate $\alpha$, i.e.~$x \leftrightarrow \op{X}_{{\bf k},\sigma}$ and $y \leftrightarrow \op{Y}_{{\bf k},\sigma}$. The corresponding quadrature fluctuations reduce to
\begin{eqnarray}
  \Delta X^2_{{\bf k},\sigma} 
  &\equiv& 
  \avex{\op{X}_{{\bf k},\sigma}\,\op{X}_{{\bf k},\sigma}}
  -
  \avex{\op{X}_{{\bf k},\sigma}}
  \avex{\op{X}_{{\bf k},\sigma}}
  = \frac{1}{4}+\frac{f_{\bf k} + \sigma |s_{\bf k}| }{2}
\nonumber\\
  \Delta Y^2_{{\bf k},\sigma} 
  &\equiv& 
  \avex{\op{Y}_{{\bf k},\sigma}\,\op{Y}_{{\bf k},\sigma}}
  -
  \avex{\op{Y}_{{\bf k},\sigma}}
  \avex{\op{Y}_{{\bf k},\sigma}}
  = \frac{1}{4}+\frac{f_{\bf k} - \sigma |s_{\bf k}|}{2}
\label{eq:XY_fluct}\;,
\end{eqnarray}
which follows directly from definition \eqref{eq:XY_quadrature} and correlations \eqref{eq:D_expect}. 
The $\sigma$ that appears defines the sign of the $ \sigma |s_{\bf k}|$ contribution. Whenever the squeezing amplitude is nonvanishing, the $\Delta X_{{\bf k},\sigma}$ and $\Delta Y_{{\bf k},\sigma}$ fluctuations become different indicating that the phase-space fluctuations have a different magnitude in the $x$ and $y$ directions. 
The maximum and minimum fluctuation then become 
\begin{eqnarray}
  \Delta X_{\rm max} 
  \equiv
  \frac{1}{2} \sqrt{1+2\left( f_{\bf k} + |s_{\bf k}| \right)}\,,
\qquad
  \Delta X_{\rm min} 
  \equiv
  \frac{1}{2} \sqrt{1+2\left( f_{\bf k} - |s_{\bf k}| \right)}
\label{eq:XY_maxmin}\;,
\end{eqnarray}
respectively, based on $\sigma=\pm$ signs in Eq.~\eqref{eq:XY_fluct}. Figure \ref{HBEsqueezing}{\bf b} shows schematically the extent of fluctuations in the phase space for the cosine ($\sigma=+$, blue area) and sine ($\sigma=-$, red area) mode. Since the singlets $\avex{\op{X}_{{\bf k},\sigma}} = \avex{\op{Y}_{{\bf k},\sigma}}=0$ vanish, the fluctuations are centered at the origin while they have a different extension in the $x$ and $y$ direction. 
We also observe that the cosine mode has increased fluctuations in the $x$ direction whereas the $y$ quadrature contains squeezing. The roles of fluctuation increase and squeezing are reversed for the sine mode. Since quantum depletion always generates $s_{\bf k} \neq 0$ due to source \eqref{eq:s_source}, it induces squeezing in a form where the two bidirectional atom modes are squeezed in opposite directions.

\begin{figure}[ht]
\includegraphics*[scale=0.5]{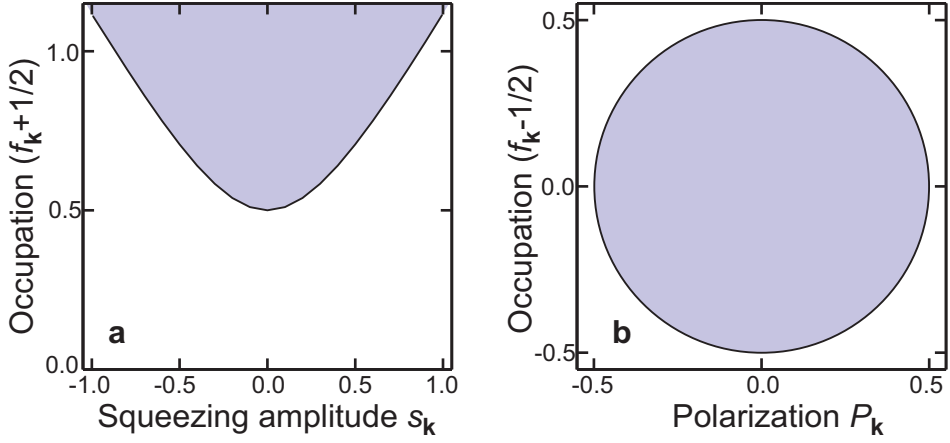}
\caption{
Geometric representation of excitation. {\bf a} Interacting Bose gas is limited to the $(s_{\bf k},\,f_{\bf k}+\frac{1}{2})$ region (shaded area) that is above  hyperbola set by the Heisenberg uncertainty principle \eqref{eq:Heisenmberg-ns}. {\bf b} In semiconductors, $(P_{\bf k},\,f_{\bf k}-\frac{1}{2})$ reside within the Bloch sphere based on Eq.~\eqref{eq:coherent-limit}. The $x$ axis corresponds to the real or imaginary part of $s_{\bf k}$ or $P_{\bf k}$.
}\label{HBEregion}
\end{figure}

Since the quadrature operators also satisfy the canonical commutation relation 
\begin{eqnarray}
  \comm{\op{X}_{{\bf k},\sigma}}{\op{Y}_{{\bf k}',\sigma'}} = -\frac{i}{2} \delta_{{\bf k}',{\bf k}} \, \delta_{\sigma,\sigma'} 
\label{eq:Canonical-comm}\;,
\end{eqnarray}
the fluctuations in the $x$ and $y$ direction satisfy the Heisenberg uncertainty relation, $\Delta X_{{\bf k},\sigma} \Delta Y_{{\bf k},\sigma} \ge \frac{1}{4}$. As we insert Eq.~\eqref{eq:XY_fluct} into this, we find that the Heisenberg uncertainly principle limits $(f_{\bf k},\,s_{\bf k})$ combinations by
\begin{eqnarray}
  \left(f_{\bf k} +{\textstyle \frac{1}{2}} \right)^2 - |s_{\bf k}|^2 \ge {\textstyle \frac{1}{4}}
\label{eq:Heisenmberg-ns}\;.
\end{eqnarray}
This means that all physically allowed $(f_{\bf k},\,s_{\bf k})$ combinations are restricted. Figure \ref{HBEregion}{\bf a} shows the set of allowed $(f_{\bf k},\,s_{\bf k})$ combinations as shaded area for real-valued $s_{\bf k}$. We see that {\it any physical} $(f_{\bf k},\,s_{\bf k})$ {\it excitations must reside at or above the hyperbola} defined by a minumum uncertainty state satisfying the Heisenberg uncertainty relation \eqref{eq:Heisenmberg-ns} with the equal sign.

Condition \eqref{eq:Heisenmberg-ns} resembles greatly the so-called coherent limit for interacting two-component fermions.\cite{Book:2011} For example, the simplest clusters in semiconductors are described by the electron occupation $f_{\bf k}$ at momentum state $\hbar {\bf k}$ and transition amplitude $P_{\bf k}$ between the two electron bands. Like for bosons, these excitations have the general restriction 
\begin{eqnarray}
  \left(f_{\bf k} -{\textstyle \frac{1}{2}} \right)^2 + |P_{\bf k}|^2 \le {\textstyle \frac{1}{4}}
\label{eq:coherent-limit}\;,
\end{eqnarray}
that defines the so-called coherent limit, cf.~Refs.~\cite{Kira:2006,Book:2011} for an extended discussion. Indeed, the simplest boson and fermion excitations resemble each other despite the sign and inequality-direction changes between the Heisenberg relation \eqref{eq:Heisenmberg-ns} and coherent limit \eqref{eq:coherent-limit}. Figure \ref{HBEregion}{\bf b} shows the allowed $(f_{\bf k},\,P_{\bf k})$-excitation values as shaded area when $P_{\bf k}$ is assumed to be real-valued. 
We see that the fermion excitations are limited within a circle. For the complex-valued $P_{\bf k}$, the parameter space becomes three dimensional limited by a sphere, commonly referred to as the Bloch sphere.\cite{Book:2011} In the same way, the complex-valued $s_{\bf k}$ introduces a three-dimensional $(f_{\bf k},\,s_{\bf k})$-parameter space for bosons. Following the fermionic name tradition, we call the limiting surface the Bloch hyperbola. 
For semiconductors, the $(f_{\bf k},\,P_{\bf k})$-excitation dynamics can be self-consistently solved from the semiconductor Bloch equations (SBEs)\cite{Lindberg:1988,Haug:2008,Book:2011} summarized in Sec.~\ref{sec:SBE}.

\subsection{Hyperbolic Bloch equations (HBEs)}
\label{sec:HBE}

For interacting bosons, the $(f_{\bf k},\,s_{\bf k})$ dynamics follows explicitly by writing down Eqs.~\eqref{eq:clusterDYN}--\eqref{eq:cubic-FIN} for all possible doublets $\DI{K}{P}$ with $K+P=2$ satisfying the homogeneous conditions discussed at the beginning of Sec.~\ref{sec:QK_doublets}: this procedure yields the hyperbolic Bloch equations (HBE) that describe the self-consistent coupling of the normal-component doublets
\begin{eqnarray}
  \ihddt s_{\bf k} 
  &=& 
  2 E^{\rm ren}_{\bf k} \, s_{\bf k} + \left( 1 + 2 f_{\bf k} \right) \Delta^{\rm ren}_{\bf k}
\nonumber\\
  &+&
  \sqrt{\NC} 
  \sum_{\bf l}
  \left[
    {\textstyle \frac{1}{2}} V_{\bf l} \left(T^{{\bf l},-{\bf k}}_{0,3} +  T^{{\bf l},{\bf k}}_{0,3}\right)
    + 
    V_{{\bf l},{\bf k}}
    \left(
      T^{{\bf k}-{\bf l},-{\bf k}}_{1,2}
      +
      T^{{\bf l}-{\bf k},{\bf k}}_{1,2}
    \right)
  \right]
\nonumber\\
  &+&
  \sum_{{\bf k}',{\bf q}} V_{\bf q}
  \left[ 
    Q^{{\bf k}'+{\bf q},{\bf k}-{\bf q},-{\bf q}}_{1,3}
    +
    Q^{{\bf q}-{\bf k}',-{\bf k}-{\bf q},{\bf k}}_{1,3}
  \right] 
\label{eq:HBE-s}\;,
\\
  \ihddt f_{\bf k} 
  &=& 
  2i\,{\rm Im}
  \left[
    \Delta^{\rm ren}_{\bf k}\,s^\star_{\bf k}
  \right] 
\nonumber\\
  &+&
  2i\,
  {\rm Im}
  \left[
  \sqrt{\NC} 
  \sum_{\bf l}
  \left(
    {\textstyle \frac{1}{2}} V_{\bf l} \, T^{{\bf l},{\bf l}-{\bf k}}_{1,2} 
    - 
    V_{{\bf l},{\bf k}} \, T^{{\bf l},{\bf k}}_{1,2}
  \right)
  +
  \sum_{{\bf k}',{\bf q}} V_{\bf q} \,Q^{{\bf q},{\bf k}',{\bf k}}_{2,2}
  \right]
\label{eq:HBE-n}\;,
\end{eqnarray}
with the BEC statistics defined by transformation \eqref{eq:BEC-QS}. Its lowest order contributions follow from the BEC number and the BEC number fluctuations
\begin{eqnarray}
  \NC&=&{\cal N} - \NN\,, \qquad \NN = \sum_{\bf k} f_{\bf k}\,
\nonumber\\
  \Delta\NC^2 &=& 
  2 \NN \, c_{\rm shape} \,,
  \qquad
  c_{\rm shape} 
  =
  \frac{1}{2}
   + \frac{1}{2\NN}
    \sum_{{\bf k}} 
    \left(
      f_{\bf k}^2+|s_{\bf k}|^2 + \sum_{{\bf k}'} Q^{0,{\bf k}',{\bf k}}_{2,2}
    \right)
\label{eq:HEB-BEC}\;,
\end{eqnarray}
respectively, obtained directly from Eqs.~\eqref{eq:NC}--\eqref{eq:c_shape}. The triplet and quadruplet terms that appear are determined by definitions \eqref{eq:triplets}--\eqref{eq:quadruplets}. 
The naming of Eqs.~\eqref{eq:HBE-s}--\eqref{eq:HEB-BEC} as the HBEs is motivated in Sec.~\ref{sec:SBE}.

The principal structure of the HBE is described by the first lines of Eqs.~\eqref{eq:HBE-s}--\eqref{eq:HBE-n} that contain the renormalized transition energy
\begin{eqnarray}
  E^{\rm ren}_{\bf k} 
  \equiv
  \frac{\hbar^2 {\bf k}^2}{2m}
  +
  \NC \, V_{\bf k}
  +
  \sum_{{\bf k}'} V_{{\bf k}'-{\bf k}}\, f_{{\bf k}'}
  -
  \sum_{{\bf k}'}{}' V_{{\bf k}'} 
    \left(
      f_{{\bf k}'}
      +
      {\rm Re} \left[s_{{\bf k}'} \right] 
    \right)
  -
   \sum_{{\bf k}',{\bf k}''}{}' 
   \frac{V_{{\bf k}'}}{\sqrt{\NC}}
   \,
  {\rm Re} 
  \left[
    T^{{\bf k}',{\bf k}''}_{1,2}
  \right]
\label{eq:E-ren-transition}\;,
\end{eqnarray}
as well as the renormalized quantum-depletion source
\begin{eqnarray}
  \Delta^{\rm ren}_{\bf k} 
  \equiv
  \NC V_{\bf k}
  +
  \sum_{{\bf k}'} V_{{\bf k}'-{\bf k}}\, s_{{\bf k}'}
\label{eq:ren-source}\;.
\end{eqnarray}
If the system has initially no normal component, $\Delta^{\rm ren}_{\bf k}$ directly generates only $s_{\bf k}$ transition amplitude whenever $\NC V_{\bf k}$ exists. 
As discussed in connection with Eq.~\eqref{eq:s_source}, the presence of BEC and interactions excites $s_{\bf k}$, and the $\NC V_{\bf k}$ is the only true source term for the quantum depletion. Once $s_{\bf k}$ exists, it modifies the source through the renormalization term that appears in Eq.~\eqref{eq:ren-source}. 

As mention in connection with Eq.~\eqref{eq:squeezing_s}, a quantity like $s_{\bf k}$ is often identified as the anomalous density, e.g.~in the of the Hartree-Fock Bogoliubov equations\cite{Baranger:1961,Goodman:1974,Mang:1975,Zaremba:1999,Bender:2003, Milstein:2003, Wuster:2005}  (HFBs), instead of the squeezing amplitude presented in Sec.~\ref{sec:bidirectional}. Due to this connection, the HBEs and HFBs partially share the same structure.
As major differences, (i) the Gross-Pitaevskii equation of HFB is replaced by the BEC-statistics transformation \eqref{eq:BEC-QS} that defines all quantum aspects of the BEC, (ii) the HFBs do not contain the two last energy-renormalization sums of Eq.~\eqref{eq:E-ren-transition}, as shown in Ref.~\cite{Kira:Xpic}, and (iii) the HFB do not contain the hierarchical coupling to the triplet and quadruplet correlations. 
The HBEs and HFBs essentially produce the same quantum kinetics for a weakly interacting Bose gas because then the renormalization and cluster effects are small. 
However, the HBE and HFB-based approach start to deviate for a strongly interacting Bose gas where both renormalization and cluster effects become decisive.

Since the HBEs do not have further corrections due to interactions, their structure is exact. In other words, the HBEs are exact long as the triplet and quadruplet contributions are formally known. One possibility is to solve, e.g., the quantum kinetics of triplets whose dynamics is presented in   \ref{app:triplets}. 
As more clusters are fully included to the analysis, one can systematically study the the sequential build up of clusters further in time, due to quantum depletion in a strongly interacting Bose gas; compare discussion in Sec.~\ref{sec:sequential}. However, the semiconductor investigations show \cite{Khitrova:1999,Kira:2006,Smith:2010,Book:2011} that such a hierarchical coupling can be included approximatively as long as the high-rank clusters mainly mediate scattering events, i.e., 
before they form new bound clusters. This reduces the computational HBE effort considerably while producing accurate description of excitation dynamics for a wide range of excitation conditions\cite{Kira:2006,Smith:2010,Book:2011}. We develop the HBE into this direction in Sec.~\ref{sec:dephasing} and the connection to the Bogoliubov excitations is made in Sec.~\ref{sec:Bogoliubov}.

\subsection{Comparison with the semiconductor Bloch equations (SBEs)}
\label{sec:SBE}

From a formal point of view, the principal structure of HBEs greatly resemble the semiconductor Bloch equations (SBEs)\cite{Lindberg:1988,Kira:2006,Book:2011}
\begin{eqnarray}
  \ihddt P_{\bf k} 
  &=& 
  E^{\rm eh}_{\bf k} \, P_{\bf k} - \left( 1 - f^e_{\bf k}-f^h_{\bf k}  \right) \Omega_{\bf k}
  - i \Gamma^{\rm pol}_{\bf k}\,,
\nonumber\\
  \hbar \frac{\partial}{\partial t} f^\lambda_{\bf k} 
  &=&
  2\,{\rm Im}
  \left[
    \Omega_{\bf k}\,P^\star_{\bf k}
  \right] 
  - \Gamma^\lambda_{\bf k}\,, \qquad \lambda = e,\,h
\label{eq:SBE}\;,
\end{eqnarray}
which describe the quantum kinetics of polarization $P_{\bf k}$  as well as electron ($\lambda=e$) and hole ($\lambda=h$) density $f^\lambda_{\bf k}$ excited in a two-band semiconductor by a laser field; a cluster-expansion-based derivation of SBEs is presented, e.g.~in Refs.~\cite{Kira:2006,Book:2011}. For two-component fermions, $P_{\bf k}$ and $f^\lambda_{\bf k}$ are singlets and $\Gamma^{\rm pol}_{\bf k}$ and $\Gamma^\lambda_{\bf k}$ describe the hierarchical coupling to the doublets. 
In semiconductors, pairwise interactions stem from the Coulomb force among electrons and holes as well as couplings of charge carriers with two different boson fields: the quantized light (photons) and lattice vibrations (phonons). As a result of these interactions, $\Gamma^{\rm pol}_{\bf k}$ and $\Gamma^\lambda_{\bf k}$ include Coulomb-interaction induced, quantum-optical, and phonon-interaction induced correlations; for a detailed form cf.~Ref.~\cite{Kira:2006}. 

Since electrons and holes interact via a pairwise Coulomb interaction, the kinetic energy of the electron--hole pair becomes renormalized\cite{Kira:2006,Book:2011}
\begin{eqnarray}
  E^{\rm eh}_{\bf k} 
  \equiv
  \frac{\hbar^2 {\bf k}^2}{2\mu} +E_g
  -
  \sum_{{\bf k}'}{}' V_{{\bf k}'} 
    \left(
      f^e_{{\bf k}'}
      +
      f^h_{{\bf k}'}
    \right)
\label{eq:E-eh-ren}\;,
\end{eqnarray}
where $\mu$ is the reduced electron--hole mass, $E_g$ is the bandgap of the semiconductor, and $V_{\bf k}$ is the Fourier transform of the Coulomb potential; for the sake of comparison, we use here the same symbol as for the atom--atom interactions. The particle density renormalizes the energy of a free electron--hole pair, $\frac{\hbar^2 {\bf k}^2}{2\mu}+E_g$, via the Coulomb interaction very similarly as the atom--atom interaction and atom densities 
modify the free-atom energy in Eq.~\eqref{eq:E-ren-transition} defining the renormalized energy $E^{\rm ren}_{\bf k}$. Expression \eqref{eq:E-eh-ren} is strictly speaking valid close to the high-symmetry point of the band structure where the effective-mass approximation produces a quadratic dispersion used here. In a more general situation, $\frac{\hbar^2 {\bf k}^2}{2\mu} +E_g$ should be replaced with a general electron--hole pair energy whose full ${\bf k}$ dependence is defined by the actual energy difference between the conduction- and valence-band energies.

The $P_{\bf k}$ part of the SBEs becomes generated only if $\Omega_{\bf k}$ is nonzero. The HBE--SBE analogy also touches this source, the so-called renormalized Rabi energy\cite{Kira:2006,Book:2011}
\begin{eqnarray}
  \Omega_{\bf k} 
  \equiv
  d_{\rm eh} \, E(t)
  +
  \sum_{{\bf k}'} V_{{\bf k}'-{\bf k}}\, P_{{\bf k}'}
\label{eq:Rabi-ren}\;,
\end{eqnarray}
where $d_{\rm eh}$ is the dipole-matrix element for an electron--hole transition and $E(t)$ is the electric field driving the transition. In practice, $E(t)$ is defined by the laser that excites the semiconductor whereas $d_{\rm eh}$ is material dependent. By comparing $\Omega_{\bf k}$ with Eq.~\eqref{eq:ren-source}, we conclude that the laser pulse corresponds to the atom BEC. 
For both $\Delta_{\bf k}$ and $\Omega_{\bf k}$, the generated transition amplitudes renormalize the direct source via $\sum_{{\bf k}'} V_{{\bf k}'-{\bf k}}\, s_{{\bf k}'}$ and $\sum_{{\bf k}'} V_{{\bf k}'-{\bf k}}\, P_{{\bf k}'}$, respectively. As a practical difference, $E(t)$ can range from continuous-wave to ultrafast laser pulse whereas $\NC$ is defined by the lifetime of the BEC, which is typically long.

Despite the structural similarity of the HBEs and SBEs, these two approaches are bound to have obvious differences because the SBEs characterize fundamentally fermionic excitations while HBEs describe the quantum kinetics of the interacting Bose gas. By comparing the structure of the SBEs \eqref{eq:SBE} with the 
HBEs \eqref{eq:HBE-s}--\eqref{eq:HBE-n}, we recognize a critical sign flip; in Eq.~\eqref{eq:HBE-s}, the source is multiplied by $(1+2f_{\bf k})$ whereas the SBE contains $-1 +f^e_{\bf k} +f^h_{\bf k}$. Note that this term reduces to $(-1+2f_{\bf k})$ for equal electron and hole distributions $f^e_{\bf k} = f^h_{\bf k} \equiv f_{\bf k}$. In other words, the direct source of the transition amplitudes has an opposite sign in the SBE and HBE. This seemingly innocent difference, i.e.~$(\pm1+2f_{\bf k})$, changes the equation class so that the excitation stays within the Bloch sphere ($-$ sign) for the SBEs whereas HBE excitations remain at the upper exterior of the Bloch hyperbola ($+$ sign). 
As shown in Fig.~\ref{HBEregion}, the corresponding excitations look geometrically completely different even though the major part of the SBE/HBE structure is identical. Since the SBEs are a well-established concept in semiconductor optics\cite{Haug:2008,Kira:2011}, I decide to call the principle structure of doublet dynamics in interacting Bose gas the ``hyperbolic Bloch equations'', based on the fermion-excitation tradition.

Regardless of the obvious similarity between the HBEs and SBEs, $P_{\bf k}$ and $f^\lambda_{\bf k}$ are singlets while the corresponding Bose-gas quantities, $s_{\bf k}$ and $f_{\bf k}$, are doublets; the correlation rank is halved in the SBEs. This is a pure identification issue because fermionic expectation values and correlations vanish if the sum of creation- and annihilation-operator numbers is odd\cite{Kira:2006,Book:2011}, unlike for bosons. 
Therefore, a fermionic correlation with $K$ creation and $P$ annihilation operators is labeled as a $\frac{K+P}{2}$-particle correlation. For the same number of operators, an equivalent amount of boson operators would count as a $(K+P)$-particle correlation. Despite this difference in correlation classification, the hierarchical coupling structure is analogous for the interacting fermion and boson systems, and even when bosons and fermions are mixed via pairwise interactions. For a textbook discussion, cf.~Ref.~\cite{Book:2011} 

Current state of the art in SBE-based studies solves the quantum kinetics up to three hierarchical levels of clusters\cite{Hoyer:2004,Kira:2006,Schneebeli:2008}, i.e.~dynamics up to triplets can be solved numerically; for an interacting Bose gas, and an equivalent numerical effort would include quantum kinetics up to six-atom correlations. Such an analysis is clearly beyond the scope of this paper because we first need to explore how the doublets, $s_{\bf k}$ and $f_{\bf k}$, explain the elementary properties of the strongly interacting Bose gas. 
Therefore, we focus on solving the doublet HBEs numerically even through we  have already derived the general quantum dynamics of any clusters in Sec.~\ref{sec:cluster_kinetics}. Predicting from the historic evolution of the SBE-based studies, identification of higher-order correlation effects will be a keen topic of future research for the interacting Bose gas.

\subsection{Dephasing in the HBEs}
\label{sec:dephasing}

The SBEs establish one of the most accurate, and microscopically sound, descriptions of the optical excitations in semiconductors. The SBE computations are already at a level where one can precisely explain and isolate which intriguing many-body effects are responsible for features appearing in high-precision measurements\cite{Smith:2010}, especially when the doublets are systematically included to the SBEs. We can directly benefit from this research tradition by meaningfully focusing the HBE-based investigations to conditions matching the SBE-based studies. 
We can then introduce phenomenological approximations --- thoroughly tested with the SBEs --- to the HBEs in order to simplify the numerical effort. Since the properties of the HBEs are currently unexplored, we concentrate next on determining the principal effects resulting from the quantum kinetics of the doublets. This will set the basis to understand overall nonequilibrium quantum kinetics following an interaction switch-on. 

In the SBEs, the hierarchical coupling $\Gamma^\lambda_{\bf k}$ ($\lambda={\rm pol},\, e,\, h$) induce scattering among clusters that generates effective decay, energy shifts, and relaxation of $P_{\bf k}$ and $f^\lambda_{\bf k}$, as shown in Ref.~\cite{Jahnke:1996,Khitrova:1999,Kira:2006,Smith:2010,Book:2011}. This is not entirely unexpected based on Fig.~\ref{Cnumbers} showing that the nonlinear contributions produce literally a scatter of cluster contributions; for a detailed identification of Boltzmann-type scattering, see the   \ref{app:triplets}. 
The microscopic origin of dephasing, energy shifts, and relaxation indeed stems from such nonlinear terms, as shown in Ref.~\cite{Kira:2006}. These scattering effects depend then nonlinearly on $P_{\bf k}$ and $f^\lambda_{\bf k}$ such that one can detect excitation-induced dephasing, energy shifts, and relaxation in semiconductor experiments.\cite{Peyghambarian:1984,Wang:1993,Hu:1994,Jahnke:1996,Chow:1997,Manzke:1998,
Sieh:1999,Khitrova:1999,Shacklette:2002,Smith:2010} 
Similar excitation-induced effects are also to be expected in the interacting Bose gas. 

For the SBEs, the principal effects of $\Gamma^\lambda_{\bf k}$ can be described by replacing the microscopic $\Gamma^\lambda_{\bf k}$ by phenomenalogical dephasing and relaxation constants discussed, e.g.~in Ref.~\cite{Kira:2006}. When they are defined either from microscopic computations\cite{Jahnke:1996,Kira:2006,Smith:2010} or experiments\cite{Tomaino:2012,Rice:2013}, 
one can indeed explain nontrivial details of the experimentally observed quantum-kinetic phenomena, as long the effects depend directly on polarization and density dynamics. Naturally, this approximation scheme should not be used when one studies direct effects of higher-order clusters. 

In the HBEs, the quantum-depletion source \eqref{eq:ren-source} generates normal-component excitations in the form of doublets. Once the doublets are created, they start to sequentially generate higher-order correlations. As discussed in Sec.~\ref{sec:sequential}, this process has several competing channels yielding a diffusion-like scattering; the scattering structure is particularly clear in the triplet dynamics presented in   \ref{app:triplets}. 
Therefore, the hierarchical coupling of the HBEs indeed yields similar excitation-induced dephasing and energy shifts as well as relaxation of the atom distribution. 
Motivated by the SBE insights above, we simply replace all triplets and quadruplets within the Eqs.~\eqref{eq:HBE-s}--\eqref{eq:HBE-n} by the phenomenological dephasing constants $\gamma_{\bf k}$ and  $\Gamma_{\bf k}$ for squeezing amplitude and atom distribution, respectively. At this point, we can also include energy and particle exchange phenomenologically by introducing a relaxation rate $r$ toward a quasi-equilibrium distribution $n^{\rm eq}_{{\bf k}}$. With these substitutions, the HBEs \eqref{eq:HBE-s}--\eqref{eq:HEB-BEC} cast into form
\begin{align}
  \ihddt s_{\bf k} 
  &= 
  2 \left( E^{\rm ren}_{\bf k} - i \gamma_{\bf k} \right) s_{\bf k} + \left( 1 + 2 f_{\bf k} \right) \Delta^{\rm ren}_{\bf k}\;,
\nonumber\\
  \hbar \frac{\partial}{\partial t} f_{\bf k} 
  &=
   -\Gamma_{\bf k}\, f_{\bf k}
  -r \left(f_{\bf k} - n^{\rm eq}_{\bf k} \right)
  + 2\,{\rm Im}
  \left[
    \Delta^{\rm ren}_{\bf k}\,s^\star_{\bf k}
  \right]\,,
\qquad
  \NC = {\cal N} - \sum_{\bf k} f_{\bf k}
\label{eq:HBE-FIN}\;.
\end{align}
By letting $\gamma_{\bf k}$ be complex valued and excitation dependent, we can estimate the effect of both excitation-induced dephasing and shifts on the quantum depletion. 

Based on the Heisenberg relation~\eqref{eq:Heisenmberg-ns}, quantity
\begin{eqnarray}
  \eta_{\bf k} \equiv
  \left(f_{\bf k} +{\textstyle \frac{1}{2}} \right)^2 - |s_{\bf k}|^2
\label{eq:eta-quantitiy}\;
\end{eqnarray}
defines how close the system is to a minimum uncertainty state having $\eta_{\bf k} = \frac{1}{4}$ that is the lower bound of the Heisenberg relation. As shown in Ref.~\cite{Kira:2006b,Book:2011}, any boson-field that is described by doublets has the lowest entropy when it is exactly a minimum uncertainty state.  
For a low-temperature atom gas, the entropy tends to approach its minimum value (zero), which helps us to connect $\gamma_{\bf k}$ and $\Gamma_{\bf k}$. For an unexcited system, the entropy vanishes because it produces $\eta_{\bf k} = \frac{1}{4}$, i.e.~a minimum uncertainty state. If we ignore the coupling of the interacting Bose gas to the outside world by setting $r=0$, 
Eq.~\eqref{eq:HBE-FIN} produces
\begin{eqnarray}
  \ihddt \eta_{\bf k} =
  i 
  \left(
    4 {\rm Re} \left[ \gamma_{\bf k} \right] |s_{\bf k}|^2
    - \Gamma_{\bf k} (1+2 f_{\bf k})\, f_{\bf k}
  \right)
\label{eq:eta-dyn}\;.
\end{eqnarray}
Interestingly the imaginary part of $\gamma_{\bf k}$, that defines the energy shift, does not influence $\eta_{\bf k}$ dynamics.  In general, the system remains in a minimum uncertainty state if the right-hand side of Eq.~\eqref{eq:eta-dyn} vanishes, which sets up the {\it minimum uncertainty condition} for the dephasing:
\begin{eqnarray}
  \Gamma_{\bf k} =
  4 {\rm Re} \left[ \gamma_{\bf k} \right] \frac{|s_{\bf k}|^2}{(1+2 f_{\bf k})\, f_{\bf k}}
  =
   2{\rm Re} \left[ \gamma_{\bf k} \right] \frac{2+2 f_{\bf k}}{(1+2 f_{\bf k})}
\label{eq:Gamma-gamma}\;.
\end{eqnarray}
The last step follows when we replace $|s_{\bf k}|^2$ by $f_{\bf k} (1+f_{\bf k})$ that is satisfied by any  minimum uncertainty state.

As a general feature, low excitations ($f_{\bf k} \ll 1$) produce $\Gamma_{\bf k} = 4 \gamma_{\bf k}$ whereas high excitations ($f_{\bf k} \gg 1$) yield $\Gamma_{\bf k} =  2\gamma_{\bf k}$. Therefore, highly excited parts of the normal component decay {\it slower} than the low-excitations parts. This tendency is opposite to that observed for two-component fermions, which again highlights the fundamental difference of boson and fermion systems. 
More specifically, the squeezing amplitude decays with a time constant $T_{2,{\bf k}} = \frac{\hbar}{2 \gamma_{\bf k}}$ whereas the density decays with a time constant $T_{1,{\bf k}} \equiv \frac{\hbar}{\Gamma_{\bf k}}$. The population decay is two times faster, i.e.~$T_{1,{\bf k}} =  T_{2,{\bf k}}/2$, for dilute $f_{\bf k}$ occupations. This is the well-known connection between population and transition decay time of many physically different two-level systems \cite{Levitt:2008,Book:2011}. 
Interestingly, elevated excitations convert the decay relation to be equally slow, i.e.~$T_{1,{\bf k}} =  T_{2,{\bf k}}$.

Based on common wisdom of the SBE studies, we can expect that model \eqref{eq:HBE-FIN} describes how the switch-on investigations produce a nonequilibrium many-body system through the formation of dimers. Obviously, trimer dynamics requires a further analysis that includes at least the triplets. This follows straightforwardly from the general cluster-expansion framework developed in this work.

\subsection{Hyperbolic Wannier equation}
\label{sec:Wannier}

The dynamics of the transition amplitude is typically much faster than the density dynamics such that the densities can often be treated as quasi-stationary. In this situation, the principal structure of both SBEs and HBEs  identifies an eigenvalue problem whose solutions explain how individual particles become bound pairs\cite{Book:2011}, as discussed below. More specifically, the right-hand side of HBE \eqref{eq:HBE-FIN}, without the scattering terms, contains a homogeneous part for $s_{\bf k}$ that establishes an eigenvalue problem 
\begin{eqnarray}
  2 E^{\rm ren}_{\bf k} \phi^R_\nu({\bf k}) + (1+2 f_{\bf k}) \sum_{{\bf k}'} V_{{\bf k}'-{\bf k}} \phi^R_\nu({\bf k}')
  = E_\nu \, \phi^R_\nu({\bf k})
\label{eq:H-Wannier}\;,
\end{eqnarray}
where $E_\nu$ is the eigenvalue and $\phi^R_\nu({\bf k})$ is the right-handed eigenfunction associated with $s_{\bf k}$. We will show below that $\phi^R_\nu({\bf k})$ defines the relative wave function of the bound atom pairs, the so-called dimers.

Structurally, the eigenvalue problem \eqref{eq:H-Wannier} is almost identical to the generalized Wannier equation\cite{Wannier:1937,Book:2011} that defines the wave functions of bound electron--hole pairs in semiconductors. Only the population factor $(1+2 f_{\bf k})$ should be changed into $(-1+2 f_{\bf k})$ to get the usual Wannier equation. This is exactly the same sign change that makes the Bose-gas dynamics hyperbolic compared to the SBEs, as discussed in Sec.~\ref{sec:HBE}. 
In other words, the sign in the occupation factor $(\pm1+2 f_{\bf k})$ stems from the ``elementary'' particles being either bosons $(+)$ or fermions $(-)$. Due to the formal analogy and geometrical difference between Eq.~\eqref{eq:H-Wannier} and the generalized Wannier equations, we call the eigenvalue problem \eqref{eq:H-Wannier} as the {\it hyperbolic Wannier equation}. Next, we directly convert the properties already known for the generalized Wannier equations to be applicable for the interacting Bose gas.

At low excitation levels $f_{\bf k} \ll 1$, we may ignore $f_{\bf k}$ in Eq.~\eqref{eq:H-Wannier} as well as the energy renormalizations, yielding $E^{\rm ren}_{\bf k} \rightarrow \frac{\hbar^2 {\bf k}^2}{2m}$. This limit converts the hyperbolic Wannier equation into
\begin{eqnarray}
  2\frac{\hbar^2 {\bf k}^2}{2m}\, \phi_\nu({\bf k}) + \sum_{{\bf k}'} V_{{\bf k}'-{\bf k}} \phi_\nu({\bf k}')
  = E_\nu \, \phi_\nu({\bf k})
\label{eq:H-Wannier-low}\;,
\end{eqnarray}
which becomes a symmetric eigenvalue problem; we have therefore dropped the ``R'' index from the eigenfunction. We may now Fourier transform Eq.~\eqref{eq:H-Wannier-low} into the real space, yielding
\begin{eqnarray}
  -2\frac{\hbar^2 \nabla^2}{2m}\, \phi_\nu({\bf r}) + V({\bf r})\, \phi_\nu({\bf r})
  = E_\nu \, \phi_\nu({\bf r})
\label{eq:H-Wannier-low-r}\;,
\end{eqnarray}
after having applied the convolution theorem\cite{Arfken:1985} to Fourier transform the interaction term. This equation matches exactly with the relative-motion Schr\"{o}dinger equation of two atoms that interact via a pairwise potential $V({\bf r})$. Therefore, the solutions to the hyperbolic Wannier equation, indeed, describe the properties of the dimers. 
As the atom density increases, we obviously must both include the $f_{\bf k}$ and renormalize the kinetic energy, as presented in Eq.~\eqref{eq:H-Wannier}. The resulting {\it hyperbolic Wannier equation describes the simplest many-body renormalizations to the dimer problem}. 
Even though it is more typical to solve the low-density dimer problem \eqref{eq:H-Wannier-low-r} in real space, the high-density problem becomes significantly simpler in the momentum space \eqref{eq:H-Wannier}. Specific examples of how the dimer states are changed by the exitation level are given in Sec.~\ref{sec:numerics}.

Equation~\eqref{eq:H-Wannier} can technically be viewed as a matrix eigenvalue problem where the matrix is asymmetric due to the $(1+2 f_{\bf k})$ factor. Consequently, it has different left-handed and right-handed solutions $\phi^L_\nu({\bf k})$ and $\phi^R_\nu({\bf k})$, respectively. It is straightforward to show that $\phi^L_\nu({\bf k})$ and $\phi^R_\nu({\bf k})$ have the same eigenvalues while they are connected via
\begin{eqnarray}
  \phi^L_\nu({\bf k}) = \frac{\phi^R_\nu({\bf k})}{1+2 f_{\bf k}}
\label{eq:LR-solutions}\;.
\end{eqnarray}
In analogy to the generalized Wannier problem\cite{Kira:2006,Book:2011}, these form a complete set orthonormal functions satisfying normalization and orthogonality relations,
\begin{eqnarray}
  \sum_{\bf k} \left[\phi^L_\lambda({\bf k})\right]^\star \phi^R_\nu({\bf k}) = \delta_{\lambda,\nu}\,,
  \qquad
  \sum_\nu \left[\phi^L_\nu({\bf k})\right]^\star \phi^R_\nu({\bf k}') = \delta_{{\bf k}',{\bf k}}
\label{eq:LR-completeness}\;,
\end{eqnarray}
respectively. Once all $\phi^R_\nu({\bf k})$ are known, we can convert the squeezing amplitude into the {\it dimer basis} by using a one-to-one mapping:
\begin{eqnarray}
  s_{\bf k} \equiv \sum_\nu s_\nu \, \phi^R_\nu({\bf k})
  \quad \Leftrightarrow \quad
  s_\nu = \sum_{\bf k} \left[\phi^L_\nu({\bf k}) \right]^\star s_{\bf k}
\label{eq:dimer-basis}\;,
\end{eqnarray}
which follows from the orthogonality and completeness relations \eqref{eq:LR-completeness}. By studying the squeezing amplitude in the dimer basis, we may determine which dimer states the normal-component atoms occupy at a given moment.

\section{Numerical results}
\label{sec:numerics}

To solve the HBEs for a realistic Bose gas, we analyze $^{85}$Rb-type systems because they are routinely applied to study the Bose-Einstein condensation\cite{Anderson:1995,Bradley:1995,Davis:1995,Dalfovo:1999,Leggett:2001,Andersen:2004,Blakie:2008,Proukakis:2008}. More specifically, we model fast-switch experiments\cite{Donley:2001,Altin:2011,Makotyn:2014} where the strength of the atom--atom interaction is abruptly jumped from weak to strong interactions. 
For ultracold conditions studied here, the system with weak interactions contains virtually no normal-component atoms, let alone correlations. The sudden jump to strong interactions initiates sequential build up of clusters to the normal component, which can efficiently followed with the cluster-expansion approach as discussed in Sec.~\ref{sec:sequential}.
The resulting quantum kinetics is largely unexplored at the moment, which makes the HBE-based switch-on investigation interesting. Next, we briefly summarize the central atom and interaction parameters as well as connect the basic concepts of interacting Bose gas with semiconductor optics. 
 	
The bosonic $^{85}$Rb atoms have a mass\cite{Emsley:1998} $m= 4.9118\,u$ where $u$ is the standard atomic weight. The individual $^{85}$Rb atoms have a covalent radius\cite{Cordero:2008} $r_0=220$\,pm that defines the typical length scale of atomic bonding, see Fig.~\ref{Vpotentials}. Bound atoms are typically referred to as dimers. For semiconductors, the SBEs also contain bound states --- called excitons --- that are Coulomb-bound electron--hole pairs\cite{Book:2011}. Due to the analogy between HBEs and SBEs, dimers and excitons are formally equivalent. When the atoms are not bound to each other, they form a boson plasma. Equivalently, unbound electrons and holes form an ionized plasma of charge carriers. 

Fesbach resonances\cite{Kohler:2006,Chin:2010} can be utilized in a situation where atoms have two energetically nearby hyperfine levels, one with a bound state (closed channel) and another one with continuum states (open channel) coupled with an external magnetic field. This coupling produces a dispersive resonance in the atom--atom interaction strength whose sign and magnitude can be freely adjusted by detuning through the resonance.
Analogous dispersive Fano resonances also exist in solids whenever a discrete resonances is coupled to a continuum of states.\cite{Kobayashi:2004,Golde:2009,Lukyanchuk:2010}  
In particular, the mixed state produces total atom--atom interactions that can be continuously and abruptly tuned from weak to strong interactions with a time-dependent external magnetic field. 

As shown in Refs.~\cite{Stwalley:1976,Tiesinga:1993,Kokkelmans:2002} and applied in Refs.~\cite{Inouye:1998,Courteille:1998,Roberts:1998,Cornish:2000,Takemura:2013,
Partridge:2005}, 
Feshbach-resonance modified interactions can be described with an effective single-atom level interaction whose interaction strength is freely adjustable. We adopt this approach to eliminate the bookkeeping related to the multiple hyperfine atom levels; the full multilevel extension is discussed in Ref.~\cite{Kira:Xpic}. In the effective single-level approach used here, the external magnetic field simply changes the magnitude of
$V({\bf r})$ to jump the atom--atom interaction from weak to strong interactions. We will next study how to perform this systematically in order to connect the analysis to realistic fast-switch experiments.

In semiconductors, Coulomb interaction cannot be changed  much externally, at least not to a degree one can control $V({\bf r})$ in boson gas through the Feshbach resonance. However, one can adjust the electron--hole interaction of semiconductors to change from attractive to repulsive. 
The simplest way is to control the electron--hole concentration via optical pumping; in the polarization dynamics \eqref{eq:SBE}, Coulomb interaction is multiplied by a so-called phase-space filling factor $(1-f^e_{\bf k}-f^h_{\bf k})$ that changes sign when the sum of the electron--hole occupation exceeds one. This border defines the onset of exciton ionization into electron--hole plasma because a negative-valued effective potential yields electron--hole repulsion. The equivalent density renormalization is always positive in the HBE \eqref{eq:HBE-FIN} such that one cannot switch the sign of total atom--atom interactions by changing the atom density. 
Nonetheless, one can analogously study dimer dissociation and exciton ionization by using the Feshbach resonance and the excitation level, respectively, to control the overall sign of the particle--particle interactions; for systematic semiconductor studies cf.~Refs.~\cite{Kira:2006,Book:2011}. 
Currently, interesting development toward Feshbach-like control in semiconductors is made in microcavities where the interaction strength is controlled by coupling between exciton-polaritons in a microcavity (open channel) and an exciton molecule (closed channel)\cite{Takemura:2013}.

\subsection{Dimer scattering length and the Morse potential}
\label{sec:scattering length}

In many cases, the specific details of the atom--atom interactions do not influence the many-body properties.\cite{Dalfovo:1999,Leggett:2001,Andersen:2004,Blakie:2008,Proukakis:2008}
Instead, $V({\bf r})$ can be reduced to a contact-potential interaction\cite{Leggett:2001,Giorgini:2008,Bloch:2008} containing a single parameter, the so-called scattering length $a_{\rm scatt}$. As summarized in Ref.~\cite{Giorgini:2008}, the dimer scattering length follows from the free atom--atom Schr\"{o}dinger equation \eqref{eq:H-Wannier-low} via\cite{Giorgini:2008}
\begin{eqnarray}
  \lim_{r \rightarrow 0} \frac{d}{dr} {\rm ln}\left[ r\,\phi_\nu(r) \right]  \equiv - \frac{1}{a_{\rm scatt}}
\label{eq:a_scatt}\;.
\end{eqnarray}
In case $V(r)$ is a real atomic interaction potential, one should take the limiting value approaching atomic scale, not exactly zero because $r\phi_\nu(r)$ always decays to zero for $r \rightarrow 0$ due to the repulsive interaction core depicted in Fig.~\ref{Vpotentials}. 

For ultracold atom gas, the most relevant atom-scattering events involve atoms that have an energy close to zero. For example, unbound atoms can become bound pairs only if the dimer binding is not too large. In fact, scattering of atoms to a strongly bound dimer state requires participation of three or more atoms. 
Such a scattering produces often an atom loss when the excess energy of the binding releases so large kinetic energy that one of the scattering partners exits the atom trap. Such a process resembles Auger scattering\cite{Klimov:2000,Hader:2005,Robel:2009} in semiconductors where a multi-carrier scattering/recombination event releases so much kinetic energy that either electrons or holes exit the quantum-confined states they initially occupy. In this paper, we focus the analysis on homogeneous excitation without the atom losses.

When the description is already reduced to a single atom level with an effective $V({\bf r})$, we choose the zero level such that unbound atoms have a positive energy while dimers have a negative energy. In BEC experiments, the cooling tends to bring the atoms close to the zero energy such that the cold-atom collisions mainly involve ionized states and a dimer state with the highest energy. 
In case the highest laying dimer state $\nu=\nu_{\rm high}$ is sufficiently close to the zero energy, the corresponding $s$-wave scattering is characterized by $a_{\rm scatt}$ that is defined for $\nu=\nu_{\rm high}$. In case the highest laying dimer is deeply bound, atoms cannot efficiently scatter to that state such that the normal component remains essentially in plasma form and the $s$-wave scattering is defined by $a_{\rm scatt}$ of the lowest laying unbound state. 

While the deeply bound dimer states are often inaccessible in ultracold Bose gas, various interaction processes of semiconductors can access even the exciton ground state. Therefore, many semiconductor phenomena can be qualitatively understood via the exciton ground-state properties characterized by the exciton Bohr radius $a_B$, which depends strongly on the electron--hole excitation level of the system. 
Since exciton's $\phi_\nu(r)$ does not diverge at $r=0$, $a_B$ follows from Eq.~\eqref{eq:a_scatt} when $r\,\phi_\nu(r)$ is replaced by $\phi_\nu(r)$ that is a solution of the generalized Wannier equation\cite{Book:2011} with the phase-space filling modifications to the Coulomb potential. Besides this technical difference between $a_B$ and  $a_{\rm scatt}$, both of them nicely classifies how the system is expected to behave in different situations. In other words, $a_B$ and $a_{\rm scatt}$ can be formally interpreted as scattering lengths. 
For a slightly positive scattering length, the many-body dynamics is often dominated by bound particle pairs, i.e.~dimers or excitons. For a slightly negative scattering length, dimers and excitons have only unbound solutions such that the many-body interactions take place in a gaseous plasma.

The limit where $|a_{\rm scatt}|$ approaches infinity is often referred to as unitarity in the interacting atom gas, including fermionic atom systems.\cite{Cornish:2000,Braaten:2006,Giorgini:2008, Papp:2008,Navon:2011,Smith:2012,Wild:2012} In this situation, the pair-state wave function has an infinite extension. The connection of unitarity and strong atom interactions becomes particularly clear when $V({\bf r})$ is replaced by the contact potential because $V({\bf r})$ becomes proportional to $a_{\rm scatt}$, see discussion related to Eq.~\eqref{eq:V_NC_eff}. 
Since $a_{\rm scatt}$ diverges (or becomes strongly renormalized\cite{Kokkelmans:2002}) at unitarity, the interactions are clearly maximized. However, the Coulomb interaction cannot be substituted by the contact potential such that semiconductor's ``unitarity'' only means the exciton-ionization threshold. In other words, the size of ground-state exciton diverges at the ionization threshold such that the system cannot anymore sustain bound pairs, i.e.~the pairwise correlations become weak. Nevertheless, both semiconductors and interacting atom gas can show intriguing new many-bopy configurations close to the ionization threshold/unitarity. 
A recent study\cite{Hunter:2014} demonstrates that electrons and holes can form a quantized electron--hole droplet --- the so-called dropleton ---  where three to seven electrons and and holes are bound within a liquid bubble. Analogously, formation of Efimov trimer\cite{Efimov:1973}, consisting of three bound atoms, have been demonstrated\cite{Kraemer:2006,Williams:2009,Castin:2010,Stecher:2009,Pollack:2009} close to unitarity.

An effective contact potential is very useful when one seeks to derive analytic properties of detail-independent thermodynamic properties.\cite{Dalfovo:1999,Giorgini:2008,Tan:2008a} Since $V(r)$ itself then becomes a nonanalytic function, the resulting analytic investigations can be systematically carried out only by carefully analyzing the properties of the nonanalytic functions and renormalizations that follow.\cite{Tan:2008a,Tan:2008b,Tan:2008c} 
The description of nonanalytic functions is cumbersome in numeric applications such as quantum Monte Carlo studies of interacting Bose or Fermi gas; for example, Refs.~\cite{Holzmann:1999,Giorgini:1999,Carlson:2003,Astrakharchik:2004, Burovski:2008,Bulgac:2008,Bulgac:2006,Lee:2006,Pilati:2008} use a simple analytic function for $V(r)$ to model the atom--atom interactions. 

Since also the HBEs can only be solved numerically, it clearly is desirable to use an analytic $V(r)$ also in this context. Despite the added numerical demands, the inclusion of a realistic $V({\bf r})$ allows us to inspect how many subtle effects --- not included by the contact potential --- emerge in experiments. For example, a realistic $V({\bf r})$  can hosts several deeply bound dimer states while the contact potential contains only one. 
Additionally, solutions with a realistic $V({\bf r})$ automatically includes all renormalizations to the description because the atom--atom repulsion and attraction at short and long distances, respectively, are fully included, unlike for the contact potential. To include the essential features of $V(r)$ described in Fig.~\ref{Vpotentials}, we use the Morse potential\cite{Morse:1929}
\begin{eqnarray}
  V(r)= D\, 
    \left(
      e^{-2 \frac{r-r_0}{\Delta r}}
      -
      2\,
      e^{-\frac{r-r_0}{\Delta r}}
    \right)
\label{eq:Morse-r}\;,
\end{eqnarray}
where $D$ is the energy depth of the attractive part of the potential, and the potential dips to its minimum value at $r_0$. For elevated $r$, the attractive dip decays towards zero exponentially with a decay constant defined by $\Delta r$ that also sets the range of the potential. At origin, the potential becomes $D\, \left( e^{2 \frac{r_0}{\Delta r}} - 2\, e^{\frac{r_0}{\Delta r}} \right)$ that is positive for $\Delta r \le r_0\,{\rm ln} \, 2$ implying that $V(r)$ also correctly includes the atom--atom repulsion at short distances. 
The short-range repulsion and long-range attraction of the Morse potential is consistent with the typical form of atom--atom interactions depicted in Fig.~\ref{Vpotentials}. Therefore, we expect that the Morse potential describes the interaction effects of interacting Bose gas very well although the $-e^{-\frac{r}{r_0}}$ decay is faster than the expected $-\frac{1}{r^6}$ dependence for the van der Waals tail induced by the dipole-dipole interaction among atoms. 
 
In the HBE computations, one eventually needs the Fourier transformation of the atom--atom potential. By integrating Eq.~\eqref{eq:V-Fourier} with potential \eqref{eq:Morse-r}, we obtain
\begin{eqnarray}
  V_{\bf k} 
  = 
  \left(\frac{2\pi \Delta r}{\cal L}\right)^3
  2 D\,e^{ \frac{r_0}{\Delta r}}\, F_{\bf k}\,,
  \qquad
  F_{\bf k}
  \equiv
    \frac{e^{\frac{r_0}{\Delta r}}}{\left(4+|\Delta r \, {\bf k}|^2\right)^2 }
    -
    \frac{1}{\left(1+|\Delta r \, {\bf k}|^2 \right)^2}
\label{eq:Morse-k}\;,
\end{eqnarray}
where we have identified a form factor $F_{\bf k}$. The positive part of $F_{\bf k}$ stems from the atom--atom repulsion whereas its negative part describes the atom--atom attraction. To have a simple scalability, we choose $r_0 =3 \Delta r \, {\rm ln}\, 2$. 
For this choice, the form factor becomes $-\frac{1}{2}$ at ${\bf k}=0$, making $V_{{\bf k}=0}$ negative. Consequently, $V_{\bf k}$ is dominated by the atom--atom attraction at small momenta. The form factor is then maximized at momentum $k_{\rm peak} =\frac{\sqrt{2}}{\Delta{}r}$ producing $\frac{1}{9}$ at the peak. Therefore, $V_{\bf k}$ is dominated by the atom--atom repulsion at large momenta. 

The Schr\"{o}dinger equation \eqref{eq:H-Wannier-low} with the Morse-potential \eqref{eq:Morse-r} can be solved analytically\cite{Morse:1929}, yielding
\begin{eqnarray}
  \phi_n(r) = {\cal A}_n \, L^{2\alpha_n}_{n-2\alpha_n}\left( 2\lambda e^{-\,\frac{r-r_0}{\Delta r}} \right) \,
  {\rm exp}\left(-\lambda\,e^{-\frac{r-r_0}{\Delta r}}\right)\,
  \frac{e^{-\alpha_n\,\frac{r-r_0}{\Delta r}}}{r} 
\label{eq:Morse-solution}\;,
\end{eqnarray}
where ${\cal A}_n$ is the normalization constant, $L^\alpha_n(x)$ is the Laguerre polynomial\cite{Arfken:1985}, and the quantum number $n$ is an integer from zero to ${\rm Trunc}\left[\lambda\right]$. In this context, $\lambda \equiv \sqrt{\frac{D}{E_M}}$, $E_M = \frac{\hbar^2}{m \Delta r^2}$, and $\alpha_n \equiv \lambda - n - \frac{1}{2}$ are constants and ${\rm Trunc}\left[\lambda\right]$ creates an integer number by truncating $\lambda$ to its integer part. Whenever $\alpha_n$ is negative for all $n$ values, the corresponding solution is unbound. 
Therefore, the system has only unbound solutions when $\lambda$ is smaller than $\frac{1}{2}$. Otherwise, the eigenenergy of a bound state is $E_n = -E_M \, \alpha^2_n$. We see that $r \, \phi_n(r)$ approaches asymptotically $e^{-\alpha_n\,\frac{r}{\Delta r}}$ at $r\rightarrow 0$; when we insert this limit into definition \eqref{eq:a_scatt}, we obtain
\begin{eqnarray}
 a_{\rm scatt,n} = \frac{\Delta r}{\alpha_n} = \frac{\Delta r}{\sqrt{\frac{D}{E_M}} - n - \frac{1}{2}}
\label{eq:a_Morse}\;,
\end{eqnarray}
as the scattering length associated with dimer state $n$. When the dimer state becomes unbound, its energy approaches zero that also implies $\alpha_n \rightarrow 0$. We observe that the scattering length correctly diverges at this limit. Interestingly, the scattering-length relation \eqref{eq:a_Morse} can be applied to all ranges of $D$. When $D$ becomes too small to produce a positive $\alpha_n$, the scattering length becomes negative.

In our computations, we use $m=84.9118\,u$ and set $r_0$ to be equal to the covalent bond length of $^{85}$Rb, i.e.~$r_0=220$\,pm=$4.16\,a_0$ where $a_0=0.529\,{\rm \AA}$ is the hydrogen Bohr radius. The range of the potential is chosen to be $\Delta r = \frac{r_0}{{\rm ln}\ 8} = 106$\,pm=$2.00\,a_0$, based on the discussion above. 
This $(r_0,\, \Delta{}r)$ pair is just one possible example because these parameters depend on the specific atom--atom interaction orbitals studied. Therefore, they should be adjusted based on the experimental specifications to provide the most accurate description. Here, we do not analyze any specific experiment, but want to illustrate the general many-body quantum kinetics of the HBEs.

Since the system is homogeneous, the doublets depend only on the magnitude $|{\bf k}|=k$ of the wave vector, which reduces the effective dimensionality of the computations. By using $(r_0=220\,{\rm pm},\, \Delta{}r= 106\,{\rm pm})$ in Eq.~\eqref{eq:Morse-k}, we find that the repulsive peak of $V_{\bf k}$ is found at the atom energy $E_{\rm peak} = \frac{\hbar^2 k^2_{\rm peak}}{2m}=\frac{\hbar^2}{m \Delta{}r^2}=43.7\,\mu$eV that corresponds to 10.6\,GHz frequency or 0.507\,K. 
Since Bose-Einstein condensation involves atoms cooled down to roughly nK temperatures\cite{Anderson:1995,Bradley:1995,Davis:1995}, the energy and momentum of ultracold atoms is typically orders of magnitude smaller than $E_{\rm peak}$ and $\hbar k_{\rm peak}$, respectively. Still, the $k$ values close to $k_{\rm peak}$ are important for the interactions because they include the atom--atom repulsion at the small distances. Therefore, the numerical implementation of the HBEs is challenging because the $k$ discretization must include wave 
vectors over several orders of magnitude. Only then, the HBEs precisely include the details of strongly interacting Bose gas, including both the ultracold gas as well as the repulsive atom--atom contributions. 
To realize this, we use a non-equidistant $k$ grid which has many points at low $k$ and a sparse grid at very large $k$. The same precision is much more difficult to realize when one describes atom--atom scattering in real space because the long length-scale dependence appears as the envelope to the fast oscillating short length-scale dependence. The $k$ representation nicely separates these scales to the opposite ends of the $k$ grid.

We focus here on a case where $V(r)$ hosts one strongly bound dimer state ($n=0$) and one state ($n=1$) whose energy is close to zero. As atoms approach each other from afar, the scattering length is determined exclusively by the $n=1$ state, as discussed above. Therefore, the dimer scattering length $a_{\rm scatt,n}$ is computed by inserting $n=1$ into Eq.~\eqref{eq:a_Morse}. More specifically, we choose $D_{\rm weak}=99.971\,\mu$eV that corresponds to 24.173\,GHz and $D_{\rm uni} =98.311\,\mu$eV (23.771\,GHz). 
These differ only by 1.7\% and yet they produce very different dimer scattering lengths: The choice $D_{\rm weak}$ yields $a_{\rm weak} \equiv a_{\rm scatt,1} = +301\,a_0$ while $D_{\rm uni}$ results in a diverging $a_{\rm uni} = \infty$. In other words, $D_{\rm weak}$ interaction corresponds to a relatively strong dimer binding, which implies weak atom--atom interactions, whereas $D_{\rm uni}$ brings the system to the unitarity, which implies strong atom--atom interactions. 

In experiments, the specific value of $a_{\rm scatt}$ can be abruptly changed by sweeping the system through a Feshbach resonance with a magnetic field. We can model this by ramping $D$ from $D_{\rm weak}$ to $D_{\rm uni}$; since the change in $D$ is very small, we indeed change only how the atoms react to each other, in full analogy to corresponding experiments. 
By sweeping $D$ from $D_{\rm weak}$ to $D_{\rm uni}$ in time, we can study the quantum kinetics that is excited to the system when it it brought from a weakly to strongly interacting Bose gas.

\subsection{Density-dependence of dimer states}
\label{sec:Wannier-state}

To illustrate the basic properties of dimer states, we solve the hyperbolic Wannier equation \eqref{eq:H-Wannier} numerically with and without densities. We discretize the momentum grid into 800 intervals to describe the HBE dynamics accurately. As discussed in Sec.~\ref{sec:scattering length}, the grid must cover a wide range of momenta in order to simultaneously describe the ultracold atom and interaction effects. 
We split the grid into two parts: the interval up to wave vector 151.2\,$/\mu{}m$ contains 400 equidistantly positioned $k$ points, constituting the low-$k$ grid. The grid is continued into high momentum with an equidistant grid that has 400 times larger momentum spacing. More specifically, the high-$k$ part of the grid contains another 400 points, reaching momentum $6.06\times10^4/{\rm \mu{}m}$ at its last grid point. 
This way, the total grid covers wave vectors that span over five orders of magnitude. We will show below that this grid accurately describes the physics of the interacting Bose gas.

\begin{figure}[t]
\includegraphics*[scale=0.385]{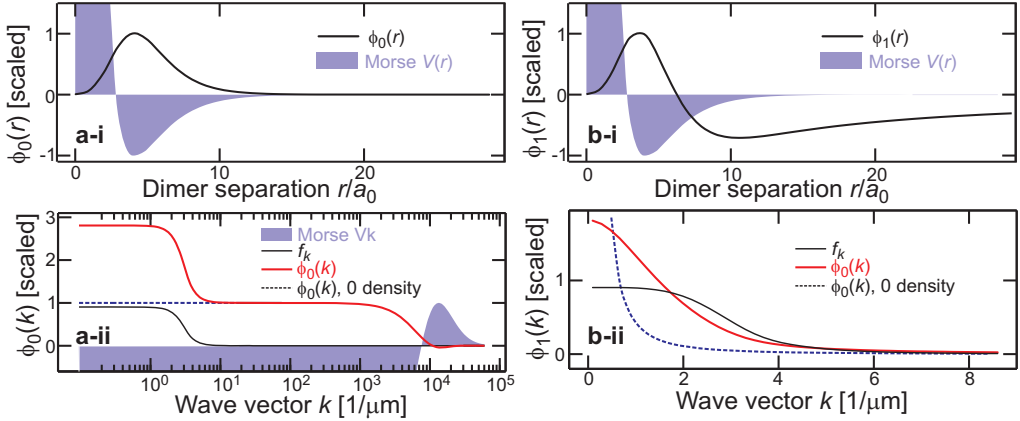}
\caption{Eigensolutions of hyperbolic Wannier equation.
Dimer wave function $\phi^R_n(r)$ (line) for {\bf a-i} ground state $n=0$ and {\bf b-i} $n=1$ as function of atom--atom separation $r$. The shaded area denotes the atom--atom interaction potential $V(r)$ (scaled).
{\bf a-ii} Full momentum dependence of $\phi^R_{n=0}(k)$ for the ground state. Result with (red line) and without (black dashed line) atom density $n_{\bf k}$ (black line) are shown together with $V_k$ (shaded area).
{\bf b-ii} The corresponding $\phi^R_{n=1}(k)$ for the excited state closest to the continuum; line styles as in {\bf a-ii}. The hyperbolic Wannier equation \eqref{eq:H-Wannier} is solved numerically for the grid and material parameters determined in the text.
}\label{HBEwannier}
\end{figure}

In our computations, we use amplitude $D_{\rm uni}$ (23.771\,GHz) for the Morse potential \eqref{eq:Morse-k}; the other system parameters are given at the end of Sec.~\ref{sec:scattering length}. The shaded area in Figs.~\ref{HBEwannier}{\bf a}-i and \ref{HBEwannier}{\bf b}-i shows the $r$ dependence of the Morse potential; as explained in Sec.~\ref{sec:scattering length}, $V(r)$ dips to its minimum value at $r=r_0=4.16\,a_0$ and it becomes large and positive as $r$ approaches zero. 
The repulsive peak (out of scale) at origin is 48.7 times higher than the attractive dip, which shows that the chosen $V(r)$ produces a significant atom--atom repulsion at short distances while attraction appears at the intermediate atom separation. Therefore, this potential describes typical atom--atom interactions very well, based on the qualitative criteria indicated in Fig.~\ref{Vpotentials}.

We use the $V_{\bf k}$ of this potential in the hyperbolic Wannier equation \eqref{eq:H-Wannier} and first solve the corresponding eigenfunctions numerically at vanishing atom density, i.e.~$f_{\bf k}=0$. Figures \ref{HBEwannier}{\bf a}-i and \ref{HBEwannier}{\bf a}-ii show the computed ground state $\phi_0(r)$ and first excited state  $\phi_1(r)$, respectively, as black line (peak normalized to one).
We observe that both solutions peak around the attractive dip of the potential (shaded area) and that they decay toward zero at the region of the repulsive interaction. However, these solutions behave very differently at the asymptotic region, roughly $r > 10 a_0$. More specifically, the ground state decays exponentially whereas $\phi_1(r)$ extends very far in the asymptotic region, implying unitarity interactions. There are no more bound states, such that the $n=1$ dimer defines the properties of the interacting Bose gas, as discussed in Sec.~\ref{sec:scattering length}.

The interaction strength $D_{\rm uni}$ (23.771\,GHz) analytically produces the energy $E_0=-10.57$\,GHz and the scattering length $a_{\rm scatt,\,0} = 2.00\,a_0$ for the ground state, as well as $E_1=0$ and $a_{\rm scatt,\,1} = \infty$ for the excited state. To check the accuracy of our numerical implementation, we compare these numbers with numerically determined $E_n$ and $a_{\rm scatt, n}$; 
the scattering length is computed by straightforward discretizing differentiation \eqref{eq:a_scatt} evaluated around $r=20\,a_0$ that is clearly within the asymptotic region. The computed energy and scattering length are $E_0=-10.57$\,GHz and $a_{\rm scatt,\,0} = 2.02\,a_0$ for the ground state and $E_1=-1.26\times10^{-4}$\,GHz and $a_{\rm scatt,\,1} = 1.53\times10^4\,a_0$ for the excited state, respectively. The excited state has nearly a vanishing energy and the computed $a_{\rm scatt}$ is extremely large, as expected at the unitarity.
Naturally numerical differentiation \eqref{eq:a_scatt} can only produce a very large $a_{\rm scatt}$, not infinite, even at the unitarity. It has been thoroughly tested that adding more $k$ points improves the accuracy of the $E_n$ and $a_{\rm scatt,\,n}$ toward the analytic limit whereas the HBE excitations remain unaltered, which validates the accuracy of our numerical approach. 

To analyze the effect of atomic densities on dimer states, we solve the hyperbolic Wannier equation \eqref{eq:H-Wannier} with atom density $f_{\bf k}$ shown as black-solid line in Figs.~\ref{HBEwannier}{\bf b}-i and \ref{HBEwannier}{\bf b}-ii. Here, $f_{\bf k}$ is the final-time result of an actual HBE computation, analyzed in Fig.~\ref{HBEdynamics}. 
Figure \ref{HBEwannier}{\bf b}-ii shows $\phi_0(k)$ with (red solid line) and without (dashed line) $f_{\bf k}$ densities; the shaded area indicates the corresponding $V_{\bf k}$. Both $\phi_0({\bf k})$ and $V_{\bf k}$
are scaled to peak to one at the asymptotic region. Since the interaction and ultracold effects span over five orders of magnitude in momentum, we have use a logarithmic scale for the $k$. 

We observe that $\phi_0({\bf k})$ becomes terminated close to the repulsive $V_{\bf k}$ peak, and it has the same asymptotic form with and without atom density. At the same time, the low-$k$ distributions are clearly modified by the presence of atom density; the $\phi_0({\bf k})$, with atom density, has an additional bump that is synchronized with the region where $f_{\bf k}$ is appreciable. 
However, these low-momenta changes do essentilly not affect the real-space $\phi_0({\bf r})$, which characterizes many of the interaction properties of the dimers. In this specific case, the $\phi_0({\bf r})$ appears to be essentially unchanged for the length scales shown in Fig.~\ref{HBEwannier}{\bf a}; indeed, the computed scattering length $a_{\rm scatt,0}$ changes by a very small factor of $2.0\times10^{-6}$ and the energy changes by an even smaller factor of $3.5\times10^{-9}$. We conclude that the atom densities do not affect the strongly bound dimer states much.

To determine how the atom density affects the weakly bound dimer states, the same analysis is repeated for the excited state in Fig.~\ref{HBEwannier}{\bf b}-ii that shows $f_{\bf k}$ (black-solid line) as well as $\phi_1({\bf r})$ solved with (red solid line) and without (dashed line) the $f_{\bf k}$ densities in the hyperbolic Wannier equation \eqref{eq:H-Wannier}. Without atom density, $\phi_1({\bf k})$ diverges toward small $k$ values, which is expected for a state that is becoming unbound. 
As atom densities are included, this divergence is removed and $\phi_1({\bf k})$ assumes a form expected for a bound state. In particular, the atom-density induced changes do not follow the shape of the $f_{\bf k}$, unlike for the deeply bound dimer states. The increased binding results from a normal-component induced enhancement of the effective atom--atom attraction by the bosonic factor $(1+2\,f_{\bf k})$. 
In this specific case, the $n=1$ dimer binding is $528\,$Hz deeper with than without the normal-component atoms. At the same time, the scattering length $a_{\rm scatt,1}$ is decreased only by $0.3\%$. Even though this change is small, we conclude that the weakly bound dimer state may experience detectable normal-component induced effects in a strongly interacting Bose gas.  

\subsection{Quantum kinetics of interacting Bose gas at unitarity}
\label{sec:unitarity}

To study pure quantum depletion, we assume that thermodynamic depletion is negligible. In other words, the system temperature is assumed to be so low that the equilibrium normal-component distribution $n^{\rm eq}_{\bf k}$ vanishes in Eq.~\eqref{eq:HBE-FIN}. In this situation, the normal component is exclusively created by the atom--atom interactions, i.e.~the quantum depletion instead of the thermal depletion. Next, we will compute how the quantum depletion proceeds when the atom interaction is switched from weak to strong interaction regime. 
In practice, we ramp the amplitude of the potential from $D_{\rm weak}$ to $D_{\rm uni}$ linearly in $5\,\mu$s; the start of the ramp assigns time zero in our computation. The BEC dynamics typically evolves on time scales longer than 100\,$\mu$s \cite{Makotyn:2014} such that the used switch can be considered abrupt. To have strong atom--atom interactions, we use 3\,$/\mu{}m^3$ atom density. 
For a spherical atom trap that has a 11.7\,$\mu{}m$ radius, this density yields ${\cal N}=2\times 10^4$ atoms in total, which has already been reached in several experiments\cite{Dalfovo:1999,Andersen:2004,Leggett:2001,Ketterle:2002,Cornell:2002,Proukakis:2008}. 

To reveal the quantum dynamics resulting from the principal structure of the HBEs, we replace the triplets and quadruplets with a constant $\gamma_{\bf k}$ and apply the minimum-uncertainty connection \eqref{eq:Gamma-gamma}. In practice, we numerically solve the HBEs \eqref{eq:HBE-FIN} with a dephasing constant ${\rm Re}[\gamma_{\bf k}]$ and a constant energy shift ${\rm Im}[\gamma_{\bf k}]$ that corresponds to frequencies 761\,Hz and 475\,Hz, or 209\,$\mu$s and 335\,$\mu$s temporal duration, 
matching roughly with the relaxation times observed in a recent switch-on experiment\cite{Makotyn:2014}. Other parameters of the system are given at the end of Sec.~\ref{sec:scattering length}.

\begin{figure}[t]
\includegraphics*[scale=0.385]{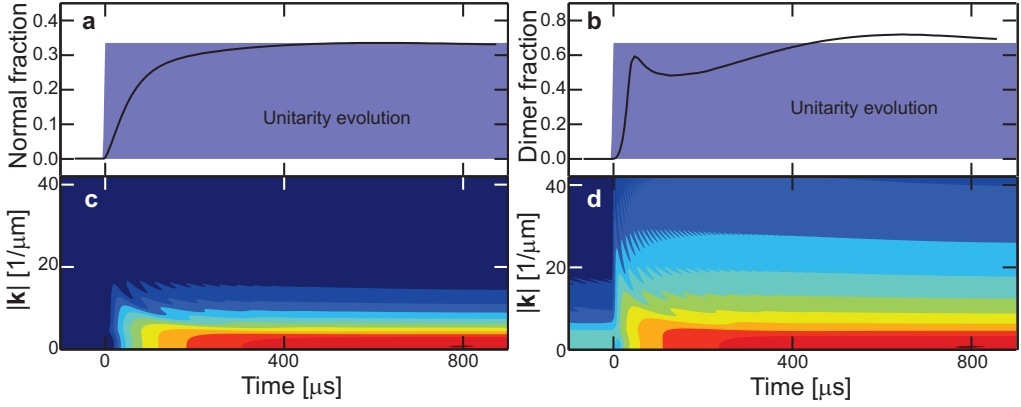}
\caption{Fast-switching from weakly to strongly  interacting Bose gas based on the HBEs. 
Time evolution of {\bf a} normal-component fraction (solid line) and {\bf b} dimer-state occupation $F^{\rm dim}_{n=1}$ (black line) are shown; The scattering length is changed at $t=0$ within $5\,\mu$s from $a_{\rm scatt,1}=301\,a_0$ to unitarity, and the shaded area indicates the unitary evolution.
Contour plot of corresponding {\bf c} $f_{\bf k}$ and {\bf d} $|s_{\bf k}|$ distributions as function of time. The contour lines correspond to are shown at $2^{-n}$ from the peak value ($n=1,2,\cdots$). The numerical parameters are the same as in Figs.~\ref{HBEwannier}.
}\label{HBEdynamics}
\end{figure}

The solid line in Fig.~\ref{HBEdynamics}{\bf a} shows how the fraction $\nuN \equiv \sum_{\bf k} f_{\bf k} /{\cal N}$ of the normal-component atoms evolves in time. In sheer numbers, the normal-component fraction approaches 32.9\%, 
which shows that {\it even a strongly interacting Bose gas can have an appreciable number of BEC atoms even at unitarity}, in agreement with recent experiments\cite{Makotyn:2014}. 
We have also projected the dimer occupation coefficient $s_{n=1}$ using Eq.~\eqref{eq:dimer-basis} for the excited state that is closet to zero. We have then computed a relative dimer-state occupation
\begin{eqnarray}
  F^{\rm dim}_{n=1} \equiv \frac{|s_{n=1}(t)|^2}{\sum_{n=0}^\infty |s_n(t=\infty)|^2}
\label{eq:dimer-occupation}\;,
\end{eqnarray}
of the dominant dimer state, $n=1$, from the squeezing amplitude as function of time. The result is normalized to the total dimer occupations at the final time, denoted here by $\infty$. The computed $F^{\rm dim}_{n=1}$ is shown in Fig.~\ref{HBEdynamics}{\bf b} as solid line.

At the regime of weakly interacting Bose gas ($t<0$), all atoms essentially occupy the BEC because $\nuN$ and $F^{\rm dim}_{n=1}$ remain extremely small. After the switch-on to strong interactions ($t>0$), both $\nuN$ and $F^{\rm dim}_{n=1}$ start to increase rapidly. The squeezing first peaks around 50\,$\mu$s, much faster than the density does, 
which indicates that the BEC first drives the squeezing amplitude, resembling the behavior in semiconductors where the optical light field first creates polarization followed by the creation of densities\cite{Kira:2006}. We note that the density approaches a constant level roughly within 200\,$\mu$s that matches well the dephasing time. 

\begin{figure}[t]
\includegraphics*[scale=0.55,angle=0]{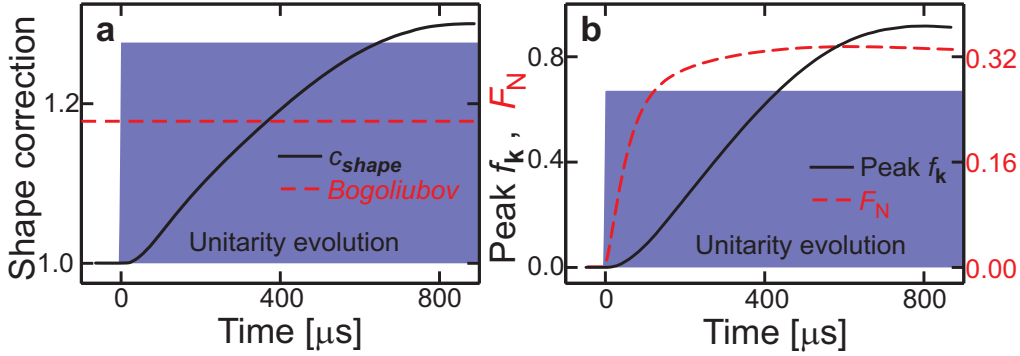}
\caption{Shape of quantum depletion $c_{\rm shape}$ based on the HBEs. {\bf a} Time evolution of $c_{\rm shape}$ (solid line) is compared with $c_{\rm shape}$ of the Bogoliubov excitations (dashed line); The shaded area indicates the evolution at unitarity after a fast switch of Fig.~\ref{HBEdynamics}. {\bf b} Peak value of $f_{\bf k}$ (solid line) is compared with the normal-component fraction $\nuN$ (dashed line). Constructed from the data shown in Fig.~\ref{HBEdynamics}.
}\label{HBEcshape}
\end{figure}

However, the squeezing amplitude shows a more complicated dynamics than density because the fast switch-on of $F^{\rm dim}_{n=1}$ is followed by a slower relaxation that extends up to 800\,$\mu$s. We also see that the $F^{\rm dim}_{n=1}$ has a nonmonotonic evolution suggesting that the squeezing amplitude exhibits a nontrivial quantum kinetics, which is studied in Fig.~\ref{HBEcshape}. 
According to Fig.~\ref{HBEdynamics}{\bf b}, $F^{\rm dim}_{n=1}$ approaches 71.9\%, which shows that dimer state $n=1$ indeed dominates the quantum depletion. The dimer basis \eqref{eq:dimer-basis} indicates that $s_{\bf k} = \sum_{n=0} s_n \, \phi^R_n({\bf k})$ can be expanded in terms of multiple dimer states. All of them have a different energy $E_n$ such that each $s_n$ evolves with their own frequency, $e^{-\frac{i}{\hbar} E_n t}$. 
As they are added up in superposition in $s_{\bf k}$, one expects to see a temporal interference pattern in $s_{\bf k}$, which also explains why $F^{\rm dim}_{n=1}$ has more complicated dynamics than $\nuN$.

To show the full evolution of doublets, Figs.~\ref{HBEdynamics}{\bf c} and \ref{HBEdynamics}{\bf d} present contour plots of $f_{\bf k}$ and $|s_{\bf k}|$ distributions as function of time, respectively. The contours are shown at 14 values corresponding to $2^{-1}$, $2^{-4}$, $\cdots 2^{-14}$ of the peak height. We notice that the tails of $s_{\bf k}$ and $f_{\bf k}$ first spread toward a large $k$ and then focus back on the low-momentum range. 
The temporal dynamics also exhibits a highly nontrivial interference pattern that appears up to 200\,$\mu$s, matching well with the dephasing time used. This confirms that the strongly interacting Bose gas can indeed produce a dynamically changing interference pattern of dimer components. In general, the squeezing amplitude contains stronger and broader oscillations than the atom occupation does.
After this quantum-kinetic transient, the system approaches quasi-equilibrium -- determined by interactions, not thermodynamics --- in roughly 400\,$\mu$s. 

Even though the squeezing amplitude and the normal-component fraction are built up fast by the quantum depletion, they both exhibit clear transients before relaxing toward their steady-state forms. This suggests that the shape of the quantum depletion is modified much longer than the creation of the normal component lasts. The shape of the quantum depletion can be characterized by a single number $c_{\rm shape}$, determining the extent of BEC number fluctuations according to Eq.~\eqref{eq:HEB-BEC}. 
The solid line in Fig.~\ref{HBEcshape}{\bf a} shows how $c_{\rm shape}$ evolves as function of time for the same computation presented in Fig.~\ref{HBEdynamics}. It starts from value $c_{\rm shape}=1$ at the weak-interaction regime ($t<0$) and grows monotonically toward 1.3037 in roughly $800\,\mu$s after switching to the regime of the strong interactions ($t>0$, shaded area). As a comparison, Bogoliubov excitations have\cite{Kira:Xpic} $c_{\rm shape}=\frac{3\pi}{8}\approx1.1781$ (dashed line), which is very different from the actual $c_{\rm shape}$. 
This result and Eq.~\eqref{eq:HEB-BEC} suggest that {\it measuring BEC number fluctuations $\Delta\NC$ can sensitively monitor the shape changes in the normal component}, and resolve when the shape of the quantum depletion deviates from the Bogoliubov excitation. These differences are studied in more detail in Sec.~\ref{sec:Bogoliubov}

According to Figs.~\ref{HBEdynamics}{\bf c}--{\bf d}, the reshaping of quantum depletion makes both $f_{\bf k}$ and $s_{\bf k}$ sharper. To compare this trend with the $c_{\rm shape}$ dynamics, Fig.~\ref{HBEcshape}{\bf b} presents the peak occupation, $f_0 = \lim_{{\bf k} \rightarrow 0} f_{\bf k}$ (red line, scale left), together with the normal-component fraction (dashed line, scale right). 
The shaded area denotes the time range of the strong interactions. We observe that $f_0$ keeps on growing long after the normal-component fraction levels toward a steady state. In fact, we find that both $c_{\rm shape}$ and $f_0$ level toward a steady state roughly in 800\,$\mu$s, which matches perfectly with the relaxation of the squeezing $F^{\rm dim}_{n=1}$ analyzed in Fig.~\ref{HBEdynamics}{\bf a}. Consequently, the {\it shape changes of quantum depletion are connected with a growing peak in $f_{\bf k}$ as well as the redistribution of dimer components within the squeezing amplitude.}

\subsection{Bogoliubov vs.~HBE excitations}
\label{sec:Bogoliubov}

\begin{figure}[t]
\includegraphics*[scale=0.37,angle=0]{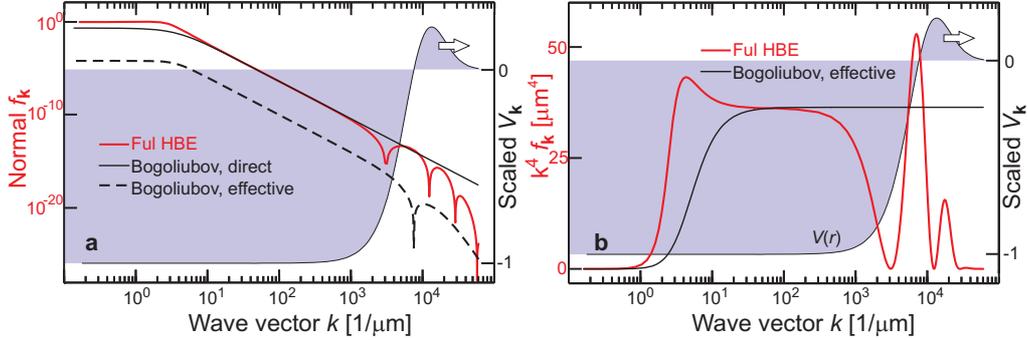}
\caption{Bogoliubov excitations vs.~steady-state atom distribution. 
{\bf a} The final $f_{\bf k}$ of Fig.~\ref{HBEdynamics}{\bf c} (red line, scale left) is compared with Bogoliubov distributions evaluated with the actual $V_{\bf k} \NC$ (dashed line) and effective $a_{\rm scatt}=2606\,a_0$ (black line). The shaded area indicates the potential $V_{\bf k}$, scale on the right.
{\bf b} Analysis of so-called Tan tails\cite{Tan:2008b}. The computed ${\bf k}^4 f_{\bf k}$ for the full HBE computation (red line, left scale) vs.~the effective Bogoliubov result (black line, left scale) and $V_{\bf k}$ (shaded area, right scale). Constructed from the final-time data of Fig.~\ref{HBEdynamics}.
}\label{HBEsnap}
\end{figure}

To study the equilibrium of quantum depletion in more detail, Fig.~\ref{HBEsnap}{\bf a} presents the normal-component distribution $f_{\bf k}$ reached at the final time (1\,ms) of the HBE dynamics shown in Fig.~\ref{HBEdynamics}. More specifically, red line shows $f_{\bf k}$ on a double logarithmic scale (scale on the left axis) and the shaded area shows the potential $V_{\bf k}$ on a semilogarithmic scale (scale on the right axis). 
We observe that $f_{\bf k}$ saturates at small $k$ whereas it decays as $\frac{1}{|{\bf k}|^4}$ at the intermediate asymptotic region between 10 and 1000$/\mu$m. In the vicinity of the repulsive $V_{\bf k}$ peak, $f_{\bf k}$ develops an interference pattern because the dimer wave function is ``reflected'' from the repulsive atom--atom interaction ``barrier''. The $f_{\bf k}$ is terminated rapidly after the potential peak.

The asymptotic $f_{\bf k} \propto \frac{1}{|{\bf k}|^4}$ behavior has been identified as a universal behavior determining many central thermodynamic relations, introduced by Tan \cite{Tan:2008a,Tan:2008b}. In Tan's derivations, one must carefully study how seemingly diverging sums, such as $\sum_{\bf k} E_{\bf k} \, f_{\bf k}$, actually converge\cite{Tan:2008a} when the class of extended functions is used together with the contact-potential description of the atom--atom interactions. 
The more realistic potential provides this universal $\frac{1}{|{\bf k}|^4}$ proportionality, but only at the intermediate asymptotic region. For exceedingly large momentum ranges beyond the repulsive atom--atom repulsion peak, $f_{\bf k}$ decays much faster than the $\frac{1}{|{\bf k}|^4}$ dependence, which guarantees that all physically relevant sums converge without the need for a further study. This fast decay appears at $k$ greater than $10^4/\mu$m corresponding to the wave length $\frac{2\pi}{k} \approx 6$\AA, i.e.~the scale of the atom size, 
which verifies that the fast decay results from the atom--atom repulsion. In the analyzed computation, the distribution becomes terminated when it already has decayed 12 orders of magnitude from its peak value. Therefore, $f_{\bf k}$ becomes extremely small, below $10^{-8}$, for $k$ greater than $10^4/\mu$m. Such atom occupation numbers seem much smaller than near-future experiments can resolve, which makes the actual detection of abrupt $f_{\bf k}$ termination challenging.

To present the $\frac{1}{|{\bf k}|^4}$ tail of the distribution directly, Fig.~\ref{HBEsnap}{\bf b} shows ${\bf k}^4 f_{\bf k}$ (red line, scale on the left axis) together with $V_{\bf k}$ (shaded area, scale on the right axis), both on a semilogarithmic scale. We observe that ${\bf k}^4 f_{\bf k}$ is essentially constant for ${\bf k}$ values from $7\,/\mu$m to $500\,/\mu$m, which shows that $f_{\bf k}$ indeed contains a $\frac{1}{|{\bf k}|^4}$ tail in the intermediate asymptotic region. 
This presentation also shows clearly how an pattern interference develops before the repulsive $V_{\bf k}$ peak. As $f_{\bf k}$ extends beyond the repulsive peak, it terminated much faster than the $\frac{1}{|{\bf k}|^4}$, confirming the conclusions made from  Fig.~\ref{HBEsnap}{\bf a}.

Reference \cite{Kira:Xpic} shows that the excitation picture includes both Bogoliubov excitations\cite{StamperKurn:1999,Jin:1996,Mewes:1996,Utsunomiya:2008} as well as interactions between them. Also the HBEs \eqref{eq:HBE-FIN} can be limited to pure, noninteracting,  
Bogoliubov excitations by eliminating the Coulomb coupling among {\it different} $\left(s_{\bf k},\, f_{\bf k}+\frac{1}{2}\right)$ pairs on ${\bf k}$-labeled hyperbolas. 
With this constraint, the renormalized quantities \eqref{eq:E-ren-transition}--\eqref{eq:ren-source} reduce to the Bogoliubov limit
\begin{eqnarray}
  E^{\rm ren}_{\bf k} \rightarrow \frac{\hbar^2 {\bf k}^2}{2m} + \NC\, V_{\bf k} \equiv E^V_{\bf k}\,,
  \qquad
  \Delta^{\rm ren}_{\bf k} \rightarrow \NC\, V_{\bf k}
\label{eq:Bogo_lim}\;.
\end{eqnarray}
In this situation, the HBEs \eqref{eq:HBE-FIN} have such a simple form that their steady state can be solved analytically, as show in   \ref{app:HBE-Bogo}. More specifically, we find a steady-state solution of the Bogoliubov excitation:
\begin{eqnarray}
  s^{\rm BG}_{\bf k} 
  =
  - \frac{1}{2} \frac{E^V_{\bf k} + i \gamma_{\bf k}}{|E^V_{\bf k} + i \gamma_{\bf k}|} 
    \frac{\NC V_{\bf k}}{E^{\rm BG}_{\bf k}}\,,
\qquad
  n^{\rm BG}_{\bf k} 
  = 
  \frac{1}{2} 
  \left(
    \frac{|E^V_{\bf k} + i \gamma_{\bf k}|}{E^{\rm BG}_{\bf k}}
    -1
  \right)
\label{eq:s_n_Bogoliubov}\;,
\end{eqnarray}
where $\gamma_{\bf k}$ is assumed to be real valued. We also have identified the energy of the Bogoliubov excitations:
\begin{eqnarray}
  E^{\rm BG}_{\bf k} \equiv \sqrt{ \left(\frac{\hbar^2 {\bf k}^2}{2m}\right)^2 +  \NC V_{\bf k} \, \frac{\hbar^2 {\bf k}^2}{m} + \gamma^2_{\bf k}}
\label{eq:E_Bogoliubov}\;.
\end{eqnarray}
Since $\NC\, V_{\bf k}$ approaches a constant value at small ${\bf k}$, the Bogoliubov excitations have a linear dispersion at small ${\bf k}$ while the dispersion becomes quadratic at large ${\bf k}$; this dispersion behavior has been measured in atomic BECs\cite{StamperKurn:1999} and with semiconductor microcavity polaritons\cite{Utsunomiya:2008}.

Results \eqref{eq:s_n_Bogoliubov}--\eqref{eq:E_Bogoliubov} show that the principal structure of the HBEs fully contains the Bogoliubov excitations as a subset of general normal-component excitations. More specifically, the Bogoliubov limit \eqref{eq:Bogo_lim} removes the interaction coupling between different-momenta $f_{\bf k}$ and $s_{\bf k}$, yielding noninteracting Bogoliubov excitations. 
Since the interaction renormalization to $E^{\rm ren}_{\bf k}$ and $\Delta_{\bf k}$ couple {\it all} $(f_{\bf k}, s_{\bf k})$, they introduce a coupling between all possible Bogoliubov excitations. Therefore, already the principal structure of the HBEs already describes physics beyond the noninteracting Bogoliubov excitations.

The identified $s^{\rm BG}_{\bf k}$ and $n^{\rm BG}_{\bf k}$ have exactly the form of Bogoliubov excitations when dephasing is taken to be zero, cf.~Eq.~{\color{red} (109)} in Ref.~\cite{Kira:Xpic}, which verifies that the HBEs indeed contain Bogoliubov excitations when the relevant interactions are switched off. 
The dashed line in Fig.~\ref{HBEsnap}{\bf a} shows the numerical evaluation of $n^{\rm BG}_{\bf k}$ for the present problem. We observe that $n^{\rm BG}_{\bf k}$ is roughly four orders of magnitude too small. Therefore, we conclude that the interactions among Bogoliubov excitations strongly enhance the quantum depletion at the unitarity.

However, we may still study whether the strongly interacting Bose gas can be described via regularized Bogoliubov excitations, motivated by the careful study in Ref.~\cite{Kokkelmans:2002}. In order to regularize the distributions, we replace the interaction by a contact-potential interaction $V({\bf r}) = 4 \pi \frac{\hbar^2}{m} a_{\rm scatt} \delta({\bf r})$\cite{Dalfovo:1999,Leggett:2001,Giorgini:2008,Bloch:2008}, where $a_{\rm scatt}$ is an effective scattering length, not the actual one. With this choice, Fourier transform \eqref{eq:V-Fourier} yields a constant interaction contribution
\begin{eqnarray}
  \NC V_{\bf k} = 4 \pi \frac{\NC}{{\cal L}^3} \frac{\hbar^2}{m}\, a_{\rm scatt}
\label{eq:V_NC_eff}\;.
\end{eqnarray}
We adjust the scattering length such that the effective interaction produces $n^{\rm BG}_{\bf k}$ whose asymptotic tail coincides with the actual $f_{\bf k}$ tail. The result of $a_{\rm scatt} = 3492\,a_0$ fit is shown in Figs.~\ref{HBEsnap}{\bf a}--{\bf b} as black solid line. Even though we are analysing quantum depletion at unitarity, we find a finite $a_{\rm scatt}$ for the contact-potential interaction \eqref{eq:V_NC_eff}. 
A similar regularization of interactions has been identified in Refs.~\cite{Kokkelmans:2002,Sinatra:2002} to account for the short-range atom--atom repulsion omitted by the contact potential; see also discussion before Eq.~\eqref{eq:Morse-r}.

We observe that the regularized $n^{\rm BG}_{\bf k}$ indeed reproduces the tails of the actual $f_{\bf k}$ roughly until the repulsive peak of $V_{\bf k}$ (shaded area). However, the regularization is not accurate at small or very large values of ${\bf k}$. Therefore, the Bogoliubov excitations capture only some qualitative aspects of the strongly interacting Bose gas, even when suitable regularization is implemented. For example, the total normal-component fraction become 13.2\% instead of the actual value 32.9\%. 
The unregularized computation, with the actual $V_{\bf k}$ in Eq.~\eqref{eq:s_n_Bogoliubov} (see dashed line in Fig.~\ref{HBEsnap}{\bf a}), yield only $\nuN=5\times10^{-5}$, which largely underestimates the actual $\nuN=32.9\%$. Even though the regularized contact-potential computation produces much better results than the unregularized computation, it still is nearly 60\% too low for this specific example.

\section{Conclusions and outlook}
\label{sec:Conclusions}

I have shown that strongly interacting Bose gas can be systematically described by using the same cluster-expansion method\cite{Wyld:1963,Fricke:1996,Kira:2006,Kira:2008,Book:2011} that provides one of the most accurate descriptions of many-body and quantum-optical effects in semiconductors. Since the BEC itself is a highly correlated state, in terms of particle clusters, the problem must first be converted into the excitation picture that shifts the investigation around the BEC as illustrated in Fig.~\ref{Qdepletion}. 
Formally, the shifted system has a vacuum as the BEC whereas all aspects of BEC as well as many-body excitation are described by the normal-component clusters excited to the system. Most importantly, the normal component contains much lower level of atom clusters than the BEC does in original picture, which makes the application of the cluster-expansion method to interacting Bose gas as efficient as for semiconductors. 

In the excitation picture, the many-body Hamiltonian has the actual BEC number and number fluctuations as driving terms that convert the BEC into normal component via the atom--atom interactions. The resulting quantum depletion evolves sequentially because it can create new correlated atom clusters only from already existing lower-lever clusters. A completely analogous cluster-generation processes appears in semiconductors following an ultrafast excitation of charge carriers by a laser pulse. 
Due to the sequential build up, one only needs to include dynamics up to a certain level of clusters, before further clusters are created. Therefore, this approach has been successfully used in semiconductors to explore and explain a wealth of many-body and quantum-optical phenomena\cite{Kira:2006,Book:2011,Jahnke:2012} and the presented work provides an equivalent cluster-expansion framework for the strongly interacting Bose gas.

I also have introduced an implicit-notation formalism (INF) that delivers the explicit quantum kinetics of {\it all} atom clusters of strongly interacting Bose gas with one derivation. Especially, the INF identifies the sequential cluster-formation dynamics. In general, the clusters are created by nonlinear processes that compete with each other; also strong diffusion among clusters classes can be identified. On the simplest level, the nonlinear processes and diffusion yield dephasing, as they do in semicondctors. 
The cluster structure also yields energy renormalizations, screening of interactions, density dependent dephasing, and creation of bound clusters such as Efimov trimers and quadmers. Such multi-atom clusters have already been studied extensively in various interacting atom gases. I expect that study of excitation induced dephasing, energy shift, or dynamical build up of screening could be as intriguing as in semiconductors where a detailed studies of such effects has indeed revealed several novel aspects about the many-body quantum kinetics\cite{Huber:2001} and even about the many-body  configurations\cite{Smith:2010}.

As the first step of the cluster-expansion approach, I have thoroughly studied how simplest of clusters are created by the quantum depletion. This effort yields the hyperbolic Bloch equations (HBEs) that describe the quantum kinetic excitation of normal-component densities and squeezing amplitude. Due to momentum conservation, atom--atom interaction cannot convert a single BEC atom into normal state. Instead, the interaction converts two BEC atoms into two normal-component atoms with opposite momenta, as illustrated in Fig.~\ref{Qdepletion}. 
From the point of view of normal component in excitation picture, this process is fundamentally connected with squeezing of boson fields, therefore, the elementary quantum-depletion processes are initiated by creation of a squeezing amplitude. I have shown that this elementary process is described microscopically by the HBEs that are formally analogous to the semiconductor Bloch equations (SBEs). As the main difference between the HBEs and SBEs, the squeezing and atom occupation must reside above a hyperbola whereas the corresponding semiconductor excitations reside  within the Bloch's sphere. 

The homogeneous part of the HBE introduces the hyperbolic Wannier equation that generalizes the Schr\"{o}dinger equation of a single dimer to include the principal effect of the other atoms in the system. We find that atom densities influence mostly those dimer states that have energy closest to zero. 
As the atom density is increased, dimer states become more bound. This behavior is opposite than expected for excitons that are bound electron--hole pairs in semiconductors. For interacting boson gas, it could be interesting to study excitation-induced dimer binding as function of atom density; analogous excitation-induced dephasing and exciton ionization studied have indeed provided invaluable insights to many-body effects in semiconductors.\cite{Peyghambarian:1984,Wang:1993,Hu:1994,Jahnke:1996,Chow:1997,Manzke:1998,
Sieh:1999,Khitrova:1999,Shacklette:2002,Kira:2006,Smith:2010,Mootz:2014}

The HBEs are particularly useful when studying the quantum kinetics of excitation created by fast switch-on of quantum depletion. The presented computations demonstrate that the normal-component exhibit intriguing nontrivial transients on a 100\,$\mu$s time scale. This computation also shows that transition to unitarity does not destroy the BEC completely, in agreement with a recent experiment.\cite{Makotyn:2014} The results also indicate that the pure Bogoliubov excitations do not explain the result because the interactions among Bogoliubov excitation strongly modify them.

The presented HBE computation fully included only contributions up to doublets. The explicit quantum kinetics of Efimov trimers and quadmers are described by the quantum kinetics of triplets and quadruplets, respectively. I have worked out the quantum kinetic equations of all clusters such that the presented inviestigation serves also as a general starting point to analyze Efimov timers and quadmers; the explicit triplet dynamics is presented in \ref{app:triplets}. 
Once these are fully included to the computations, one can evaluate dephasing, screening, and trimer effects fully microscopically. In semiconductors, an equivalent level produces a fully quantitative analysis of absorption\cite{Smith:2010}, gain measurements\cite{Hofmann:2001}, and exciton-formation studies\cite{Kira:2001,Kira:2004,Kaindl:2003}. Therefore, it clearly is an intriguing future goal to solve also the triplets together with the HBEs. This analysis should reveal the quantitative relation of trimers and dimers as function of time, atom density, and the interaction strength.

In general, the triplet dynamics has very analogous structure than doublets modifying the SBEs. Therefore, the triplets share much of the physics of correlations identified already in semiconductors. We next briefly overview the principal insights of Ref.~\cite{Kira:2006} that can be directly be used to identify different many-body effects included by triplets; their explicit dynamics is given in   \ref{sec:Conclusions}. 
For example, the Boltzmann scattering source will drive the triplets and in its simplest form introduce microscopic scattering to the HBEs. If this triplet dynamics is solved dynamically, the dephasing can be computed microscopically; a similar level investigation in semiconductors describes measured optical absorption\cite{Khitrova:1999,Smith:2010} and gain\cite{Hofmann:2001} extremely accurately. It will be interesting to see how well similar level analysis will explain the dephasing rates of corresponding atom BEC experiments.

The the simplest interaction terms involving triplets contain $V_{\bf k}$ multiplied by atom occupation differences and a sum over bare triplet, cf.~the fifth line of Eq.~\eqref{eq:app_T12-dyn}. In semiconductors, analogous terms screen the Coulomb interaction in the SBEs. In other words, the screening of SBE singlets (HBE doublets) is generated by the hierarchical coupling to doublets (triplets). Therefore, the triplet contributions should screen the atom--atom interaction of that appear in the HBEs. 
Since the atom--atom interactions are very different from the Coulomb interaction --- they have a short range, they contain both attractive and repulsive part, and they can be externally adjusted --- it clearly will be interesting to study when and how the screening will be observable in strongly interacting Bose gas. An of course, how the presence of BEC changes the screening, which add another interesting possibility to modify screening.

The triplet dynamics also contains sum terms where the correlation itself is convolved with the interaction, in the same way as squeezing amplitude renormalizes the quantum-depletion source \eqref{eq:ren-source} to create bound dimer resonances to the HBE, as discussed in Secs.~\ref{sec:Wannier} and \ref{sec:Wannier-state}. Analogously, these homogeneous parts yield new bound states among three atoms. Especially, the homogeneous part of Eq.~\eqref{eq:app_T03-dyn} is equivalent to a three-atom Schr\"{o}dinger equation, 
at the low-density limit, with all pairwise interaction possibilities included. Therefore, the trimer dynamics microscopically defines the possibility to form bound Efimov trimers\cite{Efimov:1973}. Like for the hyperbolic Wannier equation \eqref{eq:H-Wannier}, atom densities modify this problem via occupation terms such as $(1+f_{\bf k} + f_{{\bf k}'})$. In this connection, it will be interesting to study how normal-component density alters Efimov physics, i.e.~the study of excitation-induced effects should be illuminating also here.

The hierarchical coupling of triplets to duadruplets and five-atom correlations further generates, energy renormalizations, dephasing, screening, as well as Efimov multimer contributions for the trimers. It is clear that these offer very broad range of many-body effects that need to be identified and characterized in the future. 
Once this is managed systematically, one can compare the results with other many-body systems, such as solids that contain analogous cocktail of many-body effects in different relative mixes and weights. Similarities and differences of various systems will most certainly provide us with new insights and perspectives to the unsolvable, i.e.~many-body quantum physics, in the spirit of the manybodypedia.

In summary, I have presented a comprehensive analysis of interacting Bose gas where I show how the related many-body physics can be connected with the prior work done in semiconductor optics and semiconductor quantum optics. Since the cluster-expansion approach has been very successful in explaining semiconductor physics, I expect it to be equally useful for investigating the strongly interacting Bose gas. Especially, the cluster-expansion formulation is tailor made for following nonequilibrium quantum kinetics such that it should provide new many-body insights to fast-switch and Bosenova experiments. 
In principle, one can also extend the method to include thermodynamic transitions. In this case, one should carefully include bath interaction and Boltzmann scattering toward thermodynamic equilibrium. It also will be interesting to study how the HBEs must be reformulated for a Bose gas with a position-dependent density profile. 
In particular, one would like to find how the internal degrees of freedom, described by the HBEs, are connected with the spatial coherence, described by some generalization of the Gross-Pitaevskii equation. All these extensions can be systematicaly built on the theoretical framework presented in this paper.

\section*{Acknowledgements}

I wish to thank Steven Cundiff for bringing to my attention the fact that experiments on BEC and quantum-degenerate Fermi gas seem to be converging toward the same set of many-body problems as the ultrafast spectroscopy on semiconductors. I am also indebted to Deborah Jin for explaining the experimentally relevant aspects in Fermi and Bose gases, and Chris Greene and Matthew Davis for insightful discussions about how many-body BEC theory is currently perceived.  I have enjoyed very much my cooperation with Stephan Koch concerning semiconductor many-body physics, over the years. I am also grateful to Renate Schmid for a very careful proof reading of THIS manuscript.

\appendix

\section{Derivation of $\chi$ dynamics}
\label{app:chi_derivation}

Before we embark on deriving the $\chi$ quantum kinetics, we collect a few useful relations. Using the boson commutation relations \eqref{eq:boson-comm} with identifications \eqref{eq:chi_imp} several times, we find
\begin{eqnarray}
  &&
  \sumA^L\, B_{\bf k} = B_{\bf k} \sumA^L\,,
  \qquad\qquad
  \sumA^L\, B^\dagger_{\bf k} =
   B^\dagger_{\bf k} \, \sumA^L +
   L\,\alpha^\star_{\bf k} \, \sumA^{L-1}\,,
\nonumber\\
  &&
  \sumA^L\, B^\dagger_{\bf k}\, B^\dagger_{{\bf k}'}  
   =
   B^\dagger_{\bf k} \, B^\dagger_{{\bf k}'} \, \sumA^L +
   L\,\left(\alpha^\star_{\bf k}\,B^\dagger_{{\bf k}'} +\alpha^\star_{{\bf k}'}\,B^\dagger_{{\bf k}}\right) \, \sumA^{L-1}
   +
   L\,(L-1)\,\alpha^\star_{\bf k} \,\alpha^\star_{{\bf k}'} \, \sumA^{L-2}\
\label{eq:app_sumA_AS_relations}\;.
\end{eqnarray}
The corresponding hermitian-conjugated formats follow straightforwardly. We will also use the following differentiation relations
\begin{align}
  &\partial^\star_{\bf k} \sumA^L = L\, B_{\bf k} \sumA^{L-1}\,,
  \quad
  \partial^\star_{\bf k} \sumAS^L =0\,,
  \quad
  \partial^\star_{\bf k} \partial^\star_{{\bf k}'}
  \sumA^L = L\,(L-1) B_{\bf k} B_{{\bf k}'} \sumA^{L-1}\,,
\nonumber\\
  &
  \partial_{\bf k} \sumAS^L = L\,  \sumA^{L-1}\,B^\dagger_{\bf k}\,,\quad
  \partial_{\bf k} \sumA^L =0\,,
  \quad
  \partial_{\bf k} \partial_{{\bf k}'}
  \sumAS^L = L\,(L-1)  \sumA^{L-1}\,B^\dagger_{\bf k} B^\dagger_{{\bf k}'}\,,
\label{eq:app_diff-rules}\;
\end{align}
that follow directly from differentiating definition \eqref{eq:sumA_AS}.

With the help of these and quantum dynamics \eqref{eq:BoccuD}--\eqref{eq:B+occuD}, we can evaluate the Heisenberg equation of motion \eqref{eq:HEM-ex} in four separate parts:
\begin{eqnarray}
  \ihddt \op{O}^{[\lambda]} \equiv
  \comm{\op{O}}{\op{H}^{[\lambda]}}
\label{eq:app_HEM-start}\;,
\end{eqnarray}
where $\lambda$ is equal to ``${\rm ren}$'', 0, 1, or 2 corresponding to renormalized $\tilde{E}_{\bf k}$, $W^{[0]}$ as well as Coulomb $W^{[1]}$, and $W^{[2]}$ contributions in Eqs.~\eqref{eq:BoccuD}--\eqref{eq:B+occuD}, respectively. For later use, we work out the quantum dynamics of $\sumA$, yielding
\begin{align}
  &\ihddt \sumA^{[{\rm ren}]} 
  = 
  \sum_{\bf k} \alpha^\star_{\bf k} \tilde{E}_{\bf k} \, B_{\bf k}\,,
  \quad 
  \ihddt \sumA^{[0]} \sum_{{\bf k} \neq {\bf k}'} \sum_{{\bf q} \neq ({\bf k},{\bf k}')}
  \alpha^\star_{\bf k} W^{[0]}_{{\bf k}'-{\bf k}}\, 
  B^\dagger_{{\bf q}-{\bf k}} \, B_{{\bf q}-{\bf k}'} \, B_{{\bf k}'}\,,
\nonumber\\
  &\ihddt \sumA^{[2]} 
  =
  \sum_{{\bf k}} B^\dagger_{{\bf k}} \alpha^\star_{-{\bf k}} W^{[2]}_{\bf k}\,,
  \quad
  \ihddt \sumA^{[1]} 
  =
  \sum_{{\bf k},{\bf k}'}{}'\,
  \alpha^\star_{\bf k} 
  \left( 
     W^{[1]}_{{\bf k},{\bf k}'-{\bf k}} B^\dagger_{{\bf k}'-{\bf k}} B_{{\bf k}'} 
    + {\textstyle \frac{1}{2}} W^{[1]}_{{\bf k}'} B_{{\bf k}-{\bf k}'} B_{{\bf k}'} 
  \right)
\label{eq:app_HEM-parts}\;,
\end{align}
after applying Eqs.~\eqref{eq:BoccuD}--\eqref{eq:B+occuD}, definition \eqref{eq:sumA_AS}, and boson commutator relations \eqref{eq:boson-comm} several times.

The quantum kinetics of the characteristic function can be constructed from the time evolution of $C^{J,K}$ constants defined in connection with Eq.~\eqref{eq:chi_imp}. To perform the derivation contribution by contribution, we evaluate
\begin{eqnarray}
  \ihddt C^{[\lambda]}_{J,K}
  &\equiv&
  \sum_{L=1}^K
    \frac{(-1)^K}{J! \,K!}
  \,
  \avex{\sumAS^J\,\sumA^{L-1} \ihddt\sumA^{[\lambda]}  \quad \sumA^{K-L} } 
\nonumber\\
  &+&
   \sum_{L=1}^J
    \frac{(-1)^K}{J! \,K!}
  \,
  \avex{\sumAS^{L-1} \ihddt\sumAS^{[\lambda]}  \quad \sumAS^{J-L} \sumA^K } 
\label{eq:app_HEM-C1}\;.
\end{eqnarray}
The first line of Eq.~\eqref{eq:app_HEM-C1} produces
\begin{eqnarray}
  \ihddt C^{{\rm [eff]},{\rm 1st}}_{J,K}
  &\equiv&
  \sum_{L=1}^K
    \frac{(-1)^K}{J! \,K!}
  \,
  \avex{\sumAS^J\,\sumA^{L-1} \sum_{\bf k} \alpha^\star_{\bf k} \tilde{E}_{\bf k} \, B_{\bf k}  \quad \sumA^{K-L} } 
\label{eq:HEM_app-Ceff1}\;,
\end{eqnarray}
once we insert dynamics \eqref{eq:app_HEM-parts} for $\lambda={\rm ``eff''}$. Since $B_{\bf k}$ commutes with $\sumA$ according to relation \eqref{eq:app_sumA_AS_relations}, Eq.~\eqref{eq:HEM_app-Ceff1} becomes
\begin{eqnarray}
  \ihddt C^{{\rm [eff]},{\rm 1st}}_{J,K}
  &=&
  \sum_{L=1}^K
    \frac{(-1)^K}{J! \,K!}
  \,
  \avex{\sumAS^J\,\sum_{\bf k} \alpha^\star_{\bf k} \tilde{E}_{\bf k} \, B_{\bf k}  \sumA^{K-1}  } 
\nonumber\\
  &=&
    \frac{(-1)^K}{J! \,K!}
  \,
  \avex{\sumAS^J\,\sum_{\bf k} \alpha^\star_{\bf k} \tilde{E}_{\bf k} \, K\,B_{\bf k} \sumA^{K-1}  } 
\label{eq:HEM_app-Ceff2}\;,
\end{eqnarray}
where the last step appears because the $L$-sum argument is independent of $L$. We may now apply the differentiation rule \eqref{eq:app_diff-rules} to convert Eq.~\eqref{eq:HEM_app-Ceff2} into
\begin{eqnarray}
  \ihddt C^{{\rm [eff]},{\rm 1st}}_{J,K}
  =
  \sum_{\bf k}{}' \tilde{E}_{\bf k} \,\alpha^\star_{\bf k} \partial^\star_{\bf k}
   C_{J,K}
\label{eq:HEM_app-Ceff3}\;.
\end{eqnarray}
A very similar derivation using commutation relations \eqref{eq:app_sumA_AS_relations} and differentiation properties \eqref{eq:app_diff-rules} eventually produces
\begin{align}
  \ihddt C^{{\rm [eff]}}_{J,K}
  &=
  \sum_{\bf k} \, \tilde{E}_{\bf k}
  \left[
    \left(
      \alpha^\star_{\bf k} \partial^\star_{\bf k}
    \right)
    -
    \left(
      \cdots 
    \right)^\star_{\alpha \rightarrow -\alpha}
  \right]
   C_{J,K}
\nonumber\\
  \ihddt C^{[0]}_{J,K}
  &=
  \sum_{{\bf k},{\bf k}',{\bf q}}{} \, W^{[0]}_{\bf q}
  \left[
    \left(
      {\textstyle \frac{1}{2}}
      \alpha^\star_{\bf k} \, \alpha^\star_{{\bf k}'} 
      -
      \alpha^\star_{\bf k} \partial_{{\bf k}'} 
    \right)
    \partial^\star_{{\bf k}-{\bf q}}\, \partial^\star_{{\bf k}'+{\bf q}}
       -
    \left(
      \cdots 
    \right)^\star_{\alpha \rightarrow -\alpha}
  \right]
   C_{J,K}
 \nonumber\\
  \ihddt C^{[1]}_{J,K}
  &=
  \sum_{{\bf k},{\bf k}'} 
  \left[
    W^{[1]}_{{\bf k},{\bf k}'} \alpha^\star_{\bf k} \partial_{{\bf k}'} \partial^\star_{{\bf k}+{\bf k}'}
    -{\textstyle \frac{1}{2}}
    \left(
      W^{[1]}_{{\bf k}',{\bf k}} \alpha^\star_{{\bf k}+{\bf k}'} \partial^\star_{{\bf k}} \partial^\star_{{\bf k}'}
      +
      W^{[1]}_{{\bf k},{\bf k}'}
      \, \alpha^\star_{{\bf k}}  \alpha^\star_{{\bf k}'} \partial^\star_{{\bf k}+{\bf k}'}  
    \right)
    -
    \left(
      \cdots 
    \right)^\star_{\alpha \rightarrow -\alpha}
  \right]
   C_{J,K}
 \nonumber\\
  \ihddt C^{[2]}_{J,K}
  &=
  \sum_{{\bf k}}
  W^{[2]}_{\bf k} 
  \left[
    {\textstyle \frac{1}{2}}
    \alpha^\star_{\bf k} \alpha^\star_{-{\bf k}}
    -
    \alpha^\star_{\bf k} \partial_{-{\bf k}}
    -
    \left(
      \cdots 
    \right)^\star_{\alpha \rightarrow -\alpha}
  \right]
   C_{J,K}
\label{eq:app_HEM-Ceff1}\;.
\end{align}
where 
$\left( \cdots \right)^\star_{\alpha \rightarrow -\alpha}$ within each contribution denotes a complex conjugation of the first term after all $\alpha$ arguments, including those appearing in differentiation, are negated. 
We also have omitted the explicit exclusion of the zero-momentum components in the sums because it is clear from now on that $C_{J,K}$ does not have $\alpha_0$ dependency. This somewhat simplifies the notation, while not confusing the correct structure.

Based on Eqs.~\eqref{eq:chi_imp}, \eqref{eq:BoccuD}--\eqref{eq:B+occuD}, and \eqref{eq:app_HEM-start}, the $\chi$ dynamics becomes
\begin{eqnarray}
  \ihddt \chi
  &=&
  \sum_{J,K=0}^\infty \sum_\lambda \ihddt C^{[\lambda]}_{J,K}
\label{eq:app_HEM-Ceff2}\;.
\end{eqnarray}
We have already worked out the dynamics of all $\lambda$ components in Eq.~\eqref{eq:app_HEM-Ceff1} which produces
\begin{eqnarray}
  \ihddt \chi
  &=&
  \left(
    \sum_{\bf k}  \tilde{E}_{\bf k}\, \alpha^\star_{\bf k} \, \partial^\star_{\bf k}
    +
    \sum_{{\bf k},{\bf k}',{\bf q}}{} \, W^{[0]}_{\bf q}
    \left[
      {\textstyle \frac{1}{2}}
      \alpha^\star_{\bf k} \, \alpha^\star_{{\bf k}'} 
      -
      \alpha^\star_{\bf k} \partial_{{\bf k}'} 
    \right]
    \partial^\star_{{\bf k}-{\bf q}}\, \partial^\star_{{\bf k}'+{\bf q}}
  \right.
\nonumber\\
  &&
  \quad+
  \sum_{{\bf k},{\bf k}'} 
  \left[
    W^{[1]}_{{\bf k},{\bf k}'} \alpha^\star_{\bf k} \partial_{{\bf k}'} \partial^\star_{{\bf k}+{\bf k}'}
    -{\textstyle \frac{1}{2}}
    \left(
      W^{[1]}_{{\bf k},{\bf k}'} \alpha^\star_{{\bf k}+{\bf k}'} \partial^\star_{{\bf k}} \partial^\star_{{\bf k}'}
      +
      W^{[1]}_{{\bf k},{\bf k}'}
      \, \alpha^\star_{{\bf k}}  \alpha^\star_{{\bf k}'} \partial^\star_{{\bf k}+{\bf k}'}  
    \right)
  \right]
\nonumber\\
  &&
  \quad +
  \left.
    \sum_{\bf k}
    W^{[2]}_{\bf k} 
    \left[
      {\textstyle \frac{1}{2}} 
      \alpha^\star_{\bf k} \alpha^\star_{-{\bf k}}
      -  
      \alpha^\star_{\bf k} \partial_{-{\bf k}}
    \right]
   \right)
   \; \chi
   -
   \left(
      \cdots 
    \right)^\star_{\alpha \rightarrow -\alpha} \chi
\label{eq:app_HEM-Ceff3}\;.
\end{eqnarray}
This presents exactly the quantum dynamics of interacting Bose gas when its quantum statistics is represented through the normally-ordered characteristic function. To simplify the notation, we have suppressed the $\{\alpha\}$ argument of $\chi$ and have not explicitly excluded the $\alpha_{\bf k}$ and $\partial_{\bf k}$ with a vanishing ${\bf k}$ because this does not produce unambiguity for computations performed in the excitation picture. 

We observe that the $\chi$ dynamics follows from integro-differential equation \eqref{eq:app_HEM-Ceff3} which we express symbolically via
\begin{eqnarray}
  \ihddt \chi_\alpha
  \equiv
  \op{D}^{\rm tot}_\alpha \, \chi_\alpha
\label{eq:app_HEM-chiFIN1}\;,
\end{eqnarray}
where we have again denoted the implicit $\alpha$ dependence for the sake of completeness. The differential operator can be organized in orders of differentiation, eventually yielding
\begin{eqnarray}
  \op{D}^{\rm tot}_\alpha
  &\equiv&
  d^{(0)}_\alpha
  +
  \sum_{\bf k} 
  \left( 
    d^{(1,-)}_{{\bf k},\alpha} \, \partial^\star_{\bf k}
    -
    \left[d^{(1,+)}_{{\bf k},\alpha}\right]^\star \partial_{\bf k}
  \right)
\nonumber\\
  &&
  +
  \frac{1}{2} 
  \sum_{{\bf k},{\bf k}'}
    \left( 
    d^{(2,-)}_{{\bf k},{\bf k}',\alpha} \, \partial^\star_{\bf k}\, \partial^\star_{{\bf k}'}
    -
    \left[d^{(2,+)}_{{\bf k},{\bf k}',\alpha}\right]^\star \partial_{\bf k}\, \partial_{{\bf k}'}
  \right)
  +
  \sum_{{\bf k},{\bf k}'}
  d^{(2)}_{{\bf k},{\bf k}',\alpha} \, \partial^\star_{\bf k}\, \partial_{{\bf k}'}
\nonumber\\
  &&
  -
    \sum_{{\bf k},{\bf k}',{\bf q}}
    \left( 
    d^{(3)}_{{\bf k},{\bf k}',\alpha} \, \partial_{{\bf k}+{\bf q}}\,  \partial^\star_{\bf k}\, \partial^\star_{{\bf k}'}
    -
    \left[d^{(3)}_{{\bf k},{\bf k}',\alpha}\right]^\star 
    \partial^\star_{{\bf k}+{\bf q}} \partial_{\bf k}\, \partial_{{\bf k}'}
  \right)
\label{eq:app_HEM-chiFIN2}\;,
\end{eqnarray}
where the first line contains a constant
\begin{eqnarray}
  d^{(0)}_\alpha \equiv  \sum_{{\bf k}} W^{[2]}_{\bf k}\,
  \frac{\alpha^\star_{\bf k} \alpha^\star_{-{\bf k}}
    -
    \alpha_{\bf k} \alpha_{-{\bf k}}}{2}
\label{eq:app_HEM-d_constant}\;,
\end{eqnarray}
as well as a linear contribution
\begin{eqnarray}
  d^{(1,\pm)}_{{\bf k},\alpha}
  \equiv  
  \tilde{E}_{\bf k} \alpha^\star_{\bf k} 
  +W^{[2]}_{-{\bf k}} \alpha_{-{\bf k}}
  \pm 
  \frac{1}{2} \sum_{{\bf k}'} W^{[1]}_{{\bf k}-{\bf k}',{\bf k}'} \, \alpha^\star_{{\bf k}-{\bf k}'}\, \alpha^\star_{{\bf k}'}
\label{eq:app_HEM-d_linear}\;.
\end{eqnarray}
The quadratic differentiations contain constants
\begin{eqnarray}
  d^{(2,\pm)}_{{\bf k},{\bf k}',\alpha}
  &\equiv&
  \sum_{\bf q} W^{[0]}_{\bf q} \alpha^\star_{{\bf k}-{\bf q}} \,  \alpha^\star_{{\bf k}'+{\bf q}} 
  \pm 
  W^{[1]}_{{\bf k},{\bf k}'} \, \alpha^\star_{{\bf k}+{\bf k}'}\,,
\nonumber\\
  d^{(2)}_{{\bf k},{\bf k}',\alpha}
  &\equiv&
  W^{[1]}_{{\bf k}',{\bf k}-{\bf k}'} \alpha^\star_{{\bf k}-{\bf k}'} 
  +
  W^{[1]}_{{\bf k},{\bf k}'-{\bf k}} \alpha_{{\bf k}'-{\bf k}} 
\label{eq:app_HEM-d_quadratic}\;.
\end{eqnarray}
while the cubic differentiations are multiplied by
\begin{eqnarray}
  d^{(3)}_{{\bf k},{\bf q},\alpha}
  &\equiv&
  W^{[0]}_{{\bf q}} \, \alpha_{{\bf k}-{\bf q}}
\label{eq:app_HEM-d_cubic}\;.
\end{eqnarray}
Equations~\eqref{eq:app_HEM-chiFIN1}--\eqref{eq:app_HEM-d_cubic} describe the $\chi$ dynamics exactly.

\section{Relevant differentiation relations in the implicit-notation formalism}
\label{app:relevant_D}

Before we show that the INF differentiation-relation \eqref{eq:INF-diff2and3} is generally valid, we consider first  useful summation relations. Based on definition \eqref{eq:sum_def},
we can perform an pointer-selection sum as follows:
\begin{eqnarray}
  \sel^{K+1}_{\ind{\lambda+1}}
  &=&
  \sum_{j_1=1}^{K-\lambda+1} \sum_{j_2=1+j_1}^{K-\lambda+2} \cdots  \sum_{j_{\lambda+1}=1+j_{\lambda}}^{K+1}
\nonumber\\
  &=&
  \left.
    \sum_{j_1=1}^{K-\lambda+1} \sum_{j_2=1+j_1}^{K-\lambda+2} \cdots  \sum_{j_{\lambda}=1+j_{\lambda-1}}^{K}
  \right|_{j_{\lambda+1}=K+1}
  +
  \sum_{j_1=1}^{K-\lambda} \sum_{j_2=1+j_1}^{K-\lambda+1} \cdots  \sum_{j_{\lambda+1}=1+j_{\lambda}}^{K}
\label{eq:app_sum1}\;.
\end{eqnarray}
where we have split the last index selection into two parts; the first one has the single pointer selection $j_{\lambda+1}=K+1$ while the second contribution contains the remaining selections. 
Note that the second selection can has upper limit $K$ because the pointer $j_{\lambda+1}=K+1$ selection has already been treated by the first term. Both of these contributions are now expressible in terms of an $\sel$ sum, based on the definition \eqref{eq:sum_def}, producing
\begin{eqnarray}
  \sel^{K+1}_{\ind{\lambda+1}}
  =
  \left.
  \sel^{K}_{\ind{\lambda}}
  \right|_{j_{\lambda+1}=K+1}
  +
  \sel^{K}_{\ind{\lambda+1}}
  =
  \sel^{K+1}_{\ind{\lambda},K+1}
  +
  \sel^{K}_{\ind{\lambda+1}}\,,
  \qquad
  \sel^{K+1}_{\ind{\lambda},K+1} \equiv  \left.
  \sel^{K}_{\ind{\lambda}}
  \right|_{j_{\lambda+1}=K+1}
\label{eq:app_sum2}\,,
\end{eqnarray}
where the highest index selection is expressed explicitly. For $\lambda=K$, the last selection does not produce any contributions. 
For later use, we have also introduced a simplified notation for explicit selection of $j_{K+1} =K+1$ as a pointer.

We now implement the differentiation \eqref{eq:INF-diff2and3} explicitly and evaluate the full pointer-selection sum \eqref{eq:sum_def} 
\begin{eqnarray}
  \sum_\ind{\lambda}^{K+1} 
  &=&
  \sum_{\lambda=0}^{K+1} \sel^{K+1}_\ind{\lambda}
  =
  \sel^{K+1}_\emptyset
  +
  \sum_{\lambda=1}^{K+1} \sel^{K+1}_\ind{\lambda}=
  \sel^{K}_\emptyset
  +
  \sum_{\lambda=0}^{K} \sel^{K+1}_\ind{\lambda+1}
\label{eq:app_sum3}\;,
\end{eqnarray}
where we have first computed the $\lambda=0$ that selects an empty set. The last step is obtained by applying a change of summation index $\lambda \rightarrow \lambda +1$ and implementing $\sel^{K+1}_\emptyset = \sel^{K}_\emptyset$ because an empty set selection is identical regardless of how many elements it is chosen from. 
As we apply relation \eqref{eq:app_sum2} to result \eqref{eq:app_sum3}, we obtain
\begin{eqnarray}
  \sum_\ind{\lambda}^{K+1} 
  &=&
  \sel^{K}_\emptyset
  +
  \sum_{\lambda=0}^{K}   
  \sel^{K+1}_{\ind{\lambda},K+1}
  +
  \sum_{\lambda=0}^{K-1}
  \sel^{K}_{\ind{\lambda+1}}
  =
  \sum_{\lambda=0}^{K}   
  \sel^{K+1}_{\ind{\lambda},K+1}
  +
  \sel^{K}_\emptyset
  +
  \sum_{\lambda=1}^{K}
  \sel^{K}_{\ind{\lambda}}
\nonumber\\
  &=&
    \sum_{\lambda=0}^{K}   
  \sel^{K+1}_{\ind{\lambda},K+1}
  +
  \sum_{\lambda=0}^{K}
  \sel^{K}_{\ind{\lambda}}
  =
  \sum_{\lambda=0}^{K}
  \left(
  \sel^{K+1}_{\ind{\lambda},K+1}
  +
  \sel^{K}_{\ind{\lambda}}  
  \right)
\label{eq:app_sum4}\;,
\end{eqnarray}
where we first have fixed the upper limit of the second term to $K-1$ because the $\lambda=0$ term does not contribute, see discussion after Eq.~\eqref{eq:app_sum2}. The change of summation, $\lambda \rightarrow \lambda -1$, allows us to collect all contributions under one $\lambda$.

To prove that differentiation \eqref{eq:INF-diff2and3} is generally valid, we show that a simpler form
\begin{eqnarray}
  \Dcoll{\seq{K}}{\emptyset}\, F_1\,F_2
  &=&
  \sum_\ind{\lambda}^K 
  \left[
    \Dcoll{\ssel{K}{\ind{\lambda}}}{\emptyset}
    F_1
  \right]
  \left[
    \Dcoll{\srm{K}{\ind{\lambda}}}{\emptyset}
    F_2
  \right]
\label{eq:app_diff1}
\end{eqnarray}
is a general relation. We prove this by using the method of induction. For $K=1$ elements, differentiation \eqref{eq:app_diff1} becomes
\begin{align}
  \Dcoll{\seq{1}}{\emptyset}\, F_1\,F_2
  &=
  \sum_{\lambda=0}^1 \sel^{1}_{\ind{\lambda}}
  \left[
    \Dcoll{\ssel{1}{\ind{\lambda}}}{\emptyset}
    F_1
  \right]
  \left[
    \Dcoll{\srm{1}{\ind{\lambda}}}{\emptyset}
    F_2
  \right]
\nonumber\\
  &=
  \sel^{1}_{\ind{\lambda=0}}
  \left[
    \Dcoll{\ssel{1}{\ind{\lambda}}}{\emptyset}
    F_1
  \right]
  \left[
    \Dcoll{\srm{1}{\ind{\lambda}}}{\emptyset}
    F_2
  \right]
  +
  \sel^{1}_{\ind{\lambda=1}}
  \left[
    \Dcoll{\ssel{1}{\ind{\lambda}}}{\emptyset}
    F_1
  \right]
  \left[
    \Dcoll{\srm{1}{\ind{\lambda}}}{\emptyset}
    F_2
  \right]
\label{eq:app_diff2}\;,
\end{align}
which follows from definition \eqref{eq:pointer_selK}. Both of the pointer selections in Eq.~\eqref{eq:app_diff2} reduce to a simple form; $\sel^{1}_{\ind{\lambda=0}}$ does not select any pointers, such that the pointer sequence $\ind{0}$ becomes an empty set $\emptyset$. 
The  $\sel^{1}_{\ind{\lambda=1}}$ selects the only possible pointer sequence $\ind{1}$. With this information, Eq.~\eqref{eq:app_diff2} reduces into
\begin{align}
  \Dcoll{\seq{1}}{\emptyset}\, F_1\,F_2
  &=
  \left[
    \Dcoll{\ssel{1}{\emptyset}}{\emptyset}
    F_1
  \right]
  \left[
    \Dcoll{\srm{1}{\emptyset}}{\emptyset}
    F_2
  \right]
  +
  \left[
    \Dcoll{\ssel{1}{\ind{1}}}{\emptyset}
    F_1
  \right]
  \left[
    \Dcoll{\srm{1}{\ind{1}}}{\emptyset}
    F_2
  \right]
\nonumber\\
&=
  \left[
    \Dcoll{\emptyset}{\emptyset}
    F_1
  \right]
  \left[
    \Dcoll{1}{\emptyset}
    F_2
  \right]
  +
  \left[
    \Dcoll{1}{\emptyset}
    F_1
  \right]
  \left[
    \Dcoll{\emptyset}{\emptyset}
    F_2
  \right]
\nonumber\\
&=
    F_1
  \left[
    \partial_{{\bf k}_1}
    F_2
  \right]
  +
  \left[
    \partial_{{\bf k}_1}
    F_1
  \right]
    F_2
\label{eq:app_diff3}\;,
\end{align}
where the last step follows by implementing definitions \eqref{eq:empty-diff}. We now see that the explicit differentiations construct $\Dcoll{\seq{1}}{\emptyset}\, F_1\,F_2 = \partial_{{\bf k}_1} F_1 F_2$, which is identical to the original definition \eqref{eq:D_collective} for the collective differentiation. 
In other words, we have shown that recursion \eqref{eq:app_diff1} is valid at least for $K=1$.

As an induction assumption, we expect that recursion \eqref{eq:app_diff1} is valid for $K$ and then show that this necessarily implies that also the $K+1$ recursion follows the same relation. Therefore, we compute
\begin{align}
  \Dcoll{\seq{K+1}}{\emptyset}\, F_1\,F_2
  &=
  \Dcoll{\seq{K},{\bf k}_{K+1}}{\emptyset}\, F_1\,F_2
  =
  \Dcoll{\seq{K}}{\emptyset}\,  \partial_{{\bf k}_{K+1}} F_1\,F_2
\nonumber\\
  &=
  \Dcoll{\seq{K}}{\emptyset}\,  \left[\partial_{{\bf k}_{K+1}} F_1\right] \,F_2
  +
  \Dcoll{\seq{K}}{\emptyset}\,   F_1 \left[\partial_{{\bf k}_{K+1}}\,F_2 \right]
\label{eq:app_IND1}\;,
\end{align}
where we have differentiated the $\partial_{{\bf k}_{K+1}}$ explicitly, based on definition \eqref{eq:D_collective} and the product rule of differentiation. 
We may now implement relation \eqref{eq:app_diff1} on the basis of the induction assumption. This converts differentiation \eqref{eq:app_IND1} into
\begin{align}
  \Dcoll{\seq{K+1}}{\emptyset}\,& F_1\,F_2
  =
  \sum_\ind{\lambda}^K 
  \left(
  \left[
    \Dcoll{\ssel{K}{\ind{\lambda}}}{\emptyset}
    \partial_{{\bf k}_{K+1}} F_1
  \right]
  \left[
    \Dcoll{\srm{K}{\ind{\lambda}}}{\emptyset}
    F_2
  \right]
  +
  \left[
    \Dcoll{\ssel{K}{\ind{\lambda}}}{\emptyset}
    F_1
  \right]
  \left[
    \Dcoll{\srm{K}{\ind{\lambda}}}{\emptyset}
    \partial_{{\bf k}_{K+1}}\,F_2
  \right]
  \right)
\nonumber\\
&=
  \sum_\ind{\lambda}^K 
  \left(
  \left[
    \Dcoll{\ssel{K+1}{\ind{\lambda},K+1}}{\emptyset}
     F_1
  \right]
  \left[
    \Dcoll{\srm{K+1}{\ind{\lambda},K+1}}{\emptyset}
    F_2
  \right]
  +
  \left[
    \Dcoll{\ssel{K+1}{\ind{\lambda}}}{\emptyset}
    F_1
  \right]
  \left[
    \Dcoll{\srm{K+1}{\ind{\lambda}}}{\emptyset}
    \,F_2
  \right]
  \right)
\label{eq:app_IND2}\;,
\end{align}
where the last step follows because we may simplify
$\Dcoll{\ssel{K}{\ind{\lambda},K}}{\emptyset} \partial_{{\bf k}_{K+1}}
= \Dcoll{\ssel{K+1}{\ind{\lambda},K+1}}{\emptyset}$ and
$\Dcoll{\srm{K}{\ind{\lambda}}}{\emptyset} \partial_{{\bf k}_{K+1}}
= \Dcoll{\srm{K+1}{\ind{\lambda}}}{\emptyset}$ and rewrite $\Dcoll{\srm{K}{\ind{\lambda}}}{\emptyset}=\Dcoll{\srm{K+1}{\ind{\lambda},K+1}}{\emptyset}$ as well as $\Dcoll{\ssel{K}{\ind{\lambda}}}{\emptyset}=\Dcoll{\ssel{K+1}{\ind{\lambda}}}{\emptyset}$ because $\lambda$ excludes the selection of the pointer $K+1$. We then implement $\sum_\ind{\lambda}^K$ with the help of definition \eqref{eq:pointer_selK}, which converts Eq.~\eqref{eq:app_IND2} into
\begin{align}
  \Dcoll{\seq{K+1}}{\emptyset}\, F_1\,F_2
  &=
  \sum_{\lambda=0}^K
  \left(
  \sel^{K+1}_{\ind{\lambda},K+1}
  \left[
    \Dcoll{\ssel{K+1}{\ind{\rm sel}}}{\emptyset}
     F_1
  \right]
  \left[
    \Dcoll{\srm{K+1}{\ind{\rm sel}}}{\emptyset}
    F_2
  \right]
  \right. 
\nonumber\\
  &\left.
  \qquad\qquad+
  \sel^{K}_{\ind{\lambda}}
  \left[
    \Dcoll{\ssel{K+1}{\ind{\rm sel}}}{\emptyset}
     F_1
  \right]
  \left[
    \Dcoll{\srm{K+1}{\ind{\rm sel}}}{\emptyset}
    F_2
  \right]
  \right)
\nonumber\\
  &=
  \sum_{\lambda=0}^K
  \left(
    \sel^{K+1}_{\ind{\lambda},K+1}
    +
    \sel^{K}_{\ind{\lambda}}
  \right)
  \left[
    \Dcoll{\ssel{K+1}{\ind{\rm sel}}}{\emptyset}
     F_1
  \right]
  \left[
    \Dcoll{\srm{K+1}{\ind{\rm sel}}}{\emptyset}
    F_2
  \right]
\label{eq:app_IND3}\;,
\end{align}
where $\ind{\rm sel}$ indicates the pointer sequence selected by the pointer selection operation $\sel$. We now see that the $\lambda$ sum is identical to the reduction formula \eqref{eq:app_sum4}. 
Therefore, Eq.~\eqref{eq:app_IND3} reduces into
\begin{align}
  \Dcoll{\seq{K+1}}{\emptyset}\, F_1\,F_2
  &=
  \sum_{\lambda=0}^{K+1}
    \sel^{K+1}_{\ind{\lambda}}
  \left[
    \Dcoll{\ssel{K+1}{\ind{\rm sel}}}{\emptyset}
     F_1
  \right]
  \left[
    \Dcoll{\srm{K+1}{\ind{\rm sel}}}{\emptyset}
    F_2
  \right]
\nonumber\\
  &=
  \sum_{\lambda=0}^{K+1}
    \sel^{K+1}_{\ind{\lambda}}
  \left[
    \Dcoll{\ssel{K+1}{\ind{\lambda}}}{\emptyset}
     F_1
  \right]
  \left[
    \Dcoll{\srm{K+1}{\ind{\lambda}}}{\emptyset}
    F_2
  \right]
\label{eq:app_IND-FIN}\,,\quad
\end{align}
which shows that our induction assumption is correct.

It is clear that also $\Dcoll{\emptyset}{\seq{P}}$ must satisfy a relation that is completely analogous to Eq.~\eqref{eq:app_diff1} because one can simply complex-conjugate relation \eqref{eq:app_diff1} to produce it. 
In total, $\Dcoll{\seq{K}}{\seq{P}} = \Dcoll{\seq{K}}{\emptyset} \Dcoll{\emptyset}{\seq{P}}$ can then be constructed by implementing two differentiation relations separately, which eventually produces result \eqref{eq:INF-diff2and3}. In this context, the application of differentiation to three functions follows recursively from the two-function differentiation.

\section{Quantum kinetics of triplets}
\label{app:triplets}

Any correlation is uniquely defined by $\DI{\seq{K}}{\seq{P}}$ where $\seq{K}$ and $\seq{P}$ contain a sequence of atom states that appear in boson creation and annihilation operators of the correlation, respectively, compare with definition \eqref{eq:imp-corr}. The quantum dynamics of triplets can be directly unraveled from Eqs.~\eqref{eq:clusterDYN}--\eqref{eq:cubic-FIN} by choosing a specific $\seq{K}$ and $\seq{P}$ combination identified by definitions \eqref{eq:triplets}. 
We apply homogeneous conditions, i.e.~the total momentum of creation operators is equal to that of annihilation operators within all correlations that appear, as discussed in Sec.~\ref{sec:relevant clusters}. Next, we use this straightforward approach and write down the explicit quantum dynamics of $T^{{\bf k}',{\bf k}}_{1,2}$ and $T^{{\bf k}',{\bf k}}_{0,3}$ correlations.  Triplets describe correlations among three boson operators, and we first need to identify the corresponding three normal-component wave vectors. 
Based on definition \eqref{eq:triplets}, $T^{{\bf k}',{\bf k}}_{1,2}$ contains state ${\bf k}_1 ={\bf k}+{\bf k}'$ that defines sequence $\seq{K}$ as well as states ${\bf p}_1={\bf k}'$, ${\bf p}_2 = {\bf k}$ that are elements of $\seq{P}$. At the same time, $T^{{\bf k}',{\bf k}}_{0,3}$ contains no elements within $\seq{K} = \emptyset$ while $\seq{P}$ contains states ${\bf p}_1 = -{\bf k} -{\bf k}'$, ${\bf p}_2={\bf k}'$, and ${\bf p}_3 = {\bf k}$. With this information, Eqs.~\eqref{eq:clusterDYN}--\eqref{eq:cubic-FIN} produce the corresponding correlation dynamics straightforwardly.

To identify the physical structure of typical scattering source that drives triplets, we introduce
\begin{eqnarray}
  {\rm BZ}^{{\bf k}_1,{\bf k}_2,{\bf k}_3}
  \equiv
  \sqrt{\NC}\,
  f_{{\bf k}_1} 
  \left(1+f_{{\bf k}_2} \right)
  \left(1+f_{{\bf k}_3} \right)
  -
  \sqrt{\NC}
  \left(1+f_{{\bf k}_1} \right) 
  f_{{\bf k}_2}\, f_{{\bf k}_3} 
\label{eq:app_Boltzmann}\;
\end{eqnarray}
which has a typical form encountered for the bosonic Boltzmann scattering. More specifically, the $\sqrt{\NC}\, f_{{\bf k}_1} \left(1+f_{{\bf k}_2} \right) \left(1+f_{{\bf k}_3} \right)$ part can formally be viewed as an in-scattering contribution whose initial state is defined by a normal-component occupation $f_{{\bf k}_1}$ and a BEC contribution $\sqrt{\NC}$ whereas the final state consists of two normal-component atoms that will contain $\left(1+f_{{\bf k}_2} \right)$ and $\left(1+f_{{\bf k}_3} \right)$ atoms after a scattering event. 
Interestingly, the BEC does not participate in the scattering via its occupation, but via  $\sqrt{\NC}$ that is essentially BEC order parameter. Intuitively, this follows because the BEC acts as a macroscopic object whose individual atoms cannot be singled out in scattering events. The role of initial and final states is reversed in an out-scattering contribution $\sqrt{\NC} \left(1+f_{{\bf k}_1} \right) f_{{\bf k}_2}\, f_{{\bf k}_3}$; 
in this context, the final-state $\sqrt{1+\NC}$ is essentially the same as $\sqrt{\NC}$, which explains why one does not need to increase the BEC number by one. In general, ${\rm BZ}^{{\bf k}_1,{\bf k}_2,{\bf k}_3}$ greatly resembles a typical fermionic Boltzmann scattering source that drives doublet dynamics entering the SBEs, cf.~Ref.~\cite{Kira:2006} for more details.

Besides pure occupations, the normal state may contain a squeezing amplitude which also contributes to the source that drives triplets; this term is analogous to the polarization scattering that drives the doublets entering the SBEs, described in Ref.~\cite{Kira:2006,Book:2011}. In total, the scattering contribution to the $T^{{\bf k}',{\bf k}}_{1,2}$ becomes
\begin{align}
  S^{{\bf k}',{\bf k}}_{1,2}
  &\equiv
  \frac{V_{\bf k} }{2} 
  \left(
      {\rm BZ}^{{\bf k}+{\bf k}',{\bf k}',{\bf k}}
      +
      \sqrt{\NC}\,
      (f_{{\bf k}+{\bf k}'}-f_{{\bf k}'}) s_{\bf k}
      +
      \sqrt{\NC}\,s^\star_{{\bf k}+{\bf k}'}\, s_{{\bf k}'}
  \right)
\nonumber\\
  &+
  \frac{V_{{\bf k}+{\bf k}'}}{2} 
  \sqrt{\NC}\,
  \left(
    f_{{\bf k}+{\bf k}'}-f_{{\bf k}'}
    +
    s^\star_{{\bf k}+{\bf k}'}
  \right)
  s_{\bf k}
\label{eq:app_S12}\;.\;
\end{align}
The explicit form of $T^{{\bf k}',{\bf k}}_{1,2}$ dynamics then becomes
\begin{align}
  \ihddt& T^{{\bf k}',{\bf k}}_{1,2}
  =
  \left(
    E^{\rm ren}_{{\bf k}} + E^{\rm ren}_{{\bf k}'} - E^{\rm ren}_{{\bf k}+{\bf k}'}
  \right)
  T^{{\bf k}',{\bf k}}_{1,2}
  +
  \Delta^{\rm ren}_{\bf k}
  \left[ T^{{\bf k}+{\bf k}',-{\bf k}}_{1,2}\right]^\star
  +
  \Delta^{\rm ren}_{{\bf k}'}
  \left[ T^{{\bf k}+{\bf k}',-{\bf k}'}_{1,2}\right]^\star
  -
  \left[\Delta^{\rm ren}_{{\bf k}+{\bf k}'}\right]^\star
  T^{{\bf k}',{\bf k}}_{0,3}
\nonumber\\
  &+
  S^{{\bf k}',{\bf k}}_{1,2}+S^{{\bf k},{\bf k}'}_{1,2} 
\nonumber\\
  &+
  \left(1+f_{\bf k} + f_{{\bf k}'} \right) \sum_{{\bf l}} V_{{\bf l}} \, T^{{\bf k}'+{\bf l},{\bf k}-{\bf l}}_{1,2}
  +
  \left( f_{{\bf k}+{\bf k}'} - f_{{\bf k}}\right) \sum_{{\bf l}} V_{{\bf l}-{\bf k}} \, T^{{\bf k}',{\bf l}}_{1,2}
    +
  \left( f_{{\bf k}+{\bf k}'} - f_{{\bf k}'}\right) \sum_{{\bf l}} V_{{\bf l}-{\bf k}'} \, T^{{\bf l},{\bf k}}_{1,2}
\nonumber\\
  &+
  s_{{\bf k}'}
  \left[
    \sum_{{\bf l}} V_{{\bf l}+{\bf k}'} \, T^{{\bf l},{\bf k}'+{\bf k}}_{1,2}
    -
    \sum_{{\bf l}} V_{{\bf l}-{\bf k}'-{\bf k}} \, T^{-{\bf l},{\bf l}+{\bf k}}_{1,2}
  \right]^\star
  +
  s_{{\bf k}}
  \left[
    \sum_{{\bf l}} V_{{\bf l}+{\bf k}} \, T^{{\bf l},{\bf k}'+{\bf k}}_{1,2}
    -
    \sum_{{\bf l}} V_{{\bf l}-{\bf k}'-{\bf k}} \, T^{-{\bf l},{\bf l}+{\bf k}'}_{1,2}
  \right]^\star
\nonumber\\
  &+
  s^\star_{{\bf k}+{\bf k}'}
  \left[
    \sum_{{\bf l}} V_{{\bf l}-{\bf k}'} \, T^{{\bf l},{\bf k}}_{0,3}
    +
    \sum_{{\bf l}} V_{{\bf l}-{\bf k}} \, T^{{\bf k}',{\bf l}}_{0,3}
  \right]
\nonumber\\
  &+
  V_{\bf k} \left(f_{{\bf k}+{\bf k}'}-f_{{\bf k}'} \right)
  \sum_{\bf l}  T^{{\bf l},{\bf k}}_{1,2}
  +
  V_{{\bf k}'} \left(f_{{\bf k}+{\bf k}'}-f_{{\bf k}} \right)
  \sum_{\bf l}  T^{{\bf k}',{\bf l}}_{1,2}
  +
  V_{{\bf k}+{\bf k}'} 
  \left(s_{{\bf k}}+s_{{\bf k}'} \right)
  \left[
    \sum_{\bf l}  T^{{\bf l},{\bf k}+{\bf k}'}_{1,2}
  \right]^\star
\nonumber\\
  &+
  {\rm Hi}^{{\bf k}',{\bf k}}_{1,2}+{\rm Hi}^{{\bf k},{\bf k}'}_{1,2}
\label{eq:app_T12-dyn}\;,
\end{align}
where the last contribution contains the hierarchical coupling. Its explicit form is
\begin{align}
  {\rm Hi}^{{\bf k}',{\bf k}}_{1,2}
  &\equiv
  \sum_{\bf l} \sqrt{\NC}
  \left(
    V_{\bf l} \, Q^{{\bf k},{\bf l},{\bf k}'-{\bf l}}_{1,3}
    + 
    W_{{\bf l},{\bf k}} \, Q^{{\bf k}'-{\bf l},{\bf l},{\bf k}+{\bf k}'}_{2,2}
    -
    \frac{V_{\bf l}}{2}\, \left[Q^{{\bf l}'-{\bf k}',{\bf k}',{\bf k}}_{2,2}\right]^\star
    -\frac{W_{{\bf l},{\bf k}+{\bf k}'}}{2}\,  Q^{{\bf l},{\bf k}',{\bf k}}_{1,3}
  \right)
\nonumber\\
  &+
  \sum_{{\bf j},{\bf l}} V_{\bf l} 
  \left(
    \Delta 
      \avex{
	B^\dagger_{{\bf k}+{\bf k}'} B^\dagger_{{\bf j}+{\bf l}} B_{{\bf k}'+{\bf l}} B_{\bf j} B_{\bf k}
      }
     -
     {\textstyle \frac{1}{2}}
    \Delta 
      \avex{
	B^\dagger_{{\bf j}} B^\dagger_{{\bf k}+{\bf k}'+{\bf l}} B_{{\bf k}'} B_{\bf k} B_{{\bf j}+{\bf l}}
      }     
  \right)
\label{eq:app_Hi12}\;,
\end{align}
where we have used $W_{{\bf k},{\bf l}} \equiv V_{\bf k}+V_{\bf l}$, based on Eq.~\eqref{eq:W_def}. These contributions
couple the triplets to quadruplets as well as five-atom correlations. The quadruplets are defined by Eq.~\eqref{eq:quadruplets}.

The quantum kinetics of $T^{{\bf k}',{\bf k}}_{0,3}$ follows an analogous format. Its scattering source contains only terms containing a squeezing amplitude,
\begin{eqnarray}
  S^{{\bf k}',{\bf k}}_{0,3}
  \equiv
  \frac{V_{\bf k} }{2} 
  \left(
      \sqrt{\NC}\,
      (1+f_{{\bf k}}+s_{{\bf k}}) 
      (s_{{\bf k}'}+s_{{\bf k}+{\bf k}'})
      +
      \sqrt{\NC}\,f_{{\bf k}'}\,s^\star_{{\bf k}+{\bf k}'}
      +
      \sqrt{\NC}\,f_{{\bf k}+{\bf k}'}\,s^\star_{{\bf k}'}
  \right)
\label{eq:app_S03}\;.
\end{eqnarray}
The full correlation dynamics becomes
\begin{align}
  \ihddt &T^{{\bf k}',{\bf k}}_{0,3}
  =
  \left(
    E^{\rm ren}_{{\bf k}} + E^{\rm ren}_{{\bf k}'} + E^{\rm ren}_{{\bf k}+{\bf k}'}
  \right)
  T^{{\bf k}',{\bf k}}_{0,3}
  +
  \Delta^{\rm ren}_{\bf k} \, T^{{\bf k}',-{\bf k}-{\bf k}'}_{2,1}
  +
  \Delta^{\rm ren}_{{\bf k}'} \, T^{-{\bf k}-{\bf k}',{\bf k}}_{2,1}
  -
  \Delta^{\rm ren}_{{\bf k}+{\bf k}'}\, T^{{\bf k}',{\bf k}}_{2,1}
\nonumber\\
  &+
  S^{{\bf k}',{\bf k}}_{0,3}+S^{{\bf k}',{\bf k}}_{0,3} +S^{{\bf k}',-{\bf k}-{\bf k}'}_{0,3} 
  +
  \left(1+f_{\bf k} + f_{{\bf k}+{\bf k}'} \right) \sum_{{\bf l}} V_{{\bf l}-{\bf k}} \, T^{{\bf k}',{\bf l}}_{0,3}
\nonumber\\
  &+
  \left(1+ f_{{\bf k}'} + f_{{\bf k}+{\bf k}'}\right) \sum_{{\bf l}} V_{{\bf l}-{\bf k}'} \, T^{{\bf l},{\bf k}'}_{0,3}
    +
  \left(1+ f_{{\bf k}} + f_{{\bf k}'}\right) \sum_{{\bf l}} V_{{\bf l}} \, T^{{\bf k}'+{\bf l},{\bf k}-{\bf l}}_{0,3}
\nonumber\\
  &+
  s_{{\bf k}}
  \sum_{{\bf l}}
  \left[
    V_{{\bf l}+{\bf k}+{\bf k}'} \, T^{{\bf k}',{\bf l}'}_{1,2}
    +
    V_{{\bf l}-{\bf k}'} \, T^{-{\bf k}-{\bf k}',{\bf l}}_{1,2}
  \right]
  +
  s_{{\bf k}'}
  \sum_{{\bf l}}
  \left[
     V_{{\bf l}+{\bf k}+{\bf k}'} \, T^{{\bf l},{\bf k}}_{1,2}
    +
    V_{{\bf l}-{\bf k}} \, T^{-{\bf k}-{\bf k}',{\bf l}}_{1,2}
  \right]
\nonumber\\
  &+
  s_{{\bf k}+{\bf k}'}
  \sum_{{\bf l}}
  \left[
     V_{{\bf l}-{\bf k}} \, T^{{\bf k}',{\bf l}}_{1,2}
    +
    V_{{\bf l}-{\bf k}'} \, T^{{\bf l},{\bf k}}_{1,2}
  \right]
\nonumber\\
  &+
  V_{\bf k} \left(s_{{\bf k}'}+s_{{\bf k}+{\bf k}'} \right)
  \sum_{\bf l}  T^{{\bf l},{\bf k}}_{1,2}
  +
  V_{{\bf k}'} \left(s_{{\bf k}}+s_{{\bf k}+{\bf k}'} \right)
  \sum_{\bf l}  T^{{\bf k}',{\bf l}}_{1,2}
  +
  V_{{\bf k}+{\bf k}'} 
  \left(s_{{\bf k}}+s_{{\bf k}'} \right)
  \sum_{\bf l}  T^{-{\bf k}-{\bf k}',{\bf l}}_{1,2}
\nonumber\\
  &+
  {\rm Hi}^{{\bf k}',{\bf k}}_{0,3}+{\rm Hi}^{{\bf k},{\bf k}'}_{0,3}+{\rm Hi}^{{\bf k}',-{\bf k}-{\bf k}'}_{0,3}
\label{eq:app_T03-dyn}\;,
\end{align}
where the hierarchical coupling is provided by
\begin{eqnarray}
  {\rm Hi}^{{\bf k}',{\bf k}}_{1,2}
  \equiv
  \sum_{\bf l} \sqrt{\NC}
  \left(
    \frac{V_{\bf l}}{2} \, Q^{-{\bf k}-{\bf k}',{\bf l},{\bf k}}_{0,4}
    + 
    W_{{\bf l},{\bf k}} \, Q^{-{\bf k}-{\bf k}',{\bf k}',{\bf k}+{\bf l}}_{2,2}
  \right)
  +
  \sum_{{\bf j},{\bf l}} V_{\bf l} 
    \Delta 
      \avex{
	B^\dagger_{{\bf l}+{\bf j}} B_{-{\bf k}-{\bf k}'} B_{{\bf l}} B_{{\bf k}'+{\bf j}} B_{\bf k}
      }
\label{eq:app_Hi03}
\end{eqnarray}
that contains quadruplets and five-atom correlations. The other triplets $T^{{\bf k}',{\bf k}}_{2,1}$ and $T^{{\bf k}',{\bf k}}_{3,0}$ are connected with $T^{{\bf k}',{\bf k}}_{1,2}$ and $T^{{\bf k}',{\bf k}}_{0,3}$, respectively, with complex conjugation such that we do not need to elaborate their explicit dynamics here.

\section{Hyperbolic Bloch equation at Bogoliubov limit}
\label{app:HBE-Bogo}

By inserting  substitutions \eqref{eq:Bogo_lim} to the HBEs \eqref{eq:HBE-FIN}, we obtain the Bogoliubov limit of the HBEs:
\begin{eqnarray}
  \ihddt s_{\bf k} 
  &=& 
  2 \left( E^V_{\bf k} - i \gamma_{\bf k} \right) s_{\bf k} + \left( 1 + 2 f_{\bf k} \right) \NC\,V_{\bf k}
\label{eq:app_HBE-s-Bogo}\;,
\\
  \hbar \frac{\partial}{\partial t} f_{\bf k} 
  &=&
   -\Gamma_{\bf k}\, f_{\bf k}
  + 2\,\NC\,V_{\bf k} {\rm Im}
  \left[
    s^\star_{\bf k}
  \right] 
\label{eq:app_HBE-n-Bogo}\;,
\end{eqnarray}
where we have set the relaxation $r$ to zero. The renermalization of the kinetic energy now follows from $E^V_{\bf k}=E_{\bf k} + \NC\,V_{\bf k}$, as introduced by Eq.~\eqref{eq:Bogo_lim}. As the BEC number approaches a constant value, the squeezing amplitude and atom distribution approach a simple steady state:
\begin{eqnarray}
  s^{\rm ste}_{\bf k} 
  =
  -\frac{1}{2} \frac{\left(1+2f^{\rm ste}_{\bf k} \right) \NC V_{\bf k}}{E^V_{\bf k} - i \gamma_{\bf k}}\,,
  \qquad
   f^{\rm ste}_{\bf k} 
  =
  -\frac{2\,\NC\,V_{\bf k}}{\Gamma_{\bf k}} {\rm Im}
  \left[
    s^{\rm ste}_{\bf k}
  \right] 
\label{eq:app_n-steady1}\;.
\end{eqnarray}
By substituting $s^{\rm ste}_{\bf k}$ into Eq.~\eqref{eq:app_n-steady1}, we eventually obtain
\begin{eqnarray}
   f^{\rm ste}_{\bf k} 
  &=&
  \frac{\gamma_{\bf k}}{\Gamma_{\bf k}}
  \left(1+2f^{\rm ste}_{\bf k} \right)
  \left|\frac{\NC\,V_{\bf k}}{E^V_{\bf k}+i\gamma_{\bf k}}\right|^2
\label{eq:app_n-steady2}\;.
\end{eqnarray}
We may now apply the minimum uncertainly condition \eqref{eq:Gamma-gamma} (with $f_{\bf k} \rightarrow f^{\rm ste}_{\bf k}$), yielding
\begin{eqnarray}
   f^{\rm ste}_{\bf k} 
  &=&
  \frac{1}{4}
  \frac{\left(1+2f^{\rm ste}_{\bf k} \right)^2}{1+f^{\rm ste}_{\bf k}}
  \left|\frac{\NC\,V_{\bf k}}{E^V_{\bf k}+i\gamma_{\bf k}}\right|^2
\label{eq:app_n-steady3}\;.
\end{eqnarray}
By multiplying both sides with $1+f^{\rm ste}_{\bf k}$, this produces a regular quadratic equation whose roots yield only one positive atom distribution
\begin{eqnarray}
   f^{\rm ste}_{\bf k} 
  &=&
  \frac{1}{2}
  \left(
    \frac{|E^V_{\bf k}+i\gamma_{\bf k}|}{E^{\rm BG}_{\bf k}}
    -1
  \right)
\label{eq:app_n-steadyFIN}\;,
\end{eqnarray}
where we have identified the standard energy dispersion
\begin{eqnarray}
  E^{\rm BG}_{\bf k} \equiv \sqrt{|E^V_{\bf k}+\NC V_{\bf k} + i \gamma_{\bf k}|^2 -\left( \NC V_{\bf k} \right)^2 }
  = \sqrt{ \left(\frac{\hbar^2 {\bf k}^2}{2m}\right)^2 +  \NC V_{\bf k} \, \frac{\hbar^2 {\bf k}^2}{m} + \gamma^2_{\bf k}}
\label{eq:app_E_Bogoliubov}\;
\end{eqnarray}
for the Bogoliubov excitations. As result \eqref{eq:app_n-steadyFIN} is inserted back into Eq.~\eqref{eq:app_s-steady}, we obtain
\begin{eqnarray}
  s^{\rm ste}_{\bf k} 
  &=&
  -\frac{1}{2} \frac{E^V_{\bf k} - i \gamma_{\bf k}}{\left| E^V_{\bf k} - i \gamma_{\bf k} \right|}
  \frac{\NC V_{\bf k}}{E^{\rm BG}_{\bf k}}
\label{eq:app_s-steadyFIN}\;.
\end{eqnarray}
Results \eqref{eq:app_n-steadyFIN}--\eqref{eq:app_s-steadyFIN} define the standard result of the Bogoliubov excitations for vanishing dephasing. Therefore, they indeed describe how the Bogoliubov excitations behave when dephasing is included.

\bibliographystyle{elsarticle-num} 
\bibliography{ref,Seminal,Reviews,EXP,TheoKin,TheoStat}





\end{document}